\documentclass[epj,nopacs]{svjour}
\pdfoutput=1
\usepackage{graphicx}
\usepackage{atlasphysics,lineno,url}
\usepackage{fixltx2e}
\makeatletter
\g@addto@macro\bfseries{\boldmath}
\makeatother

\newcommand{\ttxval}{242.4}
\newcommand{\ttxstat}{1.7}
\newcommand{\ttxsyst}{5.5}
\newcommand{\ttxlumi}{7.5}
\newcommand{\ttxebeam}{4.2}
\newcommand{\ttxtot}{10.3}
\newcommand{\ttxrel}{4.3\,\%}

\newcommand{\dmtopval}{-0.28} 
\newcommand{\dmtoperr}{0.03}
\newcommand{\intlumi}{20.3}

\newcommand{\ttxvalw}{182.9}
\newcommand{\ttxstatw}{3.1}
\newcommand{\ttxsystw}{4.2}
\newcommand{\ttxlumiw}{3.6}
\newcommand{\ttxebeamw}{3.3}
\newcommand{\ttxtotw}{7.1}
\newcommand{\ttxrelw}{3.9\,\%}

\newcommand{\intlumiw}{4.6}

\newcommand{\LLintlumi}{20.2}
\newcommand{\LLttxval}{242.9}
\newcommand{\LLttxstat}{1.7}
\newcommand{\LLttxsyst}{5.5}
\newcommand{\LLttxlumi}{5.1}
\newcommand{\LLttxebeam}{4.2}
\newcommand{\LLttxtot}{8.8}
\newcommand{\LLttxrel}{3.6\,\%}

\newcommand{\mtopxs}{172.9}
\newcommand{\emtopxsp}{2.5}
\newcommand{\emtopxsm}{2.6}

\newcommand{\mstopexm}{177} 
\newcommand{\mstopexmlh}{175} 

\newcommand{\xtt}{\mbox{$\sigma_{\ttbar}$}}
\newcommand{\xfid}{\mbox{$\sigma_{\ttbar}^{\rm fid}$}}
\newcommand{\sxv}{\mbox{$\sqrt{s}=8$\,\TeV}}
\newcommand{\sxw}{\mbox{$\sqrt{s}=7$\,\TeV}}
\newcommand{\sxvt}{$\sqrt{s}=8$\,\TeV}
\newcommand{\sxwt}{$\sqrt{s}=7$\,\TeV}

\newcommand{\xtttheo}{\mbox{$\sigma^{\rm theo}_{\ttbar}$}}
\newcommand{\mtop}{\mbox{$m_t$}}
\newcommand{\mtpole}{\mbox{$m_t^{\rm pole}$}}
\newcommand{\mtref}{\mbox{$m_t^{\rm ref}$}}
\newcommand{\epsem}{\mbox{$\epsilon_{e\mu}$}}
\newcommand{\aem}{\mbox{$A_{e\mu}$}}
\newcommand{\gem}{\mbox{$G_{e\mu}$}}

\newcommand{\epsb}{\mbox{$\epsilon_{b}$}}
\newcommand{\epsbb}{\mbox{$\epsilon_{bb}$}}
\newcommand{\cb}{\mbox{$C_b$}}

\newcommand{\nib}{\mbox{$N_1^{\rm bkg}$}}
\newcommand{\niib}{\mbox{$N_2^{\rm bkg}$}}

\newcommand{\rtt}{\mbox{$R_{32}$}}

\newcommand{\nubar}{\mbox{$\overline{\nu}$}}
\newcommand{\rxtt}{\mbox{$R_{\ttbar}$}}
\newcommand{\dmtop}{\mbox{${\rm d}\xtt/{\rm d}\mtop$}}

\newcommand{\ndu}[2]{\mbox{$N_#1^{\rm #2}$}}
\newcommand{\njfakeos}{\ndu{j}{mis,OS}}
\newcommand{\njfakess}{\ndu{j}{mis,SS}}

\newcommand{\gauss}[3]{G(#1|#2,#3)}
\newcommand{\xttp}{\mbox{$\sigma^{\prime}_{\ttbar}$}}
\newcommand{\xerrexp}{\mbox{$\rho_{\rm exp}$}}
\newcommand{\xerrtheo}{\mbox{$\rho^\pm_{\rm theo}$}}

\newcommand{\stopi}{\mbox{$\tilde{t}_1$}}
\newcommand{\xiz}{\mbox{$\tilde{\chi}_1^0$}}
\newcommand{\mstop}{\mbox{$m_{\tilde{t}_1}$}}
\newcommand{\xstop}{\mbox{$\sigma_{\tilde{t}_1\tilde{t}_1}$}}
\newcommand{\rstop}{\mbox{$R_{\tilde{t}_1\tilde{t}_1}$}}
\newcommand{\mxiz}{\mbox{$m_{\tilde{\chi}_1^0}$}}

\newcommand{\ajinst}[4]{ATLAS Collaboration, JINST {\bf #2} #4 (#3)}
\newcommand{\aepjc}[5]{ATLAS Collaboration, Eur.\ Phys.\ J.\ C {\bf #2} #4 (#3), [arXiv:#5]}

\newcommand{\ajhep}[5]{ATLAS Collaboration, J.\ High Energy Phys.\ {\bf #2} #4 (#3), [arXiv:#5]}

\newcommand{\apubref}[3]{ATLAS Collaboration, {\em #1}, ATL-PHYS-PUB-#2, 
\hfill\linebreak[0]\mbox{\url{http://cdsweb.cern.ch/record/#3}}}
\newcommand{\aconfref}[3]{ATLAS Collaboration, {\em #1}, ATLAS-CONF-#2, 
\hfill\linebreak[0]\mbox{\url{http://cdsweb.cern.ch/record/#3}}}

\newcommand{\prd}[6]{#1, Phys.\ Rev.\ D {\bf #3} #5 (#4), [arXiv:#6]}
\newcommand{\prdna}[5]{#1, Phys.\ Rev.\ D {\bf #3} #5 (#4)}
\newcommand{\plb}[6]{#1, Phys.\ Lett.\ B {\bf #3} #5 (#4), [arXiv:#6]}
\newcommand{\plbna}[5]{#1, Phys.\ Lett.\ B {\bf #3} #5 (#4)}
\newcommand{\prl}[6]{#1, Phys.\ Rev.\ Lett.\ {\bf #3} #5 (#4), [arXiv:#6]}
\newcommand{\jhep}[6]{#1, J.\ High Energy Phys.\ {\bf #3} #5 (#4), [arXiv:#6]}
\newcommand{\epjc}[6]{#1, Eur.\ Phys.\ J.\ C {\bf #3} #5 (#4), [arXiv:#6]}

\newcommand{\npb}[6]{#1, Nucl.\ Phys.\ B {\bf #3} #5 (#4), [arXiv:#6]}
\newcommand{\npbna}[5]{#1, Nucl.\ Phys.\ B {\bf #3} #5 (#4)}
\newcommand{\arxiv}[3]{#1, [arXiv:#3]}
\newcommand{\nima}[5]{#1, Nucl.\ Instr.\ Meth.\ A {\bf #3} #5 (#4)}
\newcommand{\cpc}[6]{#1, Comp.\ Phys.\ Comm.\ {\bf #3} #5 (#4), [arXiv:#6]}
\newcommand{\zphysc}[6]{#1, Z.\ Phys.\ C {\bf #3} #5 (#4), [arXiv:#6]}
\newcommand{\npprocsup}[6]{#1, Nucl.\ Phys.\ Proc.\ Suppl.\ {\bf #3} #5 (#4), [arXiv:#6]}
\newcommand{\ijmpa}[6]{#1, Int.\ J.\ Mod.\ Phys.\ A {\bf #3} #5 (#4), [arXiv:#6]}
\newcommand{\jpgna}[5]{#1, J.\ Phys.\ G {\bf #3} #5 (#4)}
\newcommand{\phrep}[6]{#1, Phys.\ Rept.\ {\bf #3} #5 (#4), [arXiv:#6]}

\newcommand{\singlefigure}[2]{
\parbox{91mm}{
\includegraphics[width=90mm]{#1}
\vspace{-7mm}
\center{(#2)}
}
}
\newcommand{\singlefigurex}[1]{
\parbox{91mm}{
\includegraphics[width=90mm]{#1}
\vspace{-3mm}
}
}
\newcommand{\singlefigurexw}[3]{
\hspace{#3}
\parbox{#2}{
\includegraphics[width=#2]{#1}
\vspace{-3mm}
}
}
\newcommand{\splitfigure}[4]{
\parbox{85mm}{
\includegraphics[width=78mm]{#1}
\vspace{-7mm}

\center{(#3)}
}
\parbox{85mm}{
\includegraphics[width=78mm]{#2}
\vspace{-7mm}

\center{(#4)}
}
\\
}

\newcommand{\rhtitle}{Measurement of the \ttbar\ production cross-section
using $e\mu$ events with $b$-tagged jets
in $pp$ collisions at $\sqrt{s}$ = 7 and 8\,\TeV\ with the ATLAS detector}

\newcommand{\rhabstract}{
The inclusive top quark pair (\ttbar) production cross-section \xtt\ has been 
measured in proton--proton collisions at \sxw\ and \sxv\ with the ATLAS
experiment at the LHC, using \ttbar\ events 
with an opposite-charge $e\mu$ pair in the final state. The measurement was 
performed with the 2011 7\,\TeV\ dataset corresponding to an 
integrated luminosity of \intlumiw\,\ifb\ and the 2012 8\,\TeV\ dataset of 
\intlumi\,\ifb. The numbers of events 
with exactly one and exactly two $b$-tagged jets were counted and used to 
simultaneously determine \xtt\ and the efficiency to reconstruct and
$b$-tag a jet from a top quark decay, thereby minimising the associated 
systematic uncertainties. The cross-section was measured to be:
\begin{eqnarray*}
\xtt & = & \ttxvalw\pm\ttxstatw\pm\ttxsystw\pm\ttxlumiw\pm\ttxebeamw\,\rm pb\ (\sxw) {\rm\ and} \\
\xtt & = & \ttxval\pm\ttxstat\pm\ttxsyst\pm\ttxlumi\pm\ttxebeam\,\rm pb\ (\sxv),
\end{eqnarray*}
where the four uncertainties arise from data statistics, experimental and
theoretical  systematic effects, knowledge of the integrated luminosity 
and of the LHC beam energy. 
The results are consistent with recent theoretical QCD calculations
at next-to-next-to-leading order.
Fiducial measurements corresponding to the experimental acceptance of the
leptons are also reported, together with the ratio of cross-sections
measured at the two centre-of-mass energies. The inclusive cross-section
results were used to
determine the top quark pole mass via the dependence of the 
theoretically predicted cross-section on \mtpole\, giving a result of
$\mtpole=\mtopxs^{+\emtopxsp}_{-\emtopxsm}$\,\GeV. 
By looking for an excess of \ttbar\ 
production with respect to the QCD prediction, the results were also used
to place limits on the pair-production of supersymmetric top squarks $\stopi$
with masses close to the top quark mass, 
decaying via $\stopi\rightarrow t\xiz$ to predominantly right-handed
top quarks and a light neutralino \xiz, the lightest supersymmetric particle.
Top squarks with masses between the top quark mass and \mstopexm\,\GeV\ are
excluded at the 95\,\% confidence level.
}

\newcommand{\rhabstractPP}{
The inclusive top quark pair (\ttbar) production cross-section \xtt\ has been 
measured in proton--proton collisions at \sxw\ and \sxv\ with the ATLAS
experiment at the LHC, using \ttbar\ events 
with an opposite-charge $e\mu$ pair in the final state. The measurement was 
performed with the 2011 7\,\TeV\ dataset corresponding to an 
integrated luminosity of \intlumiw\,\ifb\ and the 2012 8\,\TeV\ dataset of 
\intlumi\,\ifb. The numbers of events 
with exactly one and exactly two $b$-tagged jets were counted and used to 
simultaneously determine \xtt\ and the efficiency to reconstruct and
$b$-tag a jet from a top quark decay, thereby minimising the associated 
systematic uncertainties. The cross-section was measured to be:
\begin{eqnarray*}
\xtt & = & \ttxvalw\pm\ttxstatw\pm\ttxsystw\pm\ttxlumiw\pm\ttxebeamw\,\rm pb\ (\sxw) {\rm\ and} \\
\xtt & = & \LLttxval\pm\LLttxstat\pm\LLttxsyst\pm\LLttxlumi\pm\LLttxebeam\,\rm pb\ ^{*}\ (\sxv) ,
\end{eqnarray*}
where the four uncertainties arise from data statistics, experimental and
theoretical  systematic effects, knowledge of the integrated luminosity 
and of the LHC beam energy. 
The results are consistent with recent theoretical QCD calculations
at next-to-next-to-leading order.
Fiducial measurements corresponding to the experimental acceptance of the
leptons are also reported, together with the ratio of cross-sections
measured at the two centre-of-mass energies. The inclusive cross-section
results were used to
determine the top quark pole mass via the dependence of the 
theoretically predicted cross-section on \mtpole\, giving a result of
$\mtpole=\mtopxs^{+\emtopxsp}_{-\emtopxsm}$\,\GeV. 
By looking for an excess of \ttbar\ 
production with respect to the QCD prediction, the results were also used
to place limits on the pair-production of supersymmetric top squarks $\stopi$
with masses close to the top quark mass, 
decaying via $\stopi\rightarrow t\xiz$ to predominantly right-handed
top quarks and a light neutralino \xiz, the lightest supersymmetric particle.
Top squarks with masses between the top quark mass and \mstopexm\,\GeV\ are
excluded at the 95\,\% confidence level.
}

\usepackage{preprintcover}  
\PreprintCoverPaperTitle{\rhtitle}  
\PreprintIdNumber{CERN-PH-EP-2014-124}  
\PreprintCoverAbstract{\rhabstractPP}  
\PreprintJournalName{Eur.~Phys.~J.~C}  

\usepackage[hyperindex,breaklinks]{hyperref} 

\begin{document}
\title{\rhtitle}
\author{The ATLAS Collaboration
}                     
%
%
\institute{CERN, CH-1211 Geneva 23. Switzerland}
%
\date{Received: DATE / Accepted: DATE}
%
\abstract{\rhabstract
\PACS{
      {PACS-key}{discribing text of that key}   \and
      {PACS-key}{discribing text of that key}
     } 
} 
\titlerunning{Measurement of the \ttbar\ production cross-section using $e\mu$ events with $b$-tagged jets}
\maketitle
%
\section{Introduction}\label{s:intro}

The top quark is the heaviest known fundamental particle, with a mass (\mtop)
that is much larger than any of the other quarks, and close to the scale
of electroweak symmetry breaking. The study of its production and decay 
properties forms a core part of the ATLAS physics programme
at the CERN Large Hadron Collider (LHC). At the LHC, top
quarks are primarily produced in quark--antiquark pairs (\ttbar), and the
precise prediction of the corresponding inclusive cross-section (\xtt) is a 
substantial  challenge for quantum chromodynamics (QCD) calculational 
techniques. Precise measurements
of \xtt\ are sensitive to the gluon parton distribution function (PDF),  
the top quark mass, and potential enhancements of the cross-section
due to physics beyond the Standard Model.

Within the Standard Model (SM), the top quark decays almost exclusively to a 
$W$ boson and a $b$ quark, so the final-state topologies in \ttbar\ production
are governed by the decay modes of the two $W$ bosons. This paper
describes a measurement in the dileptonic $e\mu$ channel, 
$\ttbar\rightarrow W^+bW^-\bar{b}\rightarrow e^\pm\mu^\mp\nu\nubar\bbbar$, 
selecting events with an $e\mu$ pair with opposite-sign electric
charges,\footnote{Charge-conjugate modes are implied
throughout.} and one or two hadronic jets from the $b$ quarks.
Jets originating from $b$ quarks were identified (`tagged') using a $b$-tagging
algorithm exploiting the long lifetime, high decay multiplicity, hard 
fragmentation and high mass of
$B$ hadrons. The rates of events with an $e\mu$ pair and one or two tagged 
$b$-jets were used to measure simultaneously the \ttbar\ production 
cross-section and the combined  probability to reconstruct and $b$-tag a
jet from a top quark decay. Events with electrons or muons produced
via leptonic $\tau$ decays
$t\rightarrow Wb\rightarrow\tau\nu b\rightarrow e/\mu\nu\nu\nu b$, were included as part of the
\ttbar\ signal. 

The main background is $Wt$, the associated
production of a $W$ boson and a single top quark. Other background 
contributions arise from
$Z\rightarrow\tau\tau\rightarrow e\mu$+jets ($+4\nu$) production, diboson+jets 
production and events where at least one reconstructed lepton does not arise 
from a $W$ or $Z$ boson decay.

Theoretical predictions for \xtt\ are described in Sect.~\ref{s:theo}, 
followed by the data and Monte Carlo (MC) simulation samples in 
Sect.~\ref{s:dmc}, the object and event selection in 
Sect.~\ref{s:objev}, and the extraction of the \ttbar\ cross-section in
Sect.~\ref{s:ext}. Systematic uncertainties are discussed in 
Sect.~\ref{s:syst}, the results, including fiducial cross-section
measurements, the extraction of the top quark mass from the measured 
cross-section and a limit on the production of supersymmetric top squa\-rks, are 
given in Sect.~\ref{s:res}, 
and conclusions are drawn in Sect.~\ref{s:conc}.

\section{Theoretical cross-section predictions}\label{s:theo}

Calculations of \xtt\ for hadron collisions are
now available at full next-to-next-to-leading-order (NNLO) accuracy
in the strong coupling constant $\alpha_{\rm s}$, including
the resummation of next-to-next-to-leading logarithmic (NNLL) soft gluon
terms \cite{topxtheo,topxtheot}. At a  centre-of-mass energy of \sxw\ and
 assuming $\mtop=172.5$\,\GeV, these calculations give  a prediction of
$177.3\pm 9.0\,^{+4.6}_{-6.0}$\,pb, where the first uncertainty is due to 
PDF and $\alpha_{\rm s}$ uncertainties, 
and the second to QCD scale uncertainties. The corresponding prediction
at \sxv\ is $252.9\pm 11.7\,^{+6.4}_{-8.6}$\,pb. These values were
calculated using the {\tt top++ 2.0} program \cite{toppp}. The PDF
and $\alpha_{\rm s}$ uncertainties were calculated using the PDF4LHC prescription
\cite{pdflhc} with the MSTW2008 68\,\% CL NNLO \cite{mstwnnlo}, 
CT10 NNLO \cite{cttenpdf,cttennnlo} and NNPDF2.3 5f FFN \cite{nnpdfffn} 
PDF sets, and added in quadrature to the QCD scale uncertainty. The latter was
 obtained from the envelope of predictions with the renormalisation and 
factorisation scales varied independently by factors of two up and down from 
their default values of \mtop, whilst never letting them differ by more than a 
factor of two. The ratio of cross-sections at \sxvt\ and \sxwt\ is predicted
to be $1.430\pm 0.013$ (PDF+$\alpha_{\rm s}$) $\pm 0.001$ (QCD scale). The
total relative  uncertainty is only 0.9\,\%, as the cross-section 
uncertainties at the two centre-of-mass energies are highly correlated.

The NNLO+NNLL cross-section values are about 3\,\% larger than the exact 
NNLO predictions, as implemented in {\tt Hathor} 1.5 \cite{hathor}. 
For comparison, the 
corresponding next-to-leading-order (NLO) predictions, also calculated using
{\tt top++ 2.0} with the same set of PDFs, are $157\pm 12\pm 24$\,pb  
at \sxwt\ and $225\pm 16\pm 29$\,pb 
at \sxvt, where again the first quoted uncertainties are due to PDF and
$\alpha_{\rm s}$ uncertainties, and the second to QCD scale uncertainties. The
total uncertainties of the NLO predictions are approximately 15\,\%, about 
three times larger than the NNLO+NNLL calculation uncertainties quoted above.

\section{Data and simulated samples}\label{s:dmc}

The ATLAS detector \cite{atlasdet} at the LHC covers nearly the entire
solid angle around the collision point, and consists of an inner tracking
detector surrounded by a thin superconducting solenoid magnet producing
a 2\,T axial magnetic field, electromagnetic and hadronic calorimeters, and 
an external muon spectrometer incorporating three large toroid magnet 
assemblies. The inner detector consists of a high-granularity silicon pixel
detector and a silicon microstrip tracker, together providing precision
tracking in the pseudorapidity\footnote{ATLAS uses a right-handed 
coordinate system with its origin at
the nominal interaction point in the centre of the detector, and the $z$ axis
along the beam line. Pseudorapidity is defined in terms of the polar angle
$\theta$ as $\eta=-\ln\tan{\theta/2}$, and transverse momentum and energy
are defined relative to the beamline as $\pt=p\sin\theta$ and
$\et=E\sin\theta$. The azimuthal angle around the beam line is denoted by 
$\phi$, and distances in $(\eta,\phi)$ space by 
$\Delta R=\sqrt{(\Delta\eta^2)+(\Delta\phi)^2}$.
} range $|\eta|<2.5$,
complemented by a transition radiation
tracker providing tracking and electron identification information for
$|\eta|<2.0$. A lead/liquid-argon (LAr) electromagnetic calorimeter covers the
region $|\eta|<3.2$, and hadronic calorimetry is provided by 
steel/scintillator tile calorimeters for $|\eta|<1.7$ and copper/LAr
hadronic endcap calorimeters. The forward region is covered by additional
LAr calorimeters with copper and tungsten absorbers. The muon
spectrometer consists of precision tracking chambers covering the region
$|\eta|<2.7$, and separate trigger chambers covering $|\eta|<2.4$. A 
three-level trigger system, using custom hardware followed by two
software-based levels, is used to reduce the event rate to about 400\,Hz 
for offline storage.

The analysis was performed on the ATLAS 2011--2012 proton--proton 
collision data sample, corresponding to inte\-g\-rated 
luminosities of \intlumiw\,\ifb at \sxwt\ and \intlumi\,\ifb\ at \sxvt\ 
after the application of detector status and data quality requirements. Events
were required to pass either a single-electron or single-muon trigger,
with thresholds chosen in each case such that the efficiency plateau is 
reached for leptons with $\pt>25$\,\GeV\
passing offline selections. Due to the high instantaneous luminosities
achieved by the LHC, each triggered event also includes 
the signals from on average about 9 (\sxwt) or 20 (\sxvt) additional 
inelastic $pp$ collisions in the same bunch crossing (known as pileup).

Monte Carlo simulated event samples were used to develop the analysis, 
to compare to the data and to evaluate signal and background efficiencies and
uncertainties. Samples were processed either through the full ATLAS
detector simulation \cite{simul} based on GEANT4 \cite{geant4}, or 
through a faster simulation making use of parameterised showers in
the calorimeters \cite{fastsim}. 
Additional simulated $pp$ collisions generated either with 
{\sc Pythia6} \cite{pythia6} (for \sxw\ simulation) or {\sc Pythia8} 
\cite{pythia8} (for \sxv) were overlaid to simulate the effects of both 
in- and out-of-time pileup, from additional $pp$ collisions in the same and 
nearby  bunch crossings. All simulated events were then processed using the
same reconstruction algorithms and analysis chain as the data.  Small
corrections were applied to lepton trigger and selection efficiencies
to better model the performance seen in data, as discussed further in 
Sect.~\ref{s:syst}.

The baseline \ttbar\ full simulation sample was produced using 
the NLO matrix element generator {\sc Powheg} \cite{powheg} interfaced to
{\sc Pythia6} \cite{pythia6} with the Perugia 2011C tune (P2011C) 
\cite{perugia} for parton shower, fragmentation and underlying event modelling,
and CT10 PDFs \cite{cttenpdf}, and included all
\ttbar\ final states involving at least one lepton. 
The $W\rightarrow\ell\nu$ branching ratio was set to 
the SM expectation of $0.1082$ \cite{wbrpdg}, and
\mtop\ was set to 172.5\,\GeV. Alternative \ttbar\ samples were produced 
with the NLO generator {\sc MC@NLO} \cite{mcatnlo} interfaced
to {\sc Herwig} \cite{herwig} 
with {\sc Jimmy} \cite{jimmy} for the underlying event modelling,
with the ATLAS AUET2 \cite{auet} tune and CT10 PDFs; and with 
the leading-order (LO) multileg generator {\sc Alpgen} \cite{alpgen}
interfaced to either {\sc Pythia6} or {\sc Herwig} and {\sc Jimmy}, with
the CTEQ6L1 PDFs \cite{ctsixpdf}. These samples were all normalised to the
NNLO+NNLL cross-section predictions given in Sect.~\ref{s:theo} when 
comparing simulation with data.

Backgrounds were classified into two types: those with two real prompt
leptons from $W$ or $Z$ boson decays (including those produced 
via leptonic $\tau$ decays), and those
where at least one of the reconstructed lepton candidates is
misidentified, 
i.e. a non-prompt lepton from the decay of a bottom or charm hadron, 
an electron from a photon conversion,
hadronic jet activity misidentified
as an electron, or a muon produced from an in-flight decay of a pion or
kaon. The first category with two prompt leptons
includes $Wt$ single top production, modelled using 
{\sc Powheg + Pythia6} \cite{powwtdr} with the CT10 PDFs and the P2011C tune; 
$Z\rightarrow\tau\tau$+jets modelled using {\sc Alpgen + Herwig + Jimmy} 
(\sxwt) or
{\sc Alpgen + Pythia6} including LO matrix elements for $Z\bbbar$ production,
with CTEQ6L1 PDFs; and diboson ($WW$, $WZ$, $ZZ$) 
production in association with
jets, modelled using {\sc Alpgen + Herwig + Jimmy}. 
The $Wt$ background was normalised to approximate NNLO cross-sections of 
$15.7\pm 1.2$\,pb at \sxw\ and $22.4\pm 1.5$\,pb at \sxv, determined as in 
Ref. \cite{Wttheoxsec}. The inclusive $Z$ cross-sections were set to the
NNLO predictions from FEWZ \cite{fewz}, but the normalisation of 
$Z\rightarrow\tau\tau\rightarrow e\mu 4\nu$ backgrounds with $b$-tagged jets 
were determined from data as described in Sect.~\ref{ss:back}.
The diboson background was
normalised to the NLO QCD inclusive cross-section 
predictions calculated with MCFM \cite{dibmcfm}. Production of \ttbar\ in
association with a $W$ or $Z$ boson, which contributes to the
sample with same-sign leptons, was simulated with {\sc Madgraph} 
\cite{madgraph} interfaced
to {\sc Pythia} with CTEQ6L1 PDFs, and normalised to NLO cross-section
predictions \cite{nlottwz}.

Backgrounds with one real and
one misidentified lepton include \ttbar\ events with one hadronically decaying
$W$; $W$+jets production, modelled as for $Z$+jets; $W\gamma$+jets,
modelled with {\sc Sherpa} \cite{sherpa} with CT10 PDFs; 
and $t$-chan\-nel single top production, modelled using {\sc AcerMC} 
\cite{acer} interfaced to {\sc Pythia6} with CTEQ6L1 PDFs.
Other backgrounds, including processes with two misidentified leptons, 
are negligible after the event selections used in this analysis.

\section{Object and event selection}\label{s:objev}

The analysis makes use of reconstructed electrons, muons and $b$-tagged jets.
Electron candidates were reconstructed from an isolated electromagnetic 
calorimeter energy  deposit matched to an inner detector track and
passing tight identification requirements \cite{elecperf}, with transverse
energy $\et>25$\,\GeV\ and pseudorapidity $|\eta|<2.47$. Electron candidates 
within the
transition region between the barrel and endcap electromagnetic calorimeters,
$1.37<|\eta|<1.52$, were removed. Isolation requirements were used to reduce 
background from non-prompt electrons. The calori\-meter 
transverse energy within a cone of size
$\Delta R=0.2$ and the scalar sum of track $\pt$ within a cone
of size $\Delta R=0.3$,
in each case excluding the contribution from the electron itself,
were each required to be smaller than $\et$ and $\eta$-dependent thresholds 
calibrated to separately give nominal selection
efficiencies of 98\,\% for prompt electrons from $Z\rightarrow ee$ decays.

Muon candidates were reconstructed by combining mat\-ch\-ing tracks 
reconstructed in both the inner detector and muon spectrometer \cite{muperf}, 
and were required to satisfy $\pt>25$\,\GeV\ and $|\eta|<2.5$. 
In the \sxwt\ dataset, the calorimeter transverse energy within a cone of 
size $\Delta R=0.2$, excluding the energy deposited by the
muon, was required to be less than 4\,\GeV,
and the scalar sum of track $\pt$ within a cone of size $\Delta R=0.3$, 
excluding 
the muon track, was required
to be less than 2.5\,\GeV. In the \sxvt\ dataset, these isolation requirements 
were replaced by a cut $I<0.05$, where $I$ is
the ratio of the sum of track $\pt$ in a variable-sized cone of radius
$\Delta R=10\,{\rm\GeV}/\pt^\mu$ to the transverse momentum $\pt^\mu$
 of the muon \cite{mumini}. Both sets of isolation 
requirements have efficiencies of ab\-out  97\,\% 
for prompt muons from $Z\rightarrow\mu\mu$ decays.

Jets were reconstructed using the anti-$k_t$ algorithm \cite{antikt} 
with radius parameter $R=0.4$, starting from calorimeter energy
clusters calibrated at the electromagnetic energy scale for the \sxwt\ 
dataset, or using the local cluster weighting method for \sxv\ \cite{jesx}.
Jets were calibrated using an energy- and $\eta$-dependent
simulation-based calibration scheme, with in-situ corrections 
based on data, and were required to satisfy
$\pt>25$\,\GeV\ and $|\eta|<2.5$. To suppress the contribution from low-$\pt$
jets originating from pileup interactions, a jet vertex fraction requirement 
was applied: at \sxw\ jets were required to have at least 75\,\% 
of the scalar sum of the $\pt$
of tracks associated with the jet coming from tracks associated with the 
event primary vertex. The latter was 
defined as the reconstructed vertex with the highest sum of associated 
track $\pt^2$. Motivated by the higher pileup background, in the \sxvt\ dataset
this requirement was loosened to 50\,\%, only
applied to jets with $\pt<50$\,\GeV\ and $|\eta|<2.4$, and the effects of 
pileup on the jet energy calibration were further reduced using the 
jet-area method as described in Ref. \cite{jetpile}.
Finally, to further suppress non-isolated leptons likely to have come from 
heavy-flavour decays inside jets, electrons and muons within $\Delta R=0.4$ of
selected jets were also discarded.

Jets were $b$-tagged as likely to have originated from $b$ quarks using the
MV1 algorithm, a multivariate discriminant making use of
track impact parameters and reconstructed secondary vertices 
\cite{btagcom,btagptrel}. Jets were defined to be $b$-tagged if the MV1
discriminant value was larger than a threshold (working point) corresponding 
approximately to a 70\,\% efficiency for tagging $b$-quark jets from top
decays in \ttbar\ events, with a rejection factor of about~140 
against light-quark
and gluon jets, and about five against jets originating from charm quarks.

Events were required to have at least one reconstructed primary vertex
with at least five associated tracks, and no jets failing jet quality and
timing requirements. Events with muons compatible with cosmic-ray interactions
and muons losing substantial fractions of their energy through 
 bremsstrahlung in the detector material
were also removed. A preselection requiring exactly one electron and one
muon selected as described above was then applied, with at least
one of the leptons being matched to an electron or muon object triggering
the event. Events with an
opposite-sign $e\mu$ pair constituted the main analysis sample, whilst events
with a same-sign $e\mu$ pair were used in the estimation of the background
from misidentified leptons.

\section{Extraction of the \ttbar\ cross-section}\label{s:ext}

The \ttbar\ production cross-section \xtt\
was determined by counting the numbers
of opposite-sign $e\mu$ events with exactly one ($N_1$) and exactly
two ($N_2$) $b$-tagged jets. No requirements were made on the number of 
untagged jets; such jets originate from $b$-jets from top decays which were
not tagged, and light-quark, charm-quark or gluon jets from QCD radiation. 
The two event counts can be expressed as:
\begin{eqnarray}
N_1 & = & L \xtt\ \epsem 2\epsb (1-\cb\epsb) + \nib \nonumber \\
N_2 & = & L \xtt\ \epsem \cb\epsb^2 + \niib \label{e:tags}
\end{eqnarray}
where $L$ is the integrated luminosity of the sample, \epsem\ is the 
efficiency for a \ttbar\ event to pass the opposite-sign $e\mu$ preselection
and \cb\ is a tagging correlation coefficient close to unity.
The combined probability for a jet from the quark $q$ in the $t\rightarrow Wq$ 
decay to fall within the acceptance of the detector,
be reconstructed as a jet with transverse momentum above the selection 
threshold, and be tagged as a $b$-jet, is denoted by \epsb. Although this 
quark is almost always a $b$ quark, \epsb\ thus also accounts for 
the approximately $0.2\,\%$ of top quarks that decay to $Ws$ or $Wd$ 
rather than $Wb$, slightly reducing the effective $b$-tagging efficiency. 
Furthermore, the value of \epsb\ is slightly increased by the 
small contributions to $N_1$ and $N_2$ from mistagged light-quark, charm-quark
or gluon jets from radiation in \ttbar\ events, although
more than 98\,\% of the tagged jets are expected to contain particles from
$B$-hadron decays in both the one and two $b$-tag samples.

If the decays of the two top quarks
and the subsequent reconstruction of the two $b$-tagged jets are completely
independent, the probability to tag both $b$-jets \epsbb\ is given by
$\epsbb=\epsb^2$. In practice, small correlations are present for both
kinematic and instrumental reasons, and these are taken into account via
the tagging correlation $\cb$, defined as $\cb=\epsbb/\epsb^2$ or equivalently 
$\cb=4 N^{\ttbar}_{e\mu} N_2^{\ttbar}/(N^{\ttbar}_1+2 N^{\ttbar}_2)^2$,
where $N^{\ttbar}_{e\mu}$ is the number of preselected $e\mu$ \ttbar\ events 
and $N^{\ttbar}_1$ and $N^{\ttbar}_2$ are the numbers of \ttbar\ events with 
one and two $b$-tagged jets. Values of \cb\ 
greater than one correspond to a positive correlation, where a second
jet is more likely to be selected if the first one is already selected,
whilst $\cb=1$ corresponds to no correlation. 
This correlation term also compensates for the 
effect on \epsb, $N_1$ and $N_2$ of the small number of mistagged 
charm-quark or gluon jets from radiation in the \ttbar\ events. 

Background from sources other than $\ttbar\rightarrow e\mu\nu\nubar\bbbar$
also contributes to the event counts $N_1$ and $N_2$, and is given by the terms
\nib\ and \niib. The preselection efficiency \epsem\ and tagging correlation
\cb\ were taken from \ttbar\ event simulation, and the background contributions
\nib\ and \niib\ were estimated using a combination of simulation- and 
data-based methods, allowing the two equations in Eq.~(\ref{e:tags}) to be 
solved numerically yielding \xtt\ and \epsb.

\begin{table*}[htp]
\caption{\label{t:evtcount}Observed numbers of opposite-sign $e\mu$ events
with one and two $b$-tagged jets ($N_1$ and $N_2$) for each data sample, 
together with the estimates of backgrounds and associated total
uncertainties described in Sect.~\ref{s:syst}.}
\centering
\begin{tabular}{lcccc}
\hline\noalign{\smallskip}
 & \multicolumn{2}{c}{\sxw} & \multicolumn{2}{c}{\sxv} \\
Event counts & $N_1$ & $N_2$ & $N_1$ & $N_2$ \\
\noalign{\smallskip}\hline\noalign{\smallskip}
Data & 3527 & 2073 & 21666 & 11739 \\
\noalign{\smallskip}\hline\noalign{\smallskip}
$Wt$ single top & $326\pm 36$ & $53\pm 14$ & $2050\pm 210$ & $360\pm 120$ \\
Dibosons & $19\pm 5$ & $0.5\pm 0.1$ & $120\pm 30$ & $3\pm 1$ \\
$Z(\rightarrow\tau\tau\rightarrow e\mu)$+jets & $28\pm 2$ & $1.8\pm 0.5$ & $210\pm 5$ & $7\pm 1$ \\
Misidentified leptons & $27\pm 13$ & $15\pm 8$ & $210\pm 66$ & $95\pm 29$ \\
\noalign{\smallskip}\hline\noalign{\smallskip}
Total background & $400\pm 40$ & $70\pm 16$ & $2590\pm 230$ & $460\pm 130$ \\
\noalign{\smallskip}\hline
\end{tabular}
\end{table*}

A total of 11796 events passed the $e\mu$ opposite-sign preselection in \sxw\
data, and 66453 in \sxv\ data.
Table~\ref{t:evtcount} shows the number of events with one and two $b$-tagged
jets, together with the estimates of non-\ttbar\ background and their 
systematic uncertainties discussed in detail in Sect.~\ref{ss:back} below. 
The samples with one $b$-tagged jet are expected to be about 89\,\% pure in
\ttbar\ events, with the dominant background coming from $Wt$ single top
production, and smaller contributions from events with misidentified
leptons, $Z$+jets and dibosons.
The samples with two $b$-tagged jets are expected to be about 96\% pure in 
\ttbar\ events, with $Wt$ production again being the dominant background. 

Distributions of the number of $b$-tagged jets in opposite-sign $e\mu$ events
are shown in Fig.~\ref{f:btags}, and compared to the expectations with 
several \ttbar\ simulation samples. The histogram bins with one and two 
$b$-tagged jets correspond to the data event counts shown in 
Table~\ref{t:evtcount}.
Distributions of the number of jets, the $b$-tagged jet $\pt$, 
and the electron and muon $|\eta|$ and $\pt$
are shown for opposite-sign $e\mu$ events with at least one $b$-tagged jet in
Fig.~\ref{f:dmcw} (\sxwt) and Fig.~\ref{f:dmcv} (\sxvt), 
with the simulation normalised to the same number of entries as the data.
The lepton $|\eta|$ distributions reflect the 
differing acceptances and efficiencies for electrons 
and muons, in particular the calorimeter transition region at 
$1.37<|\eta|<1.52$. In general, the agreement between data and 
simulation is good, within the range of predictions from the different 
\ttbar\ simulation samples.

\begin{figure}[htp]
\singlefigure{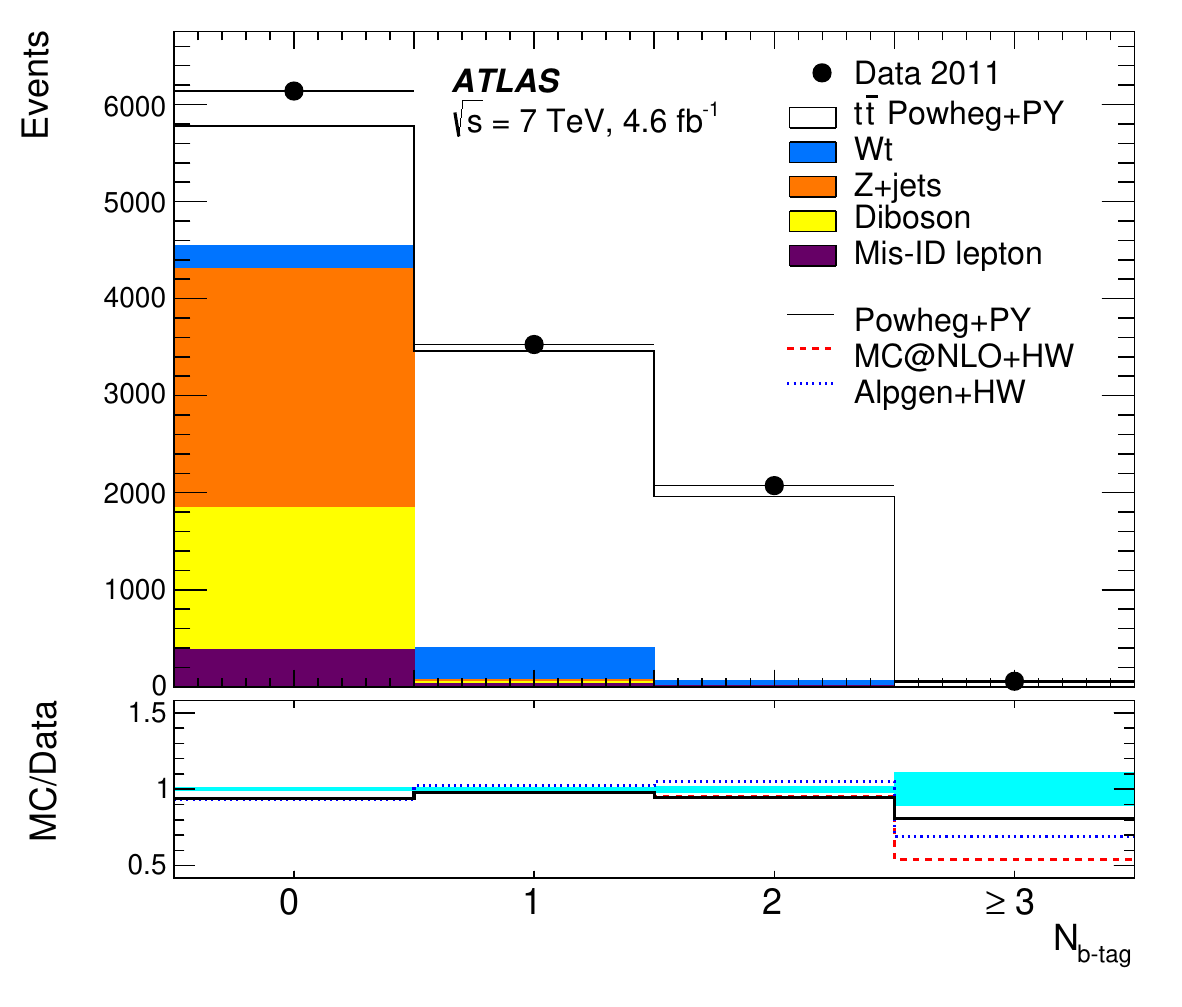}{a}
\singlefigure{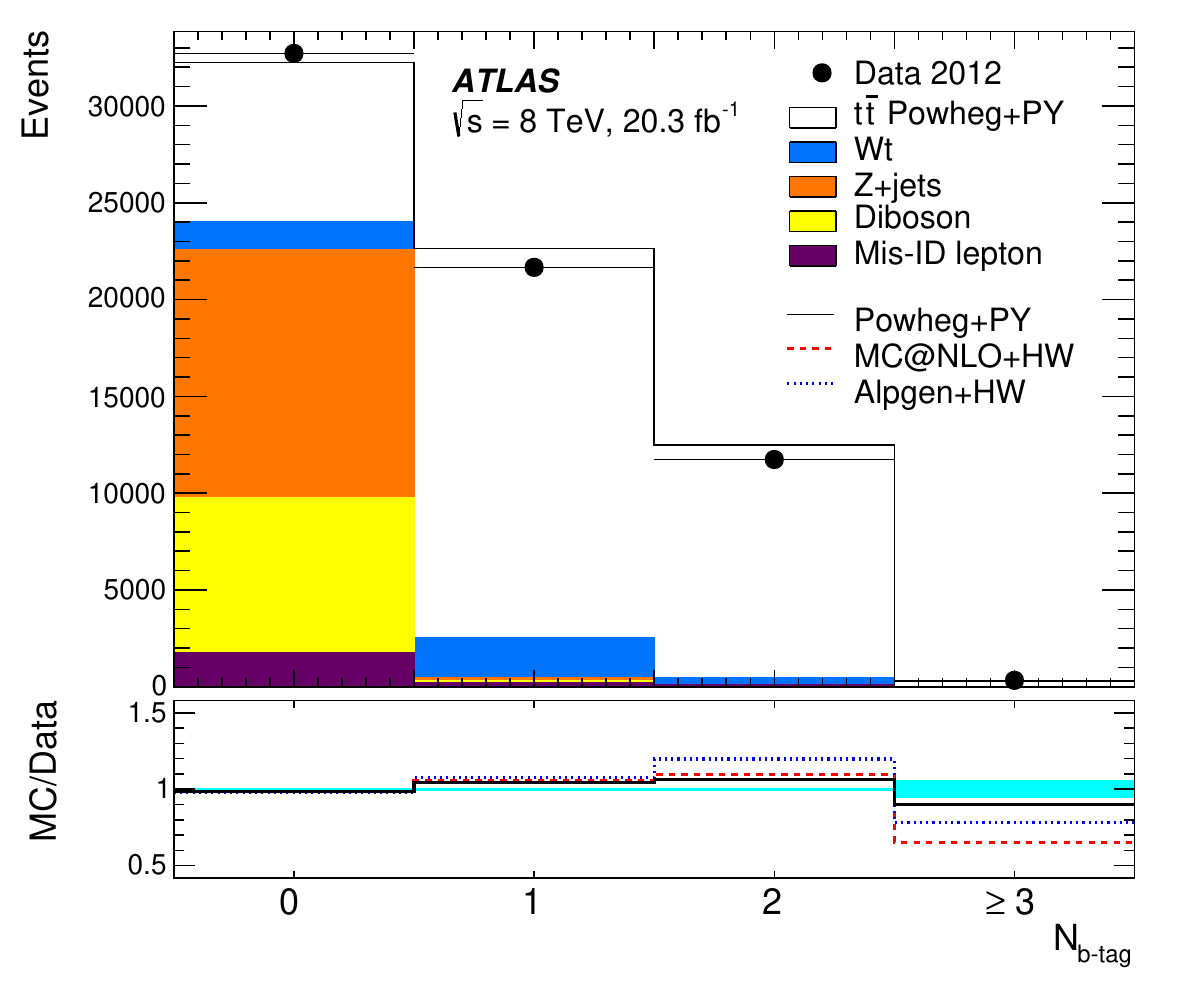}{b}
\caption{\label{f:btags}Distributions of  the number of $b$-tagged jets
in preselected opposite-sign $e\mu$ events in (a) \sxw\ and (b) \sxv\ data.
The data are shown compared
to the expectation from simulation, broken down into contributions from
\ttbar, $Wt$ single top, $Z$+jets, dibosons, and events with misidentified 
electrons or muons, normalised to the same integrated luminosity as the data.
The lower parts of the figure show the ratios of simulation to data, using
various \ttbar\ signal samples generated with {\sc Powheg + Pythia6} (PY),
{\sc MC@NLO + Herwig} (HW) and {\sc Alpgen + Herwig},
and with the cyan band indicating the statistical uncertainty.}
\end{figure}

\begin{figure*}
\splitfigure{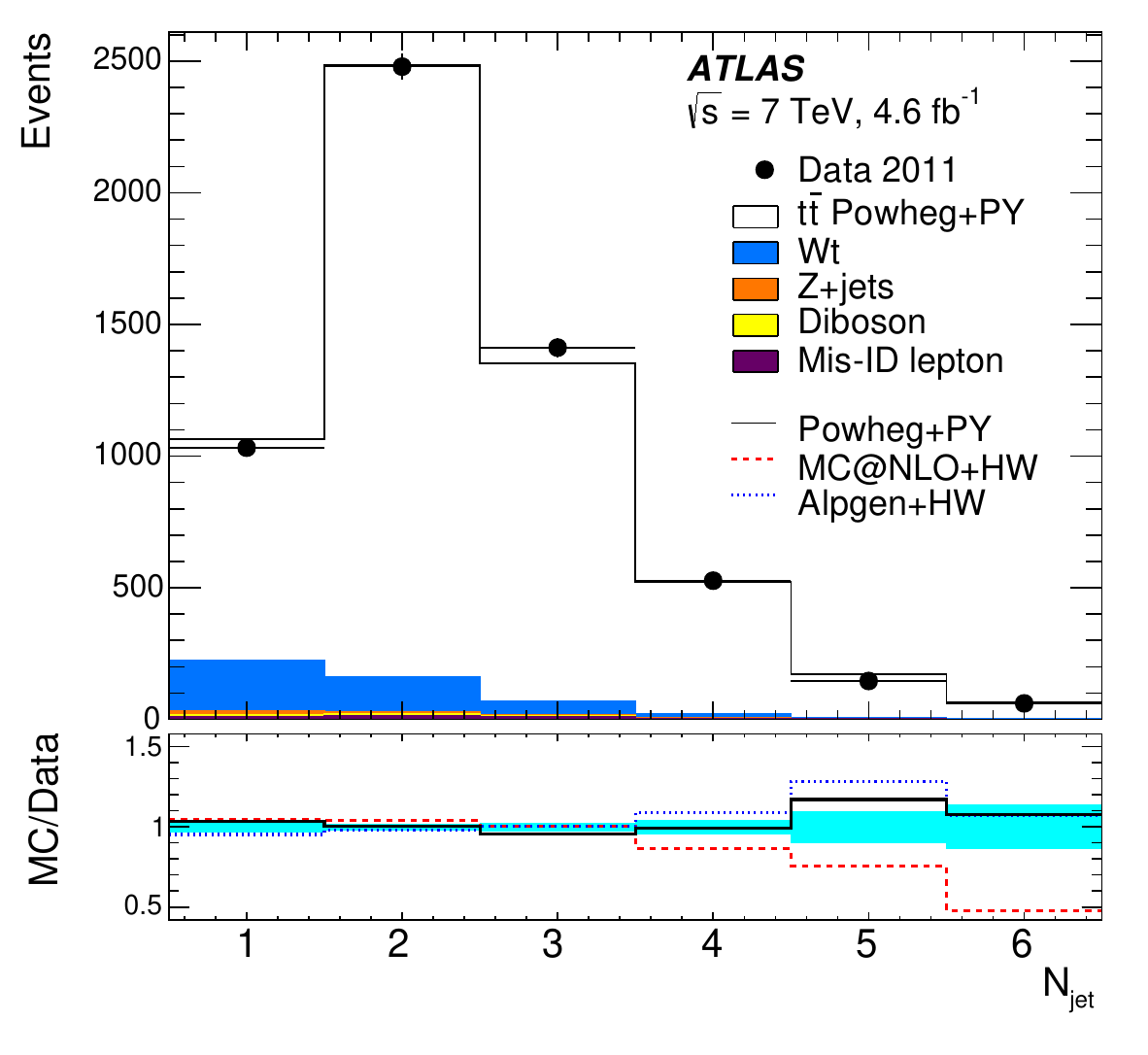}{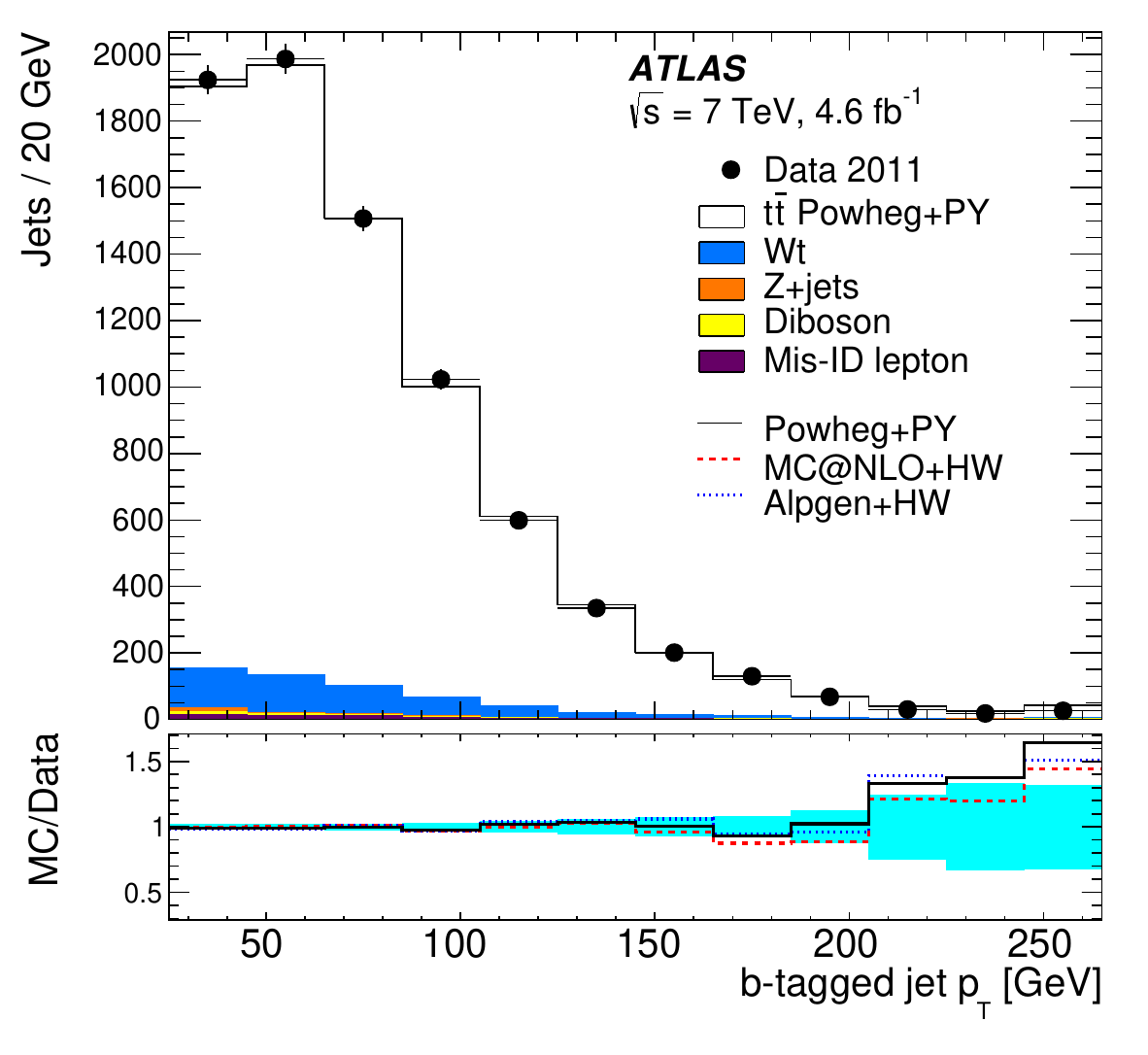}{a}{b}
\splitfigure{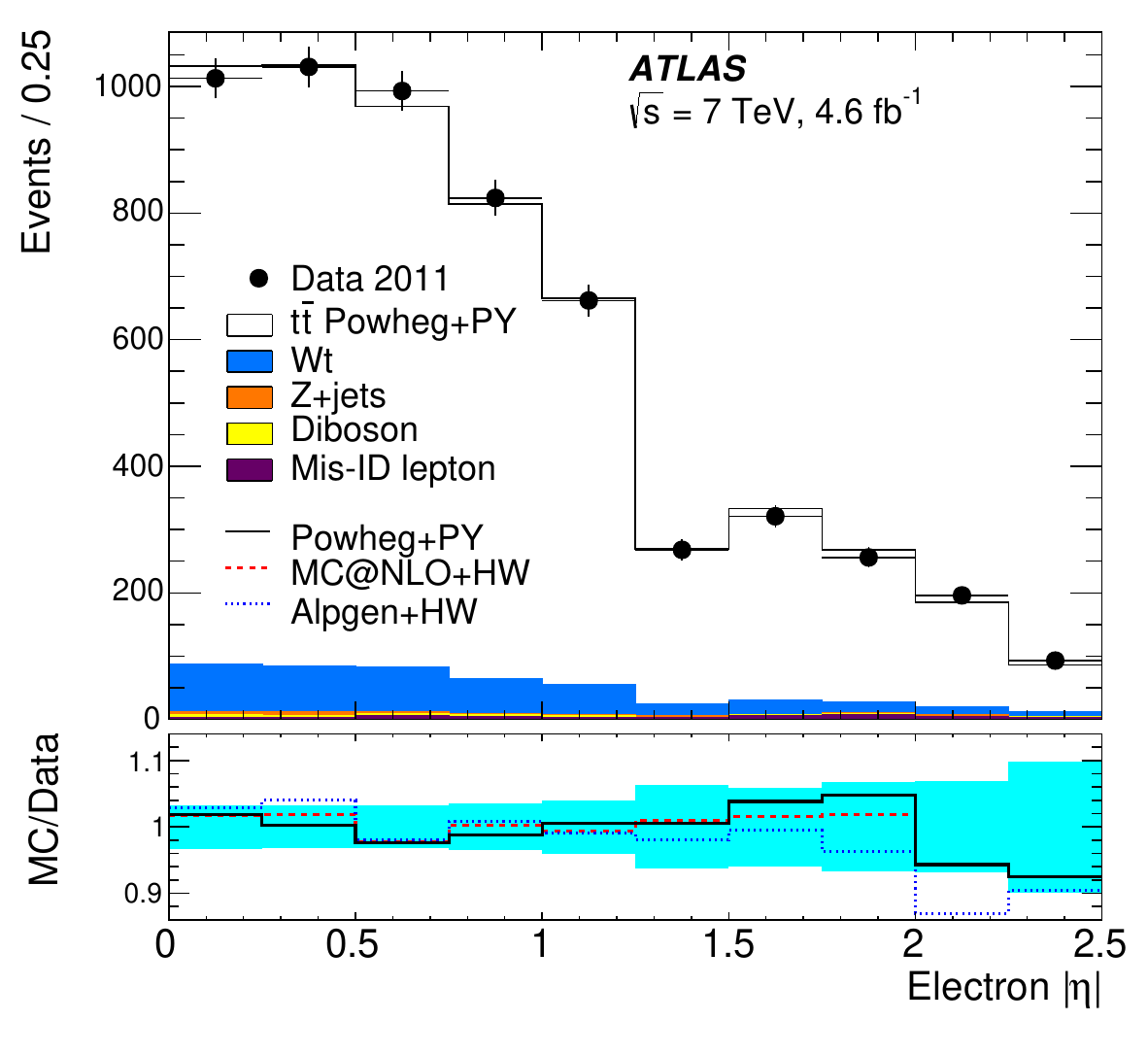}{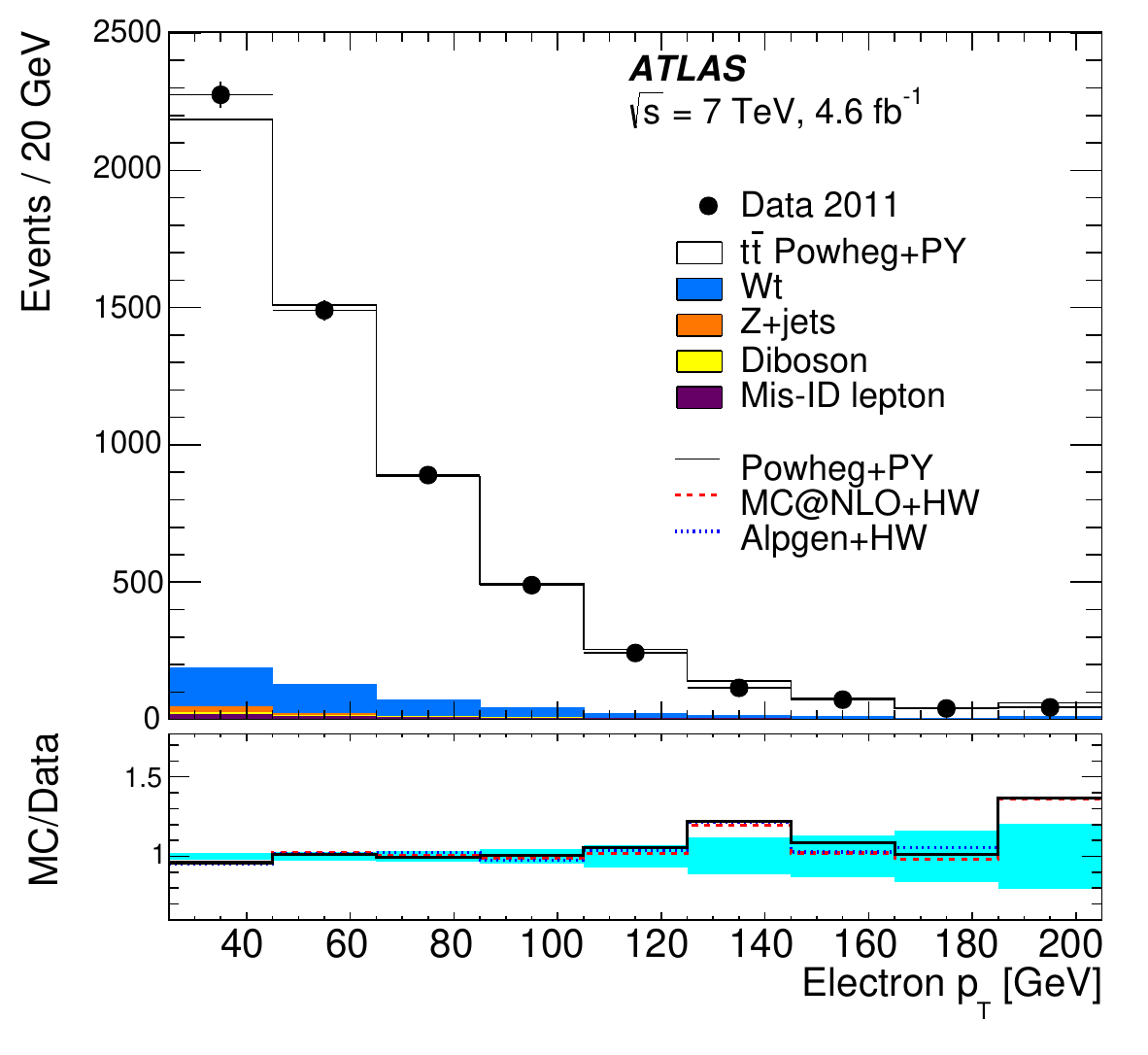}{c}{d}
\splitfigure{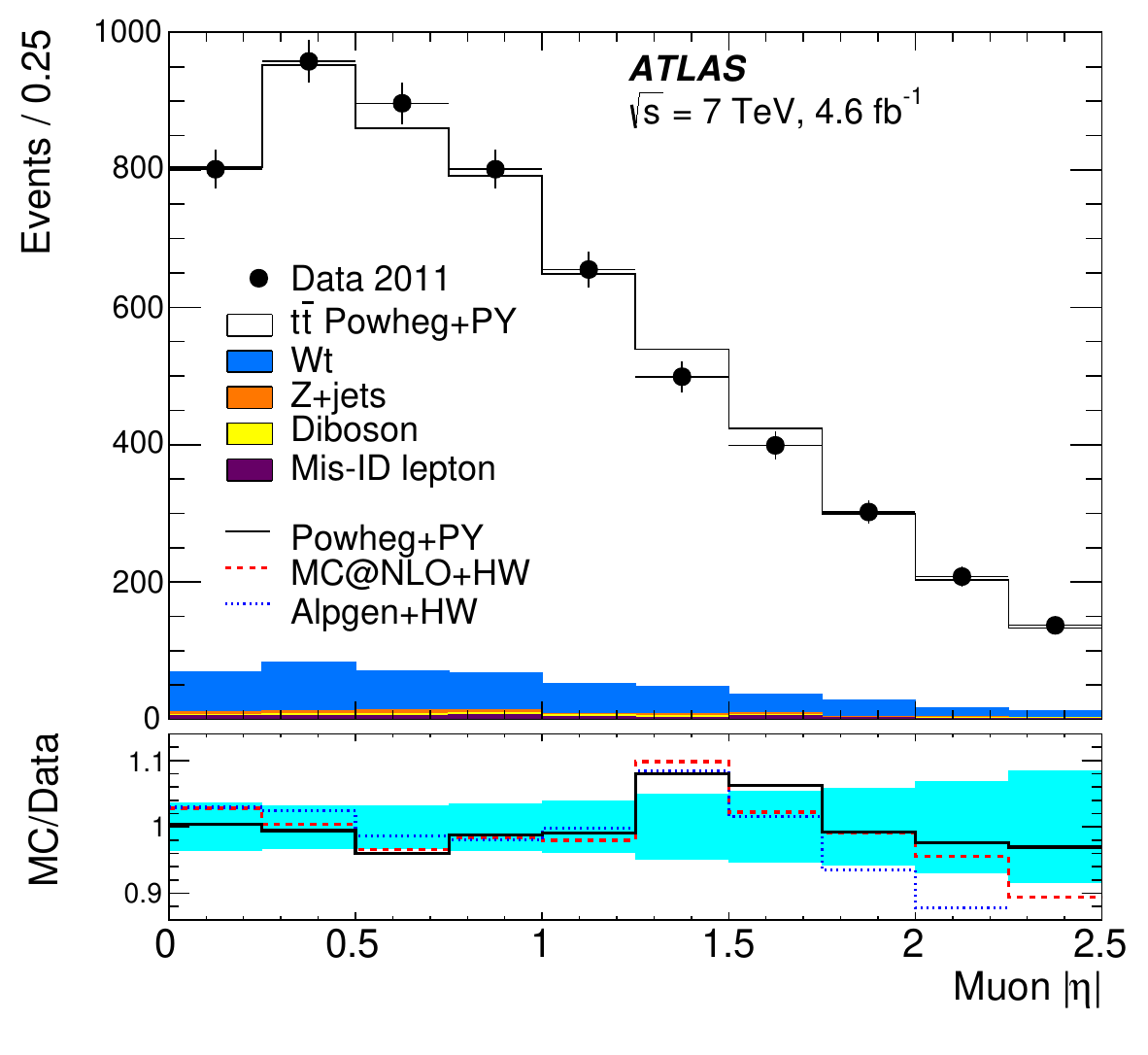}{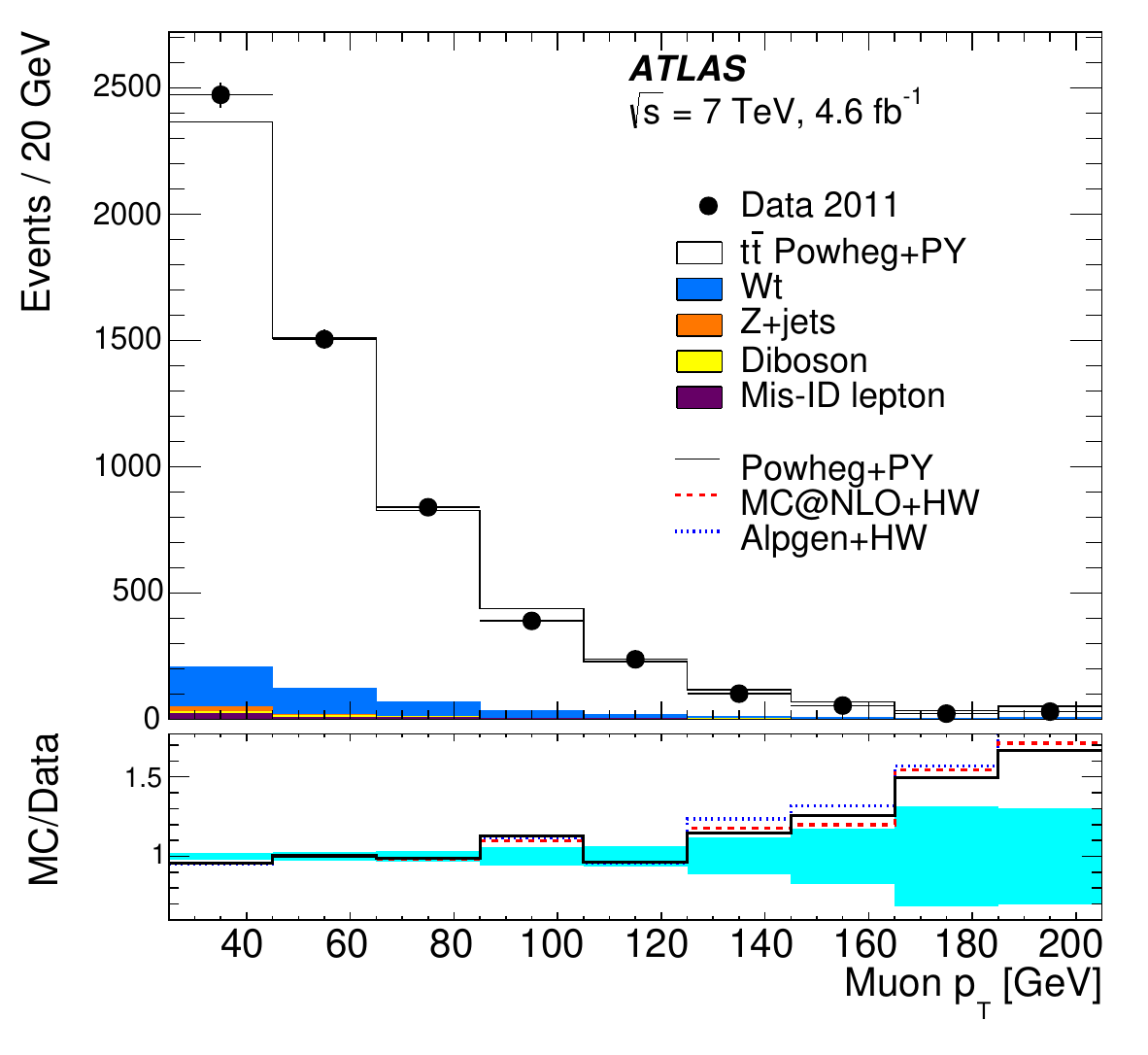}{e}{f}
\caption{\label{f:dmcw}Distributions of (a) the number of jets, (b) the
transverse momentum $\pt$ of the $b$-tagged jets, 
(c) the $|\eta|$ of the electron,
(d) the $\pt$ of the electron, (e) the $|\eta|$ of the muon and (f) the
$\pt$ of the muon, in events with an opposite-sign $e\mu$ pair and 
at least one $b$-tagged jet. The \sxw\ data are compared to the expectation 
from
simulation, broken down into contributions from \ttbar, single top, $Z$+jets,
dibosons, and events with misidentified electrons or muons, normalised to the
same number of entries as the data. 
The lower parts of the figure show the ratios of simulation to data, 
using various \ttbar\ signal samples and with the cyan band indicating the 
statistical uncertainty. The last bin includes the overflow.}
\end{figure*}

\begin{figure*}
\splitfigure{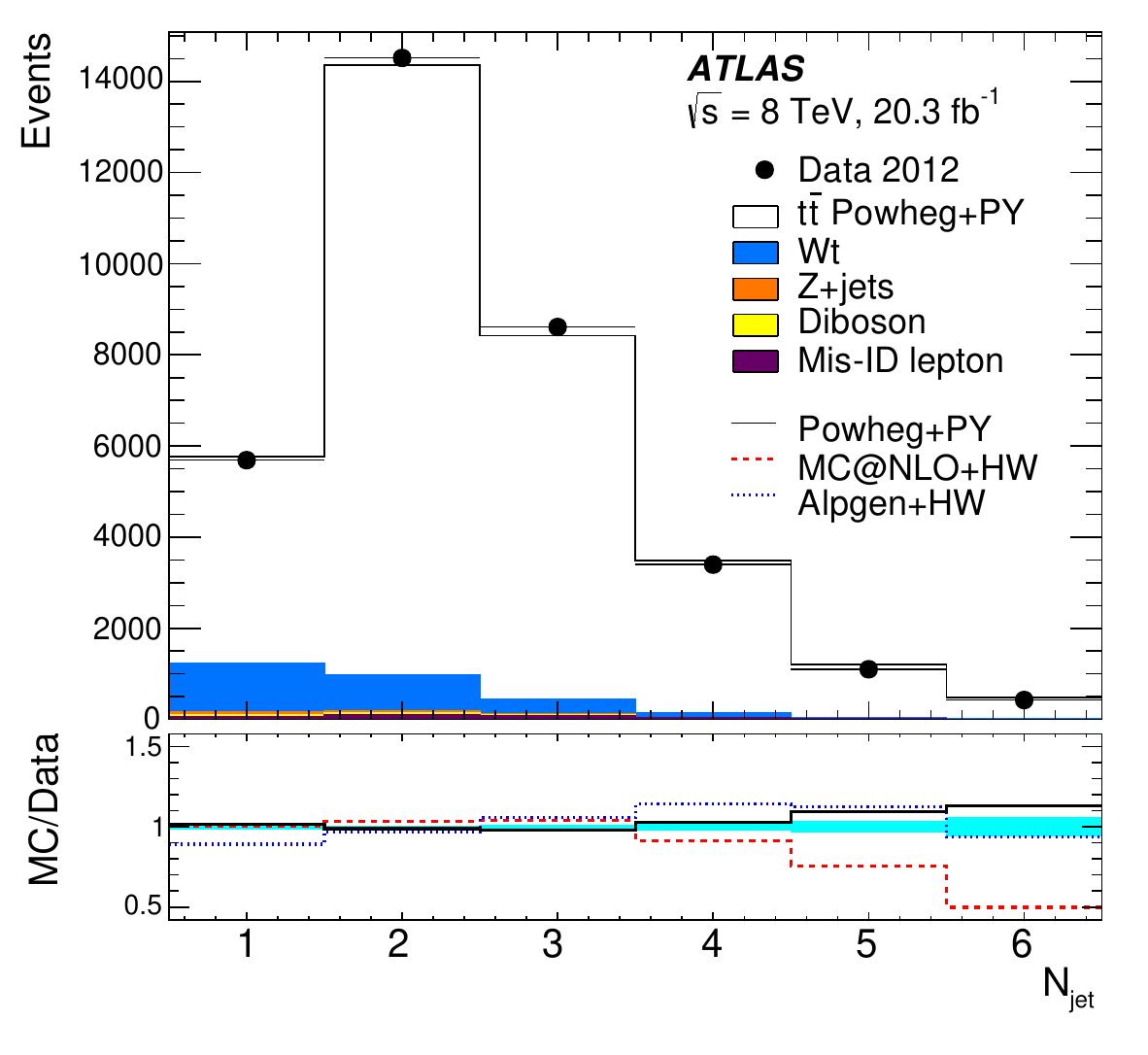}{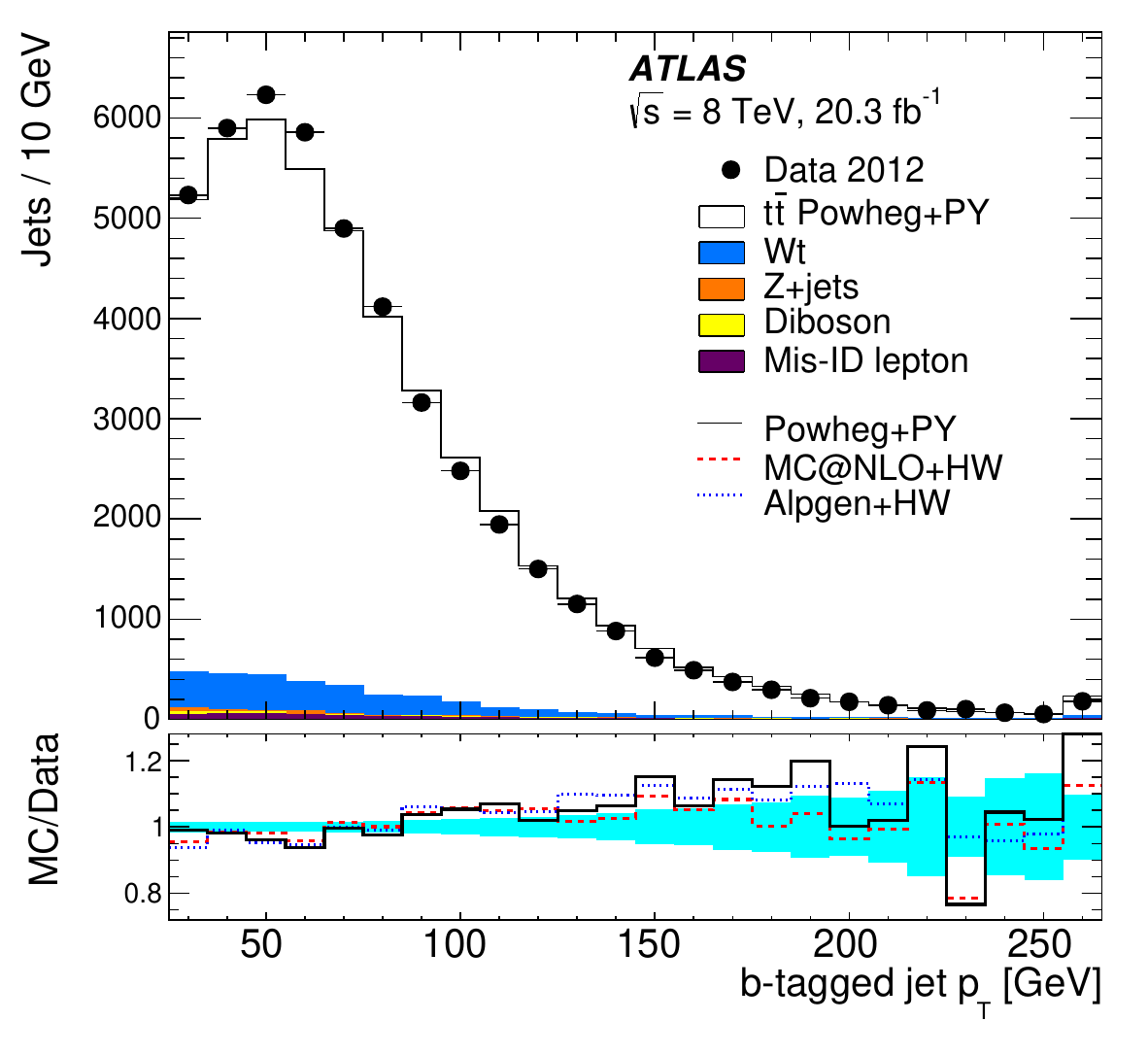}{a}{b}
\splitfigure{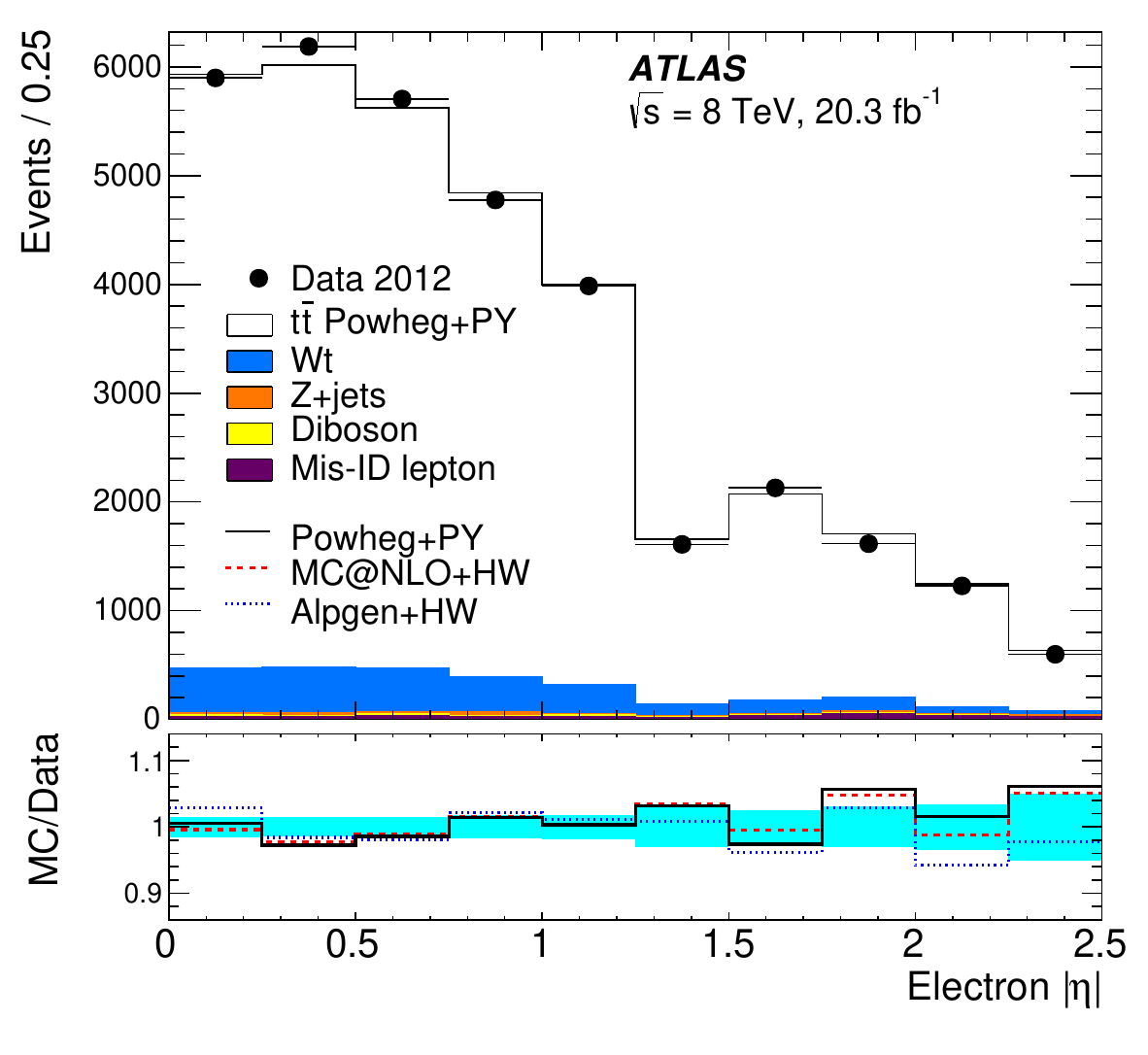}{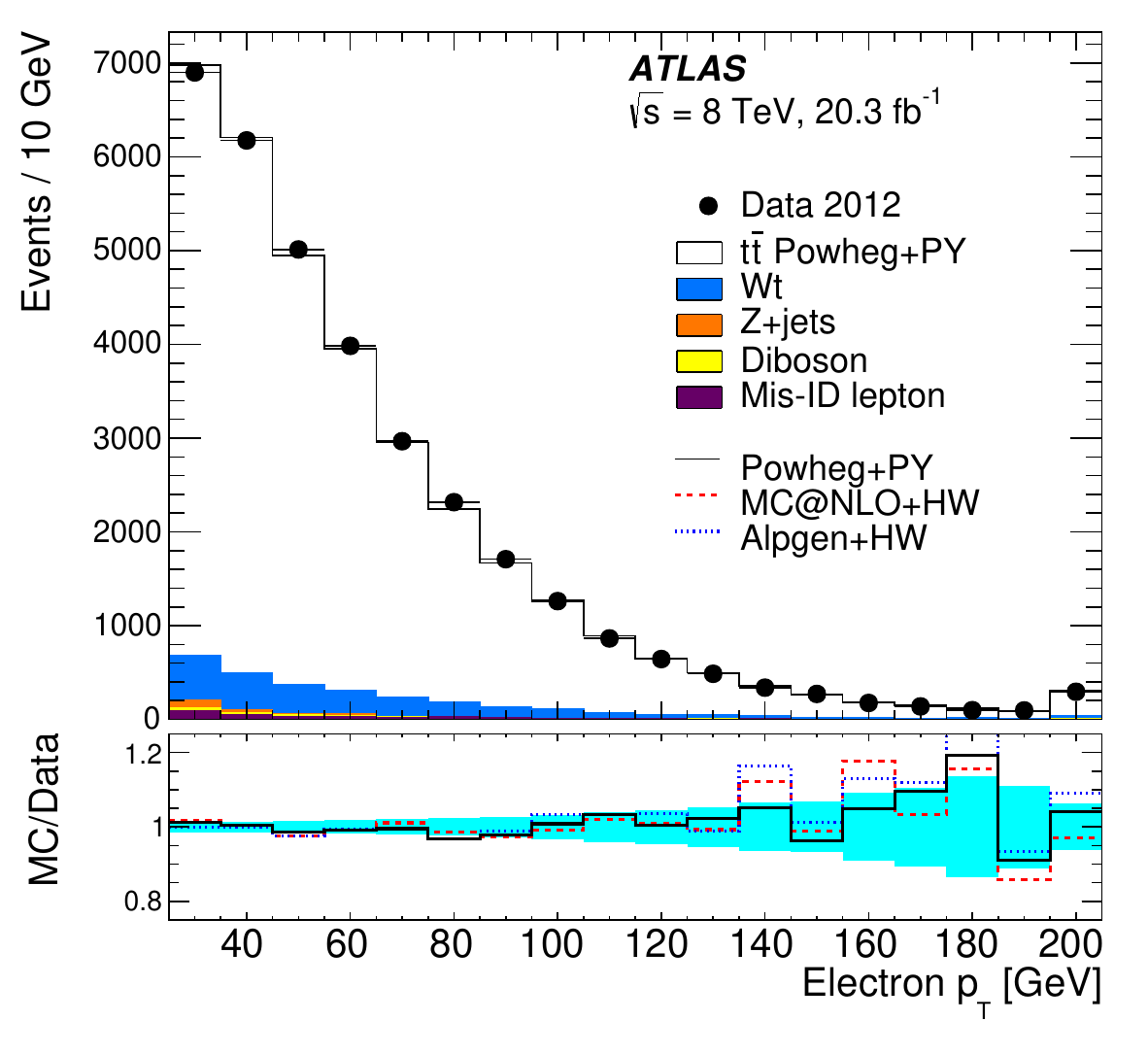}{c}{d}
\splitfigure{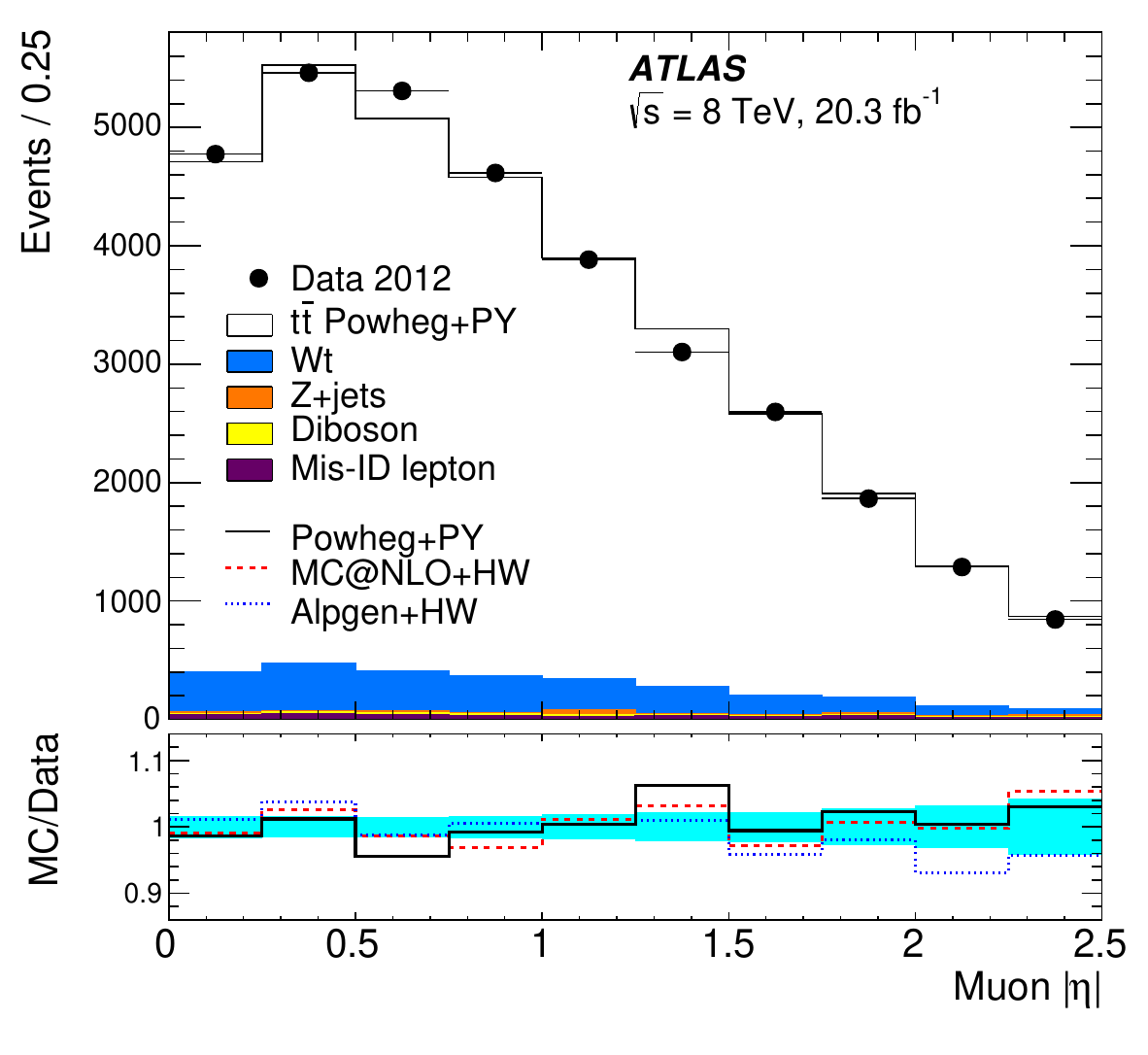}{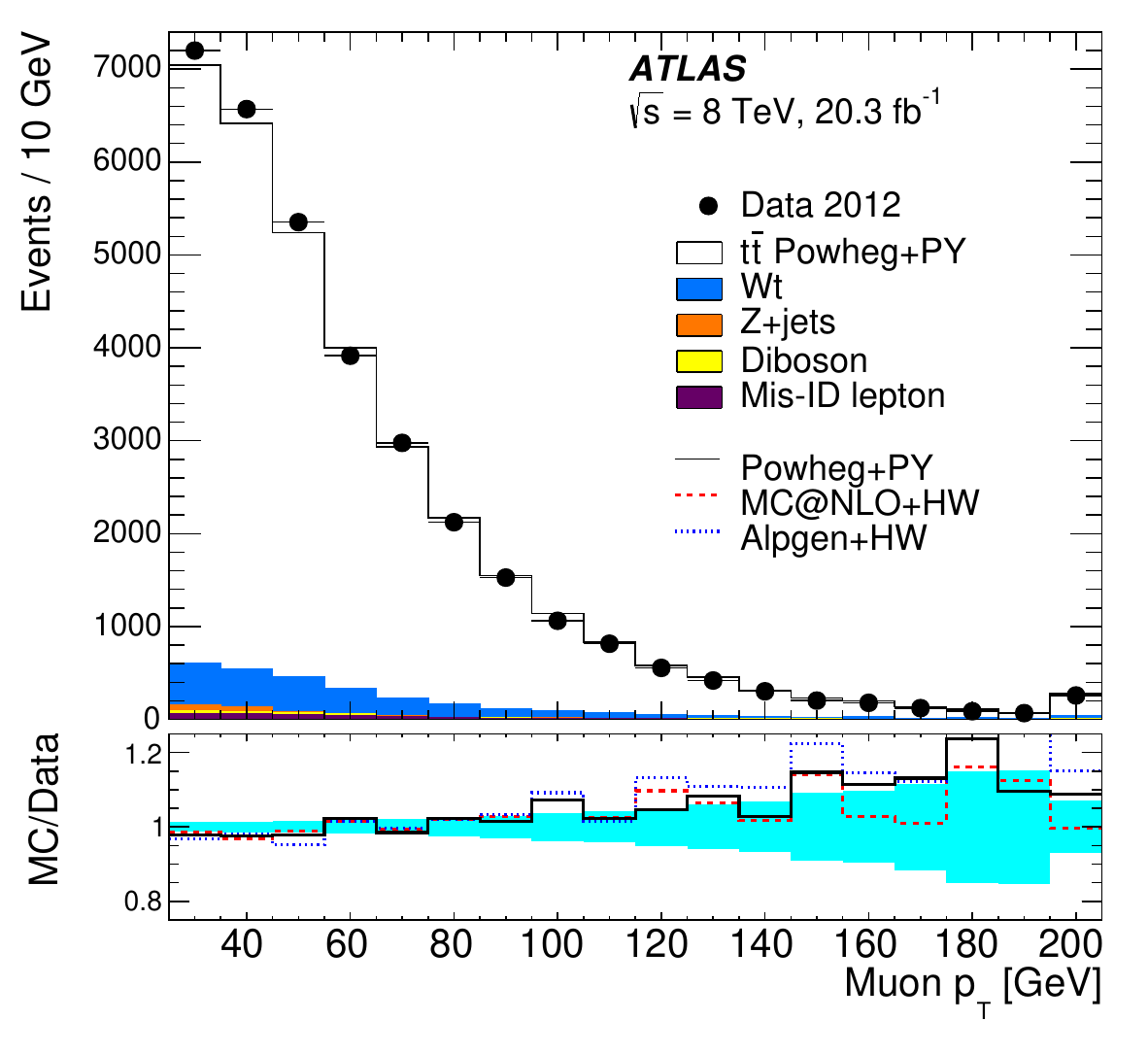}{e}{f}
\caption{\label{f:dmcv}Distributions of (a) the number of jets, (b) the
transverse momentum $\pt$ of the $b$-tagged jets, 
(c) the $|\eta|$ of the electron,
(d) the $\pt$ of the electron, (e) the $|\eta|$ of the muon and (f) the
$\pt$ of the muon, in events with an opposite-sign $e\mu$ pair and 
at least one $b$-tagged jet. The \sxv\ data are compared to the expectation 
from
simulation, broken down into contributions from \ttbar, single top, $Z$+jets,
dibosons, and events with misidentified electrons or muons, normalised to the
same number of entries as the data. 
The lower parts of the figure show the ratios of simulation to data, using
various \ttbar\ signal samples and  with the cyan band indicating the 
statistical uncertainty. The last bin includes the overflow.}
\end{figure*}

The value of \xtt\ extracted from Eq.~(\ref{e:tags}) is inversely 
proportional to the assumed value of \epsem, with 
$({\rm d}\xtt/{\rm d}\epsem)$ / $(\xtt/\epsem)=-1$. Uncertainties on \epsem\ 
therefore translate directly into uncertainties on \xtt. 
The value of \epsem\
was determined from simulation to be about 0.8\,\% for both centre-of-mass
energies, and includes the $\ttbar\rightarrow e\mu\nu\nubar\bbbar$ branching 
ratio of about 3.2\,\% including $W\rightarrow\tau\rightarrow e/\mu$ decays.
Similarly, \xtt\ is proportional to
the value of \cb, also determined from simulation, giving a dependence with
the opposite
sign, $({\rm d}\xtt$ /$ {\rm d}\cb)/(\xtt/\cb)=1$. The systematic uncertainties
on \epsem\ and \cb\ are discussed in Sect.~\ref{s:syst}.

With the kinematic cuts and $b$-tagging working point chosen for this analysis,
the sensitivities of \xtt\ to knowledge of the backgrounds \nib\ and \niib\
are given by $({\rm d}\xtt/{\rm d}\nib)$ / $(\xtt/\nib)=-0.12$ and
$({\rm d}\xtt/{\rm d}\niib)$ / $(\xtt/\niib)=-0.004$. The fitted cross-sections
are therefore most sensitive to the systematic uncertainties on \nib, whilst
for the chosen $b$-tagging working point,
the measurements of $N_2$ serve mainly to constrain \epsb. As discussed in 
Sect.~\ref{ss:corl}, consistent results were also obtained at 
different $b$-tagging efficiency working points that induce greater 
sensitivity to the background estimate in the two $b$-tag sample.

\subsection{Background estimation}\label{ss:back}

\begin{table*}[htp]
\caption{\label{t:fake} Breakdown of estimated misidentified-lepton 
contributions to the one ($1b$) and two ($2b$) $b$-tag opposite- and 
same-sign (OS and SS) $e\mu$ event samples at \sxw\ and \sxv. The different
misidentified-lepton categories are described in the text.
For the same-sign samples, the contributions from wrong-sign (where the
electron charge sign is misreconstructed) and
right-sign prompt lepton events are also shown, and the total 
expectations are compared to the data. The uncertainties shown are due to
the limited size of the simulated samples, and values and uncertainties quoted
as `0.0' are smaller than 0.05.}
\centering

\begin{tabular}{lcccccccc}
\hline\noalign{\smallskip}
 & \multicolumn{4}{c}{\sxw} & \multicolumn{4}{c}{\sxv} \\
Component & OS $1b$ & SS $1b$ & OS $2b$ & SS $2b$ & OS $1b$ & SS $1b$ & OS $2b$ & SS $2b$ \\
\noalign{\smallskip}\hline\noalign{\smallskip}
$t\rightarrow e\rightarrow\gamma$ conversion $e$ & $ 13.5\pm  0.8$ & $ 11.3\pm  0.8$ & $  6.1\pm  0.6$ & $  6.4\pm  0.6$ & $  97\pm   5$ & $  93\pm   5$ & $  67\pm   5$ & $  44\pm   4$ \\
Background conversion $e$ & $  7.2\pm  1.3$ & $  3.3\pm  0.5$ & $  1.4\pm  0.2$ & $  0.7\pm  0.2$ & $  53\pm  11$ & $  55\pm  12$ & $ 12.8\pm  2.5$ & $  8.7\pm  1.9$ \\
Heavy-flavour $e$ & $  2.9\pm  0.4$ & $  3.8\pm  0.4$ & $  0.3\pm  0.1$ & $  0.5\pm  0.1$ & $  33\pm   4$ & $  24\pm   3$ & $  5.6\pm  1.3$ & $  2.3\pm  0.8$ \\
Other $e$ & $  2.8\pm  0.7$ & $  0.0\pm  0.0$ & $  0.2\pm  0.1$ & $  0.0\pm  0.0$ & $  17\pm   7$ & $  0.5\pm  0.3$ & $  4.7\pm  1.2$ & $  0.1\pm  0.1$ \\
Heavy-flavour $\mu$ & $  3.2\pm  0.4$ & $  3.0\pm  0.4$ & $  0.5\pm  0.2$ & $  0.1\pm  0.1$ & $  26\pm   6$ & $ 17.9\pm  2.7$ & $  2.4\pm  0.8$ & $  2.8\pm  1.0$ \\
Other $\mu$ & $  0.7\pm  0.2$ & $  0.0\pm  0.0$ & $  0.2\pm  0.1$ & $  0.0\pm  0.0$ & $  2.2\pm  1.0$ & $  0.6\pm  0.4$ & $  0.8\pm  0.5$ & $  0.0\pm  0.0$ \\
\noalign{\smallskip}\hline\noalign{\smallskip}
Total misidentified & $   30\pm    2$ & $   21\pm    1$ & $    9\pm    1$ & $    8\pm    1$ & $  229\pm   16$ & $  191\pm   14$ & $   93\pm    6$ & $   58\pm    4$ \\\noalign{\smallskip}\hline\noalign{\smallskip}
Wrong-sign prompt & - & $  3.4\pm  0.4$ & - & $  1.9\pm  0.3$ & - & $  34\pm   4$ & - & $ 10.3\pm  1.9$ \\
Right-sign prompt & - & $  6.5\pm  0.5$ & - & $  2.2\pm  0.1$ & - & $ 35.4\pm  1.7$ & - & $ 12.9\pm  0.3$ \\
\noalign{\smallskip}\hline\noalign{\smallskip}
Total & - & $   31\pm    1$ & - & $   12\pm    1$ & - & $  260\pm   14$ & - & $   81\pm    5$ \\\noalign{\smallskip}\hline\noalign{\smallskip}
Data & - & 29 & - & 17 & - & 242 & - & 83 \\
\hline\noalign{\smallskip}
\end{tabular}
\end{table*}

The $Wt$ single top and diboson backgrounds were estimated from simulation 
as discussed in Sect.~\ref{s:dmc}.
The $Z$+jets background (with $Z\rightarrow\tau\tau\rightarrow e\mu 4\nu$) 
at \sxv\ was estimated from simulation using {\sc Alpgen+Pythia}, 
scaled by the ratios of $Z\rightarrow ee$ or $Z\rightarrow\mu\mu$ accompanied
by $b$-tagged jets measured in data and 
simulation. The ratio was evaluated separately in the one and two $b$-tag 
event samples. This scaling eliminates uncertainties due to the simulation
modelling of jets (especially heavy-flavour jets) produced in association with
the $Z$ bosons. The data-to-simulation ratios were measured in events with
exactly two opposite-sign electrons or muons passing the selections given in 
Sect.~\ref{s:objev} and one or two $b$-tagged jets, by fitting the 
dilepton invariant mass distributions in the range 60--120\,\GeV, accounting
for the backgrounds from \ttbar\ production and misidentified leptons.
Combining the results from both dilepton channels, the scale factors were 
determined to be $1.43\pm 0.03$ and 
$1.13\pm 0.08$ for the one and two $b$-tag backgrounds, after normalising
the simulation to the inclusive $Z$ cross-section prediction from FEWZ
\cite{fewz}. 
The uncertainties include systematic components derived
from a comparison of results from the $ee$ and $\mu\mu$ channels, and from
studying the variation of scale factors with $Z$ boson $\pt$. The average 
$\pt$ is higher in selected $Z\rightarrow\tau\tau\rightarrow e\mu 4\nu$
events than in $Z\rightarrow ee/\mu\mu$ events 
due to the momentum lost to the undetected neutrinos from the $\tau$ decays. 
The same procedure was used for the \sxw\ dataset, resulting in scale factors
of $1.23\pm 0.07$  (one $b$-tag) and $1.14\pm 0.18$ (two $b$-tags) for
the {\sc Alpgen + Herwig} $Z$+jets simulation, which predicts different
numbers of events with heavy-flavour jets than {\sc Alpgen + Pythia}.

The background from events with one real and one misidentified lepton was 
estimated using a combination of data and simulation. Simulation studies show
that the samples with a same-sign $e\mu$ pair and one or two $b$-tagged jets
are dominated by events with misidentified leptons, with rates comparable to 
those in the opposite-sign sample. The contributions of  events with
misidentified leptons
were therefore estimated using the same-sign event counts in data after
subtraction of the estimated prompt same-sign contributions, multiplied by
the opposite- to same-sign misidentified-lepton
ratios $R_j=\njfakeos/\njfakess$ estimated from simulation for events with 
$j=1$ and 2 $b$-tagged jets. 
The procedure is illustrated by Table~\ref{t:fake}, which shows the
expected numbers of events with misidentified leptons in opposite- and 
same-sign samples.
The contributions where the electron is misidentified, coming from a photon
conversion, the decay of a heavy-flavour hadron or other sources (such
as a misidentified hadron within a jet), and where the muon is misidentified, 
coming either from heavy-flavour decay or other sources (e.g. decay in
flight of a pion or kaon) are shown separately. The largest contributions
come from photon conversions giving electron candidates, and most of these
come from photons radiated from prompt electrons produced from 
$t\rightarrow Wq\rightarrow e\nu q$ in signal 
$\ttbar\rightarrow e\mu\nu\nubar\bbbar$ events. Such electrons populate
both the opposite- and same-sign samples, and are treated as 
misidentified-lepton background.

The ratios $R_j$ were estimated from simulation to be $R_1=1.4\pm 0.5$ and 
$R_2=1.1\pm 0.5$ at \sxw, and $R_1=1.2\pm 0.3$ and $R_2=1.6\pm 0.5$ at \sxv.
The uncertainties were derived by considering the range of $R_j$ values
for different components of the misidentified-lepton background, including
the small contributions from sources other than photon conversions and
heavy-flavour decays, which do not significantly populate the same-sign 
samples.  As shown in Table~\ref{t:fake}, about 25\,\% of the 
same-sign events have two prompt leptons, which come mainly from
semileptonic \ttbar\ events with an additional leptonically dec\-ay\-ing $W$ or 
$Z$ boson, diboson decays producing two same-sign leptons, and wrong-sign
$\ttbar\rightarrow e\mu\nu\nubar\bbbar$ events where the electron charge was
misreconstructed.
A conservative uncertainty of 50\,\% was assigned to this background, based
on studies of the simulation modelling of electron charge misidentification
\cite{elecperf} and uncertainties in the rates of contributing physics 
processes.

The simulation
modelling of the different components of the misidentified-lepton background
was checked by studying kinematic distributions of same-sign events, as
illustrated for the $|\eta|$ and $\pt$ distributions of the leptons in \sxv\
data in
Fig.~\ref{f:fakelept}. The simulation generally models the normalisation
and  shapes of distributions well in both the one and two $b$-tag event 
samples. The simulation modelling
was further tested in control samples with relaxed electron or muon isolation
requirements to enhance the relative contributions of electrons or muons from
heavy-flavour decays, and similar levels of agreement were observed.

\begin{figure*}
\splitfigure{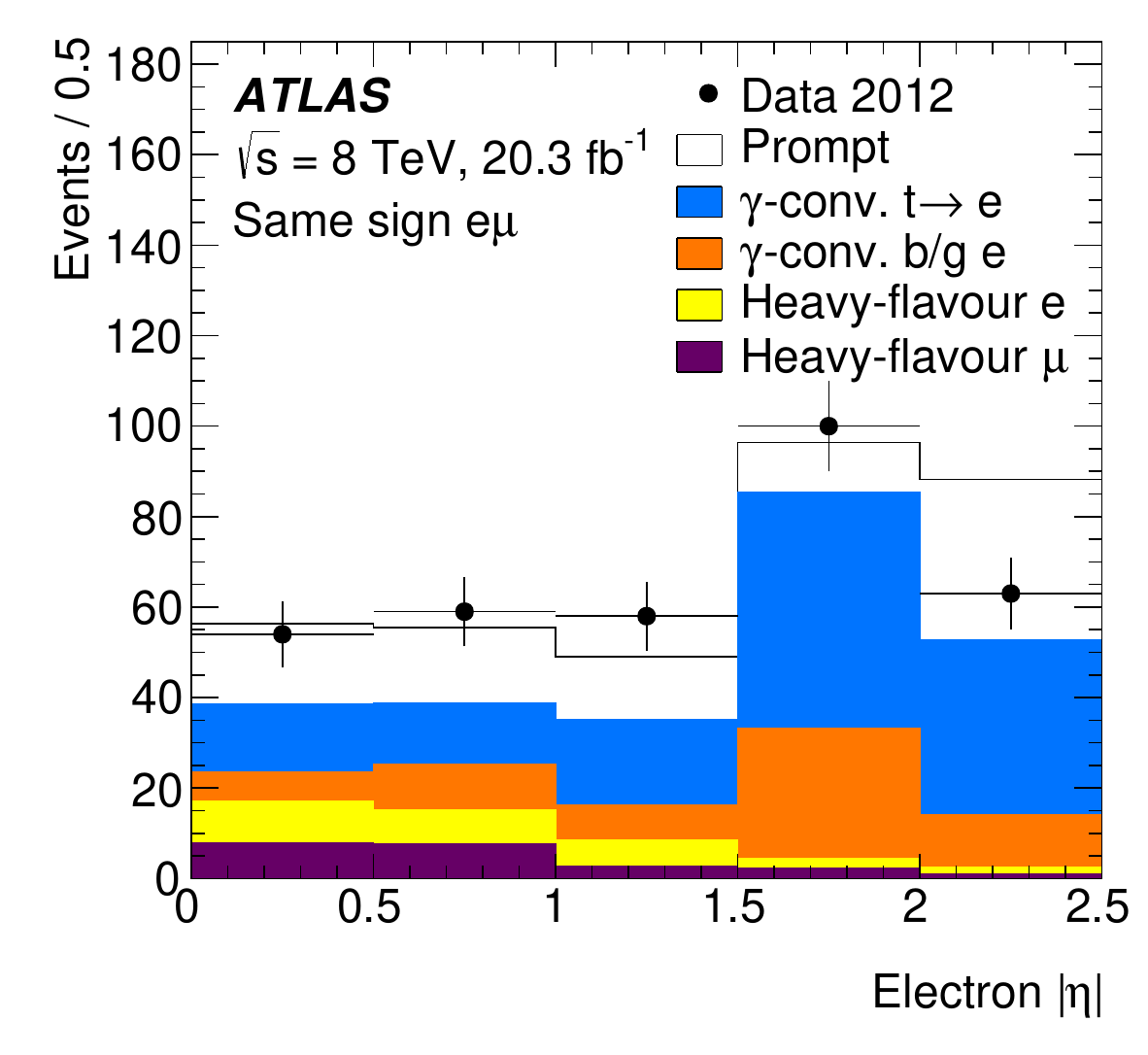}{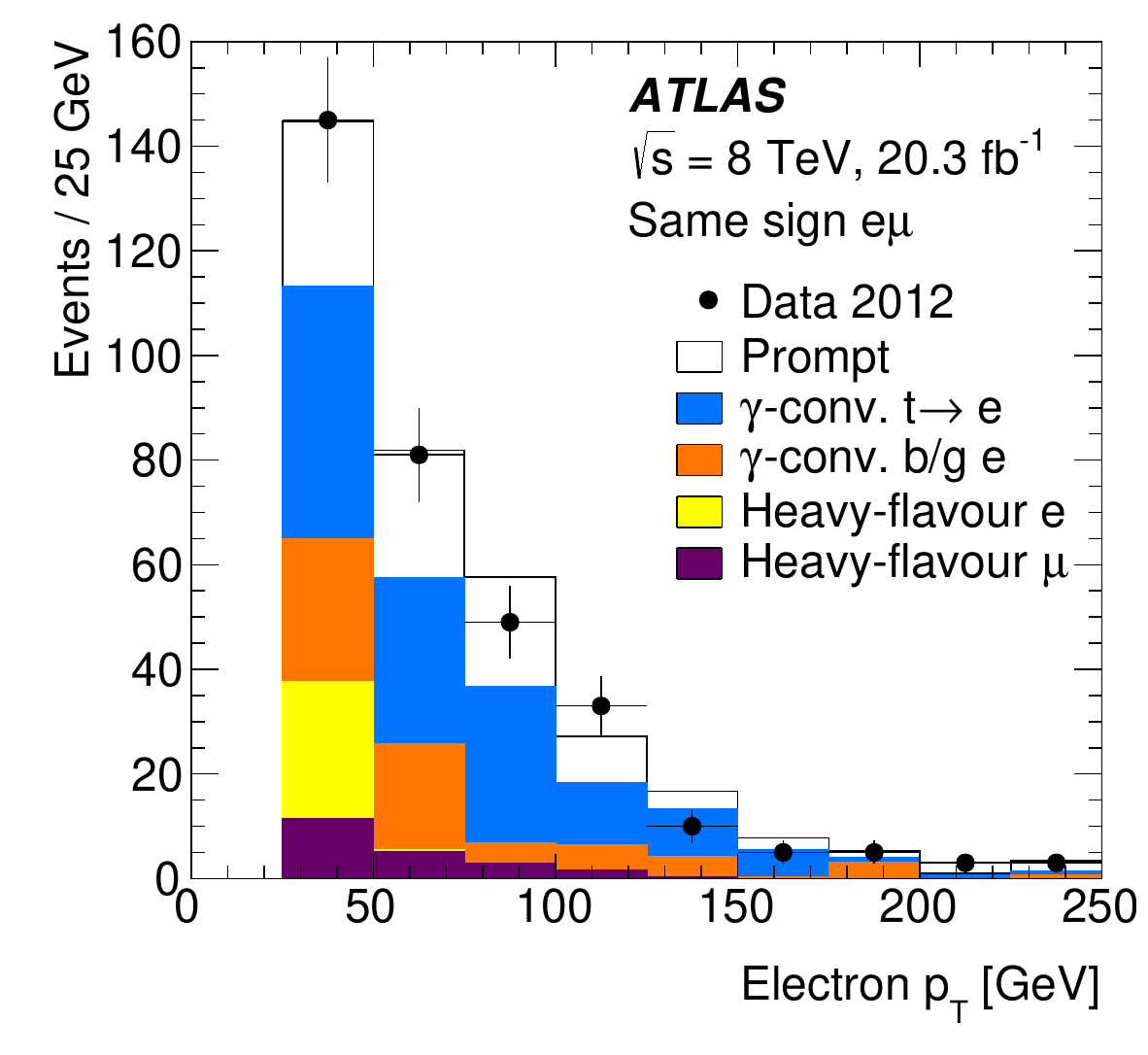}{a}{b}
\splitfigure{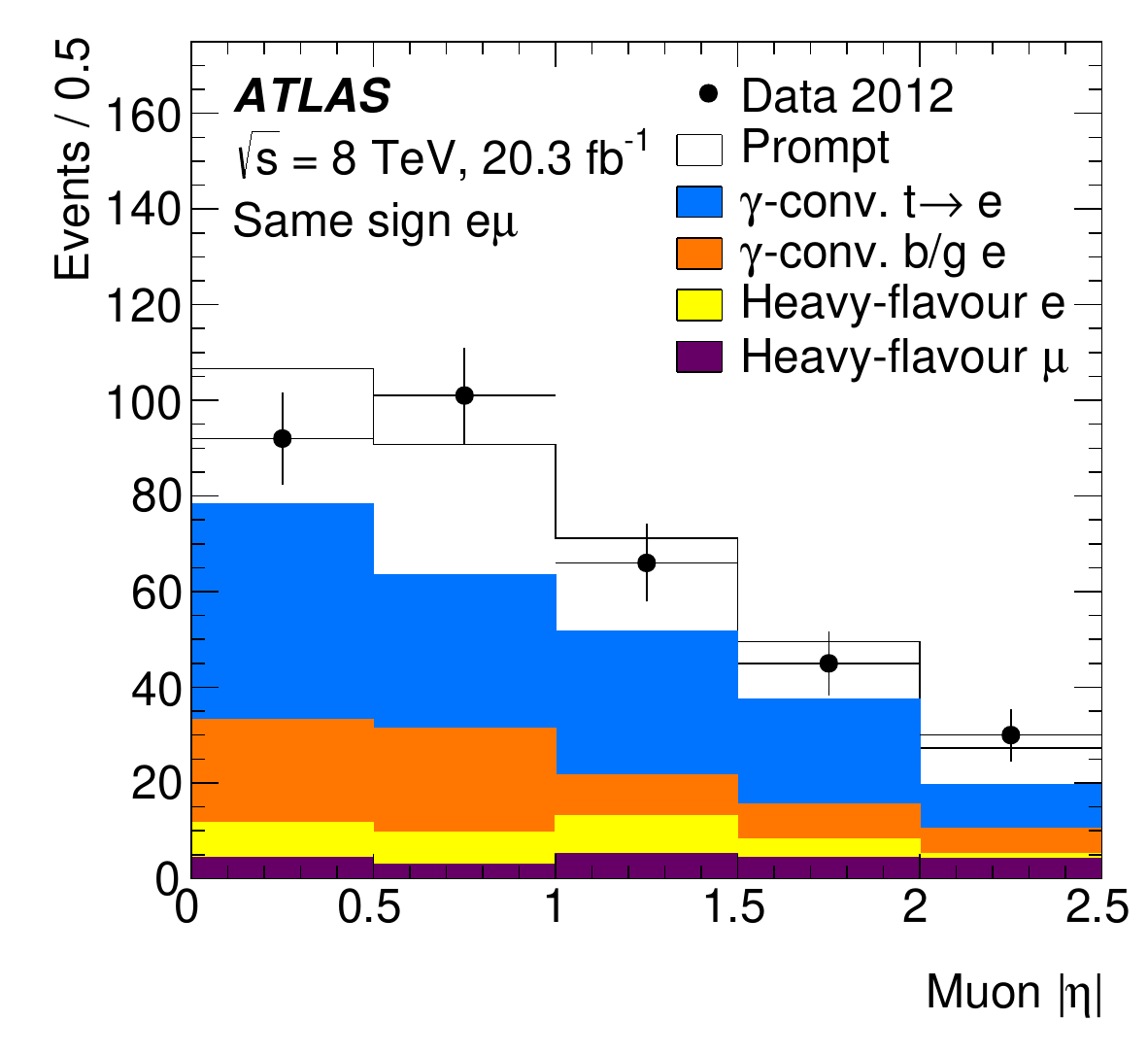}{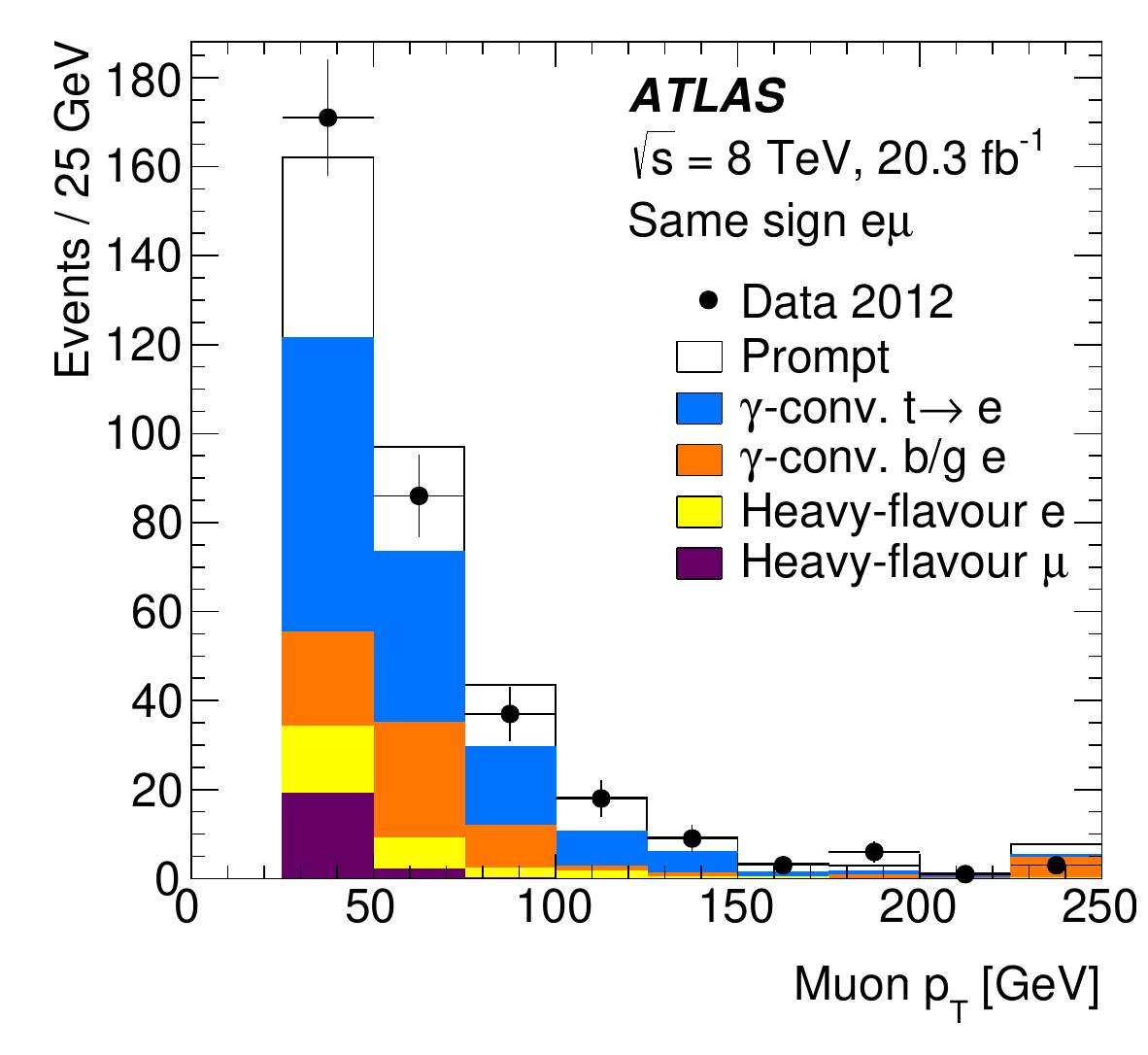}{c}{d}
\caption{\label{f:fakelept}Distributions of electron and muon $|\eta|$ and 
$\pt$ in same-sign $e\mu$ events at \sxv\ with at least one $b$-tagged jet. 
The simulation
prediction is normalised to the same integrated luminosity as the data, and
broken down into contributions where both leptons are prompt, or one is 
a misidentified lepton from a photon conversion originating from a top quark 
decay or from background, or from heavy-flavour decay.
In the $\pt$ distributions, the last bin includes the overflows.}
\end{figure*}

\section{Systematic uncertainties}\label{s:syst}

The systematic uncertainties on the measured cross-sect\-ions \xtt\
are shown in detail in 
Table~\ref{t:syst} together with the individual uncertainties on \epsem\
and \cb. A summary of the uncertainties on \xtt\ is shown in 
Table~\ref{t:systsum}. Each source of uncertainty was
evaluated by repeatedly solving Eq.~(\ref{e:tags}) with all 
relevant input parameters simultaneously changed by $\pm 1$ standard deviation.
Systematic correlations between input parameters
(in particular significant anti-correlations between \epsem\ and \cb\ which 
contribute with opposite signs to \xtt) were thus taken into account. The
total uncertainties on \xtt\ and \epsb\ were calculated by adding the effects
of all the individual systematic components in quadrature, assuming them
to be independent. The sources of systematic uncertainty are discussed
in more detail below; unless otherwise stated, the same methodology was used
for both \sxwt\ and \sxvt\ datasets.

\begin{table*}[htp]
\caption{\label{t:syst}Detailed breakdown 
of the symmetrised relative statistical, systematic
and total uncertainties on the measurements of the \ttbar\ production 
cross-section \xtt\ at \sxw\ and \sxv. Uncertainties quoted as `0.00' are 
smaller than 0.005, whilst `-' indicates the corresponding uncertainty
is not applicable. The uncertainties on \epsem\ and \cb\ are also shown, with
their relative signs indicated where relevant. They contribute with opposite
signs to the uncertainties on \xtt, which also 
include uncertainties  from estimates of the background terms
\nib\ and \niib. The lower part of the table gives the
systematic uncertainties that are different for the measurement of the 
fiducial cross-section \xfid, together with the total analysis systematic 
and total uncertainties on \xfid.}
\centering 

\begin{tabular}{lcccccc}
\hline\noalign{\smallskip}
$\sqrt{s}$ & & 7\,\TeV & & & 8\,\TeV & \\
Uncertainty (inclusive \xtt) & $\Delta\epsem/\epsem$ & $\Delta\cb/\cb$ & $\Delta\xtt/\xtt$ & 
$\Delta\epsem/\epsem$ & $\Delta\cb/\cb$ & $\Delta\xtt/\xtt$  \\
& (\%) & (\%) & (\%) & (\%) & (\%) & (\%) \\
\noalign{\smallskip}\hline\noalign{\smallskip}
Data statistics & & &  1.69 & & &  0.71 \\
\noalign{\smallskip}\hline\noalign{\smallskip}
\ttbar\ modelling &  0.71 & $-0.72$ &  1.43 &  0.65 & $-0.57$ &  1.22 \\
Parton distribution functions &  1.03 & - &  1.04 &  1.12 & - &  1.13 \\
QCD scale choice &  0.30 & - &  0.30 &  0.30 & - &  0.30 \\
Single-top modelling & - & - &  0.34 & - & - &  0.42 \\
Single-top/\ttbar\ interference & - & - &  0.22 & - & - &  0.15 \\
Single-top $Wt$ cross-section & - & - &  0.72 & - & - &  0.69 \\
Diboson modelling & - & - &  0.12 & - & - &  0.13 \\
Diboson cross-sections & - & - &  0.03 & - & - &  0.03 \\
$Z$+jets extrapolation & - & - &  0.05 & - & - &  0.02 \\
Electron energy scale/resolution &  0.19 & $-0.00$ &  0.22 &  0.46 &  0.02 &  0.51 \\
Electron identification &  0.12 &  0.00 &  0.13 &  0.36 &  0.00 &  0.41 \\
Muon momentum scale/resolution &  0.12 &  0.00 &  0.14 &  0.01 &  0.01 &  0.02 \\
Muon identification &  0.27 &  0.00 &  0.30 &  0.38 &  0.00 &  0.42 \\
Lepton isolation &  0.74 & - &  0.74 &  0.37 & - &  0.37 \\
Lepton trigger &  0.15 & $-0.02$ &  0.19 &  0.15 &  0.00 &  0.16 \\
Jet energy scale &  0.22 &  0.06 &  0.27 &  0.47 &  0.07 &  0.52 \\
Jet energy resolution & $-0.16$ &  0.08 &  0.30 & $-0.36$ &  0.05 &  0.51 \\
Jet reconstruction/vertex fraction &  0.00 &  0.00 &  0.06 &  0.01 &  0.01 &  0.03 \\
$b$-tagging & - &  0.18 &  0.41 & - &  0.14 &  0.40 \\
Misidentified leptons & - & - &  0.41 & - & - &  0.34 \\
\noalign{\smallskip}\hline\noalign{\smallskip}
Analysis systematics (\xtt) &  1.56 &  0.75 &  2.27 &  1.66 &  0.59 &  2.26 \\
\noalign{\smallskip}\hline\noalign{\smallskip}
Integrated luminosity & - & - &  1.98 & - & - &  3.10 \\
LHC beam energy & - & - &  1.79 & - & - &  1.72 \\
\noalign{\smallskip}\hline\noalign{\smallskip}
Total uncertainty (\xtt) &  1.56 &  0.75 &  3.89 &  1.66 &  0.59 &  4.27 \\
\noalign{\smallskip}\hline\noalign{\smallskip}
\noalign{\smallskip}\hline\noalign{\smallskip}
Uncertainty (fiducial \xfid) & $\Delta\epsem/\epsem$ & $\Delta\cb/\cb$ & $\Delta\xfid/\xfid$ & 
$\Delta\epsem/\epsem$ & $\Delta\cb/\cb$ & $\Delta\xtt/\xtt$  \\
& (\%) & (\%) & (\%) & (\%) & (\%) & (\%) \\
\noalign{\smallskip}\hline\noalign{\smallskip}
\ttbar\ modelling &  0.84 & $-0.72$ &  1.56  &  0.74 & $-0.57$ &  1.31  \\
Parton distribution functions &  0.35 & - &  0.38  &  0.23 & - &  0.28  \\
QCD scale choice &  0.00 & - &  0.00  &  0.00 & - &  0.00  \\
Other uncertainties (as above) &  0.88 &  0.21 &  1.40 &  1.00 &  0.17 &  1.50 \\
\noalign{\smallskip}\hline\noalign{\smallskip}
Analysis systematics (\xfid) &  1.27 &  0.75 &  2.13 &  1.27 &  0.59 &  2.01 \\
\noalign{\smallskip}\hline\noalign{\smallskip}
Total uncertainty (\xfid) &  1.27 &  0.75 &  3.81 &  1.27 &  0.59 &  4.14 \\
\hline\noalign{\smallskip}
\end{tabular}
\end{table*}

\begin{description}

\item[\bf\ttbar\ modelling:] Uncertainties on \epsem\ and
\cb\ due to the simulation of \ttbar\ events were assessed by studying the
predictions of different \ttbar\ generators and hadronisation models
as detailed in Sect.~\ref{s:dmc}. The prediction for \epsem\ was found
to be particularly sensitive to the amount of hadronic activity near the
leptons, which strongly affects the efficiency of the lepton isolation 
requirements described in Sect.~\ref{s:objev}. These isolation 
efficiencies were therefore measured directly from data, as discussed below.
The remaining uncertainties on \epsem\ relating to lepton reconstruction,
identification and lepton--jet overlap removal, were evaluated from the
differences between the predictions from the baseline {\sc Powheg + Pythia} 
\ttbar\ sample and a sample generated using {\sc MC@NLO + Herwig}, 
thus varying both the hard-scattering
event generator and the fragmentation and hadronisation model. The
{\sc MC@NLO + Herwig} sample gave a larger value of \epsem\ but a smaller value
of \cb. Additional 
comparisons of {\sc Powheg + Pythia} samples with the AUET2 rather than
P2011C tune and with {\sc Powheg + Herwig}, i.e. changing only the 
fragmentation / hadronisation model, gave smaller uncertainties.
The {\sc Alpgen + Herwig} and {\sc Alpgen + Pythia} samples gave  values of 
\epsem\ up to 2\,\% higher than
that of {\sc Powheg+Pythia}, due largely to a more central predicted 
$\eta$ distribution for the leptons. However, this sample uses a leading-order
generator and PDFs, and gives an inferior description of the 
electron and muon $\eta$ distributions (see Figs.~\ref{f:dmcv}(c) and 
\ref{f:dmcv}(e)),
so was not used to set the systematic uncertainty on \epsem. In contrast, the
{\sc Alpgen} samples were considered in setting the uncertainty on \cb, taken
as the largest difference between the predictions of {\sc Powheg + Pythia}
and any of the other generators. The effect of extra radiation in \ttbar\
events was also considered explicitly by using pairs of simulation
samples with different {\sc Pythia} tunes whose parameters span the variations
compatible with ATLAS studies of additional jet activity in \ttbar\ events
at \sxw\ \cite{atlasifsr}, generated using 
both {\sc AcerMC + Pythia} and {\sc Alpgen + Pythia}. These samples predicted
large variations in the lepton isolation efficiencies (which were instead
measured from data), but residual variations in other
lepton-related uncertainties and \cb\ within the uncertainties set from
other simulation samples.

\item[\bf Parton distribution functions:]\hfill The uncertainties on \epsem, 
\cb\ and the $Wt$ single top background due to uncertainties on the 
proton PDFs were evaluated using the error sets of
the CT10 NLO \cite{cttenpdf}, MSTW 2008 68\,\% CL NLO \cite{mstwnnlo} and 
NNPDF 2.3 NLO \cite{nnpdfffn} sets. The final uncertainty was 
calculated as half the envelope encompassing the predictions from  all three
PDF sets along with their associated uncertainties,
following the PDF4LHC recommendations \cite{pdflhc}. 

\item[\bf QCD scale choices:] The lepton $\pt$ and $\eta$ distributions, and
hence \epsem, are sensitive to the choices of QCD renormalisation and 
factorisation scales. This effect was investigated using \sxvt\ generator-level
{\sc Po\-w\-heg + Pythia} \ttbar\ samples where the two scales were separately
varied up and down by a factor of two from their default values of
$Q^2=\mtop^2+p_{{\rm T},t}^2$. The systematic uncertainty for each scale
was taken as half the difference in \epsem\ values between the samples
with increased and decreased QCD scale, and the uncertainties for the
renormalisation and factorisation scales were then added linearly to give 
a total scale uncertainty of 0.30\,\% on \epsem, assumed to be valid for both
centre-of-mass energies.

\item[\bf Single top modelling:] Uncertainties related to $Wt$ single top 
modelling were assessed by comparing the predictions from
{\sc Powheg + Pythia}, {\sc Powheg + Herwig}, {\sc MC@NLO + Herwig}, and
{\sc AcerMC + Pythia} with two tunes producing different amounts of 
additional radiation, in all cases normalising the total 
production rate to the approximate NNLO cross-section prediction. The
resulting uncertainties are about 5\,\% and 20\,\% on the one and two $b$-tag
background contributions. The background in the two $b$-tag sample
is sensitive to the production of $Wt$ with an additional $b$-jet, a NLO
contribution to $Wt$ which can interfere with the \ttbar\ final state. The
sensitivity to this interference was studied by comparing the predictions
of {\sc Powheg} with the diagram-removal (baseline) and diagram-subtraction 
schemes \cite{powwtdr,wtinter}, giving additional single-top/\ttbar\
interference uncertainties of 
1--2\,\% and 20\,\% for the one and two $b$-tag samples. The production
of single top quarks in association with a $Z$ boson gives contributions which 
are negligible compared to the above uncertainties. Production of
single top quarks via the $t$- and $s$-channels gives rise to final states
with only one prompt lepton, and is accounted for as part of the 
misidentified-lepton background.

\item[\bf Background cross-sections:] The uncertainties on the $Wt$ single top 
cross-section were taken to be 7.6\,\% and 6.8\% at \sxw\ and \sxv, based on
Ref. \cite{Wttheoxsec}. The uncertainties on the diboson cross-sections were
set to 5\,\% \cite{dibmcfm}.

\item[\bf Diboson modelling:] Uncertainties in the backgrounds from
dibosons with one or two additional $b$-tagged jets were assessed by comparing
the baseline prediction from {\sc Alpgen + Herwig} with that of 
{\sc Sherpa} \cite{sherpa} including massive $b$ and $c$ quarks, and
found to be about 20\,\%. The background from 125\,\GeV\ SM Higgs 
production
in the gluon fusion, vector-boson fusion, and $WH$ and $ZH$ associated 
production modes, with $H\rightarrow WW$ and $H\rightarrow\tau\tau$,
was evaluated to be smaller than the diboson modelling uncertainties, and 
was neglected.

\item[\bf $Z$+jets extrapolation:] The uncertainties on the extrapolation of
the $Z$+jets background from $Z\rightarrow ee/\mu\mu$ to $Z\rightarrow\tau\tau$
events result from statistical uncertainties,  comparing the results from 
$ee$ and $\mu\mu$, which have different background compositions, and
considering the dependence of the scale factors on $Z$ boson \pt.

\item[\bf Lepton identification and measurement:] The modelling of the 
electron and muon identification efficiencies, energy scales and resolutions
(including the effects of pileup)
were studied using $Z\rightarrow ee/\mu\mu$, $J/\psi\rightarrow ee/\mu\mu$
and $W\rightarrow e\nu$ events in data and simulation, using the techniques 
described in Refs. \cite{elecperf,muperf,elecperfx}.
Small corrections were applied to the simulation to better
model the performance seen in data, and the associated 
systematic uncertainties were propagated to the cross-section measurement. 

\item[\bf Lepton isolation:] The efficiency of the lepton isolation
requirements was measured directly in data, from the fraction of selected
opposite-sign $e\mu$ events with one or two $b$-tags where either the
electron or muon fails the isolation cut. The results were corrected for
the contamination from misidentified leptons, estimated using the same-sign 
$e\mu$
samples as described in Sect.~\ref{s:ext}, or by using the distributions
of lepton impact parameter significance $|d_0|/\sigma_{d_0}$, where $d_0$ is
the distance of closest approach of the lepton track to the event primary
vertex in the transverse plane, and $\sigma_{d_0}$ its uncertainty. Consistent
results were obtained from both methods, and showed that the baseline
{\sc Powheg+Pythia} simulation overestimates the efficiencies of the isolation
requirements by about 0.5\,\% for both the electrons and muons. 
These corrections were applied to \epsem, with uncertainties
dominated by the limited sizes of the 
same-sign and high impact-parameter significance samples used for background
estimation. Similar results
were found from studies in $Z\rightarrow ee$ and $Z\rightarrow\mu\mu$ events,
after correcting the results for the larger average amount of
hadronic activity near the leptons in $\ttbar\rightarrow e\mu\nu\nubar\bbbar$
events.

\item[\bf Jet-related uncertainties:] Although the efficiency to reconstruct and
$b$-tag jets from \ttbar\ events is extracted from the data, uncertainties
in the jet energy scale, energy 
resolution and reconstruction efficiency affect the
backgrounds estimated from simulation and the estimate of the tagging 
correlation \cb. They also have a small effect on \epsem\ via the 
lepton--jet $\Delta R$ separation cuts. The jet energy scale 
was varied in simulation according to the uncertainties derived from simulation
and in-situ calibration measurements \cite{jesx,jesxi}, 
using a model with 21 (\sxw) or 22 (\sxv)  separate 
orthogonal uncertainty components which were then added in quadrature. 
The jet energy 
resolution was found to be well modelled by simulation \cite{jetres},
and remaining uncertainties were assessed by applying additional smearing,
which reduces \epsem.
The calorimeter jet reconstruction efficiency was measured in data using 
track-based jets, and is also well described by the simulation;
the impact of residual uncertainties was assessed
by randomly discarding jets. The uncertainty associated with
the jet vertex fraction requirement was assessed from
studies of $Z\rightarrow ee/\mu\mu$+jets events.

\item[\bf $b$-tagging uncertainties:]\hspace{3mm}The efficiency for $b$-tagging jets
from \ttbar\ events was extracted from the data via Eq.~(\ref{e:tags}), 
but simulation was used
to predict the number of $b$-tagged jets and mistagged light-quark, gluon
and charm jets in the $Wt$ single top and diboson backgrounds. The tagging
correlation \cb\ is also slightly sensitive to the efficiencies 
for tagging heavy- and light-flavour jets. The uncertainties
in the simulation modelling of the $b$-tagging performance were assessed using
studies of $b$-jets containing muons \cite{btagptrel,systemviii}, 
jets containing $D^{*+}$ mesons \cite{btagccal} and inclusive jet events 
\cite{btagmiscal}.

%
\item[\bf Misidentified leptons:] The uncertainties on the number of events
with misidentified leptons in the one and two $b$-tagged samples were derived 
from the statistical uncertainties on the numbers of same-sign lepton events, 
the systematic uncertainties on the opposite- to same-sign ratios $R_j$,
and the uncertainties on the numbers of prompt same-sign events, as discussed
in detail in Sect.~\ref{ss:back}. The
overall uncertainties on the numbers of misidentified leptons vary from 
30\,\% to 50\,\%, dominated by the uncertainties on the ratios $R_j$.

\item[\bf Integrated luminosity:] The uncertainty on the integrated 
luminosity of the \sxw\ dataset is 1.8\,\% \cite{lumi}. Using beam-separation
scans performed in November 2012, the same methodology was applied to 
determine the \sxv\ luminosity scale, resulting in an uncertainty of 2.8\,\%.
These uncertainties are dominated by effects specific to each dataset, and so
are considered to be uncorrelated between the two centre-of-mass energies.
The relative uncertainties on the cross-section 
measurements are slightly larger than those on the luminosity measurements
because the $Wt$ single top and diboson
backgrounds are evaluated from simulation, so are also sensitive to the
assumed integrated luminosity.

\item[\bf LHC beam energy:] The LHC beam energy during the 2012 $pp$ run
was calibrated to be $0.30\pm 0.66$\,\% smaller than the 
nominal value of 4\,TeV per beam, using the revolution frequency difference
of protons and lead ions during $p$+Pb runs in early 2013 \cite{lhcenergy}. 
Since this calibration is compatible with the nominal $\sqrt{s}$ of 8\,TeV,
no correction was applied to the measured \xtt\ value. However, an 
uncertainty of 1.72\,\%, corresponding to the expected change in \xtt\ for
a 0.66\,\% change in $\sqrt{s}$ is quoted separately on the final result.
This uncertainty was calculated using {\tt top++ 2.0}, assuming that 
the relative change of \xtt\ for a 0.66\,\% change in $\sqrt{s}$ is 
as predicted by the NNLLO+NNLL calculation.
Following Ref. \cite{lhcenergy}, the same relative uncertainty on the LHC 
beam energy is applied for the \sxw\ dataset, giving a slightly larger
uncertainty of 1.79\,\% due to the steeper relative dependence of \xtt\ on 
$\sqrt{s}$
in this region. These uncertainties are much larger than those corresponding
to the very small dependence of
\epsem\ on $\sqrt{s}$, which changes by only 0.5\,\% between 7 and 8\,\TeV.

\item[\bf Top quark mass:]
The simulation samples used in this analysis were generated with
$\mtop=172.5$\,\GeV, but the acceptance for \ttbar\ and $Wt$ events, and
the $Wt$ background cross-section itself, depend on the assumed \mtop\ value. 
Alternative samples generated with \mtop\ varied in the range 165--180\,\GeV\ 
were used to quantify these effects.
The acceptance and background effects partially cancel, and the final 
dependence of the result on the assumed \mtop\ value was
determined to be $\dmtop=\dmtopval$\,\%/\GeV. 
The result of the analysis is reported
assuming a fixed top mass of 172.5\,\GeV, and the small dependence of
the cross-section on the assumed mass is not included as a systematic
uncertainty.

\end{description}

\begin{table}[htp]
\caption{\label{t:systsum}Summary of the relative statistical, systematic
and total uncertainties on the measurements of the \ttbar\ production 
cross-section \xtt\ at \sxw\ and \sxv.}
\centering 

\begin{tabular}{lcc}
\hline\noalign{\smallskip}
Uncertainty & \multicolumn{2}{c}{$\Delta\xtt/\xtt$ (\%)} \\
$\sqrt{s}$ & 7\,\TeV & 8\,\TeV \\
\noalign{\smallskip}\hline\noalign{\smallskip}
Data statistics &  1.69 &  0.71 \\
\noalign{\smallskip}\hline\noalign{\smallskip}
\ttbar\ modelling and QCD scale &  1.46 &  1.26 \\
Parton distribution functions &  1.04 &  1.13 \\
Background modelling &  0.83 &  0.83 \\
Lepton efficiencies &  0.87 &  0.88 \\
Jets and $b$-tagging &  0.58 &  0.82 \\
Misidentified leptons &  0.41 &  0.34 \\
\noalign{\smallskip}\hline\noalign{\smallskip}
Analysis systematics (\xtt) &  2.27 &  2.26 \\
\noalign{\smallskip}\hline\noalign{\smallskip}
Integrated luminosity &  1.98 &  3.10 \\
LHC beam energy &  1.79 &  1.72 \\
\noalign{\smallskip}\hline\noalign{\smallskip}
Total uncertainty &  3.89 &  4.27 \\
\hline\noalign{\smallskip}
\end{tabular}
\end{table}

As shown in Tables~\ref{t:syst} and~\ref{t:systsum}, the largest systematic
uncertainties on \xtt\
come from \ttbar\ modelling and PDFs, and knowledge of the integrated
luminosities and LHC beam energy.

\subsection{Additional correlation studies}\label{ss:corl}

The tagging correlation \cb\ was determined from simulation to be
$1.009\pm 0.002\pm 0.007$ (\sxw) and $1.007\pm 0.002\pm 0.006$ (\sxv), where
the first uncertainty is due to limited sizes of the simulated samples, 
and the second is dominated by the comparison of predictions from different
\ttbar\ generators. 
Additional studies were carried out to probe the modelling of possible
sources of correlation. One possible source is the production of
additional \bbbar\ or \ccbar\ pairs in \ttbar\ production, which tends to 
increase both \cb\ and the number of events with three or more $b$-tagged jets,
which are not used in the measurement of \xtt. The ratio \rtt\ 
of events with at least three $b$-tagged jets to events with at least two 
$b$-tagged jets was
used to quantify this extra heavy-flavour production in data. It was measured
to be $\rtt=2.7\pm 0.4$\,\% (\sxw) and $2.8\pm 0.2$\,\% (\sxv), where the
uncertainties are statistical. These values are close to the  
{\sc Powheg + Pythia} prediction of $2.4\pm 0.1$\,\% 
(see Fig.~\ref{f:btags}), and well within the spread of \rtt\ values seen 
in the alternative simulation samples.

Kinematic correlations between the two $b$-jets produced in the \ttbar\ decay
could also produce a positive tagging correlation, as the efficiency to 
reconstruct and tag $b$-jets is not uniform as a function of $\pt$ and $\eta$.
For example, \ttbar\ pairs produced with high invariant mass tend to give
rise to  two back-to-back collimated top quark decay systems 
where both $b$-jets have
higher than average $\pt$, and longitudinal boosts of the \ttbar\ system
along the beamline give rise to $\eta$ correlations between the two jets.
These effects were probed by increasing the jet $\pt$ cut in steps from
the default of 25\,\GeV\ up to 75\,\GeV; above about 50\,\GeV, the simulation
predicts strong positive correlations of up to $\cb\approx 1.2$ for 
a 75\,\GeV\ $\pt$ cut. As shown for the \sxvt\ dataset in Fig.~\ref{f:stabpt}, 
the cross-sections fitted in data after taking these
correlations into account remain stable across the full $\pt$ cut range,
suggesting that any such kinematic correlations are well modelled by the 
simulation. Similar results were seen at \sxwt.
The results were also found to be stable within the 
uncorrelated components of the statistical and systematic uncertainties
when tightening the jet and lepton $\eta$ cuts, raising the lepton
$\pt$ cut up to 55\,\GeV\ and changing the $b$-tagging working point between
efficiencies of 60\,\% and 80\,\%. No additional uncertainties were assigned 
as a result of these studies.

\begin{figure}
\singlefigurex{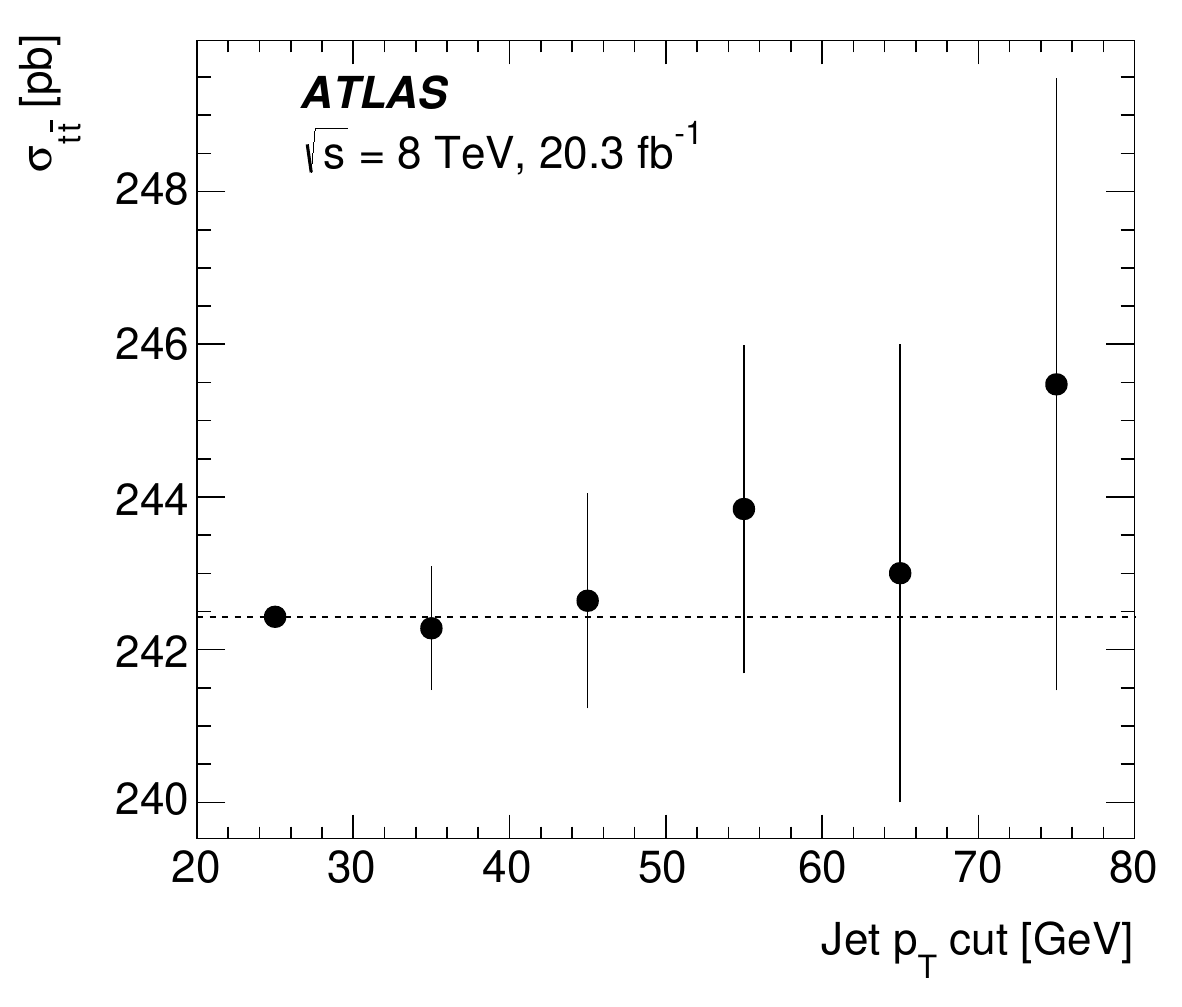}
\caption{\label{f:stabpt}Measured \ttbar\ cross-section at \sxv\ as a 
function of the $b$-tagged jet \pt\ cut. The error bars show the uncorrelated
part of the statistical uncertainty with respect to the baseline 
measurement with jet $\pt>25$\,\GeV.}
\end{figure}

\section{Results}\label{s:res}

Combining the estimates of \epsem\ and \cb\ from simulation samples, 
the estimates of the background \nib\ and \niib\ shown in 
Table~\ref{t:evtcount} and the data integrated luminosities, the
\ttbar\ cross-section was determined by solving Eq.~(\ref{e:tags}) to be:
\begin{eqnarray*}
\xtt & = & \ttxvalw\pm\ttxstatw\pm\ttxsystw\pm\ttxlumiw\pm\ttxebeamw\,\rm pb\ (\sxw) {\rm\ and} \\
\xtt & = & \ttxval\pm\ttxstat\pm\ttxsyst\pm\ttxlumi\pm\ttxebeam\,\rm pb\ (\sxv),\end{eqnarray*}
where the four uncertainties arise from data statistics, experimental
and theoretical systematic effects related to the analysis, 
knowledge of the integrated 
luminosity and of the LHC beam energy. The total uncertainties are 
\ttxtotw\,pb (\ttxrelw) at \sxwt\ and \ttxtot\,pb (\ttxrel) at \sxvt. A
detailed breakdown of the different components is given in Table~\ref{t:syst}. 
The results are reported for a fixed top quark mass of $\mtop=172.5$\,\GeV,
and have a dependence on this assumed value of $\dmtop=\dmtopval$\,\%/\GeV.
The product of jet reconstruction and $b$-tagging efficiencies \epsb\ was 
measured to be $0.557\pm 0.009$ at \sxw\ and 
$0.540\pm 0.006$ at \sxv, in both cases consistent with the values in
simulation.

The results are shown graphically as a function of $\sqrt{s}$ in 
Fig.~\ref{f:xsecsqrts}, together with previous ATLAS measurements 
of \xtt\ at \sxwt\ in the $ee$, $\mu\mu$ and $e\mu$ dilepton channels
using a count of the number of events with two leptons and at least two
jets in an 0.7\,\ifb\ dataset \cite{atlasxll}, and using a fit of
jet multiplicities and missing transverse momentum
in the $e\mu$ dilepton channel alone with the full 4.6\,\ifb\ dataset
\cite{aida}. The \sxwt\ results are all consistent, but 
cannot be combined as they are not based on independent datasets.
The measurements from this analysis at both centre-of-mass energies are 
consistent with the NNLO+NNLL QCD calculations discussed in 
Sect.~\ref{s:theo}. The \sxw\ result is
13\,\% higher than a previous measurement by the CMS collaboration 
\cite{cmsxw}, whilst the \sxv\ result is consistent with that from CMS 
\cite{cmsxv}.

\begin{figure}
\singlefigurex{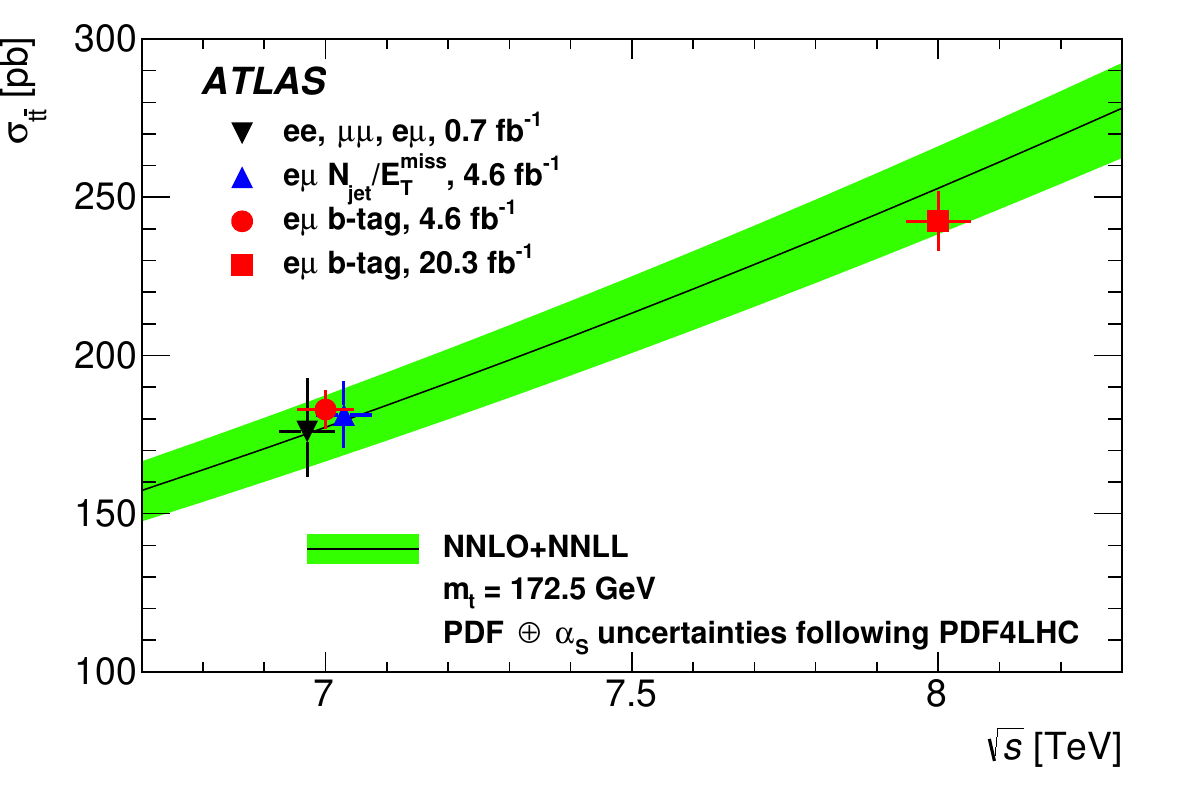}
\caption{\label{f:xsecsqrts}Measurements of the \ttbar\ cross-section at
\sxw\ and \sxv\ from this analysis ($e\mu$ $b$-tag) together with previous
ATLAS results at \sxw\ using the $ee$, $\mu\mu$ and $e\mu$ channels 
\cite{atlasxll} and using a fit to jet multiplicities and missing transverse
momentum in the $e\mu$ channel \cite{aida}. The uncertainties in 
$\sqrt{s}$ due to the LHC beam energy uncertainty are displayed as horizontal
error bars, and the vertical error bars do not include the corresponding
cross-section uncertainties. The three \sxw\ measurements
are displaced horizontally slightly for clarity. The NNLO+NNLL prediction 
\cite{topxtheot,toppp} described in Sect.~\ref{s:theo} is also shown
as a function of $\sqrt{s}$, for fixed $\mtop=172.5\,\GeV$ and with the 
uncertainties from PDFs, $\alpha_{\rm s}$ and QCD scale choices 
indicated by the green band.}
\end{figure}

From the present analysis, the ratio of cross-sections 
$\rxtt=$\xtt(8\,\TeV)/\xtt(7\,\TeV) was  determined to be:
\[
\rxtt=1.326\pm 0.024 \pm 0.015 \pm 0.049 \pm 0.001
\]
with uncertainties defined as above, adding in quadrature to a total of 
0.056. The experimental 
systematic uncertainties (apart from the statistical components of the 
lepton isolation and misidentified lepton uncertainties, which were
ev\-al\-uated independently from data in each 
dataset) and the LHC beam energy uncertainty are correlated between the 
two centre-of-mass energies. The luminosity uncertainties were taken to be 
uncorrelated between energies. The result is consistent
with the QCD NNLO+NNLL predicted ratio of 
$1.430\pm 0.013$ (see Sect.~\ref{s:theo}), which in addition to the
quoted PDF, $\alpha_{\rm s}$ and QCD scale uncertainties
varies by only $\pm 0.001$ for a $\pm 1$\,\GeV\ variation of \mtop.

\subsection{Fiducial cross-sections}\label{ss:fid}

The preselection efficiency \epsem\ can be written as the product of two
terms $\epsem=\aem\gem$, where the acceptance \aem\ represents the fraction
of \ttbar\ events which have a true opposite-sign $e\mu$ pair from 
$t\rightarrow W\rightarrow\ell$ decays (including via 
$W\rightarrow\tau\rightarrow\ell$), each with $\pt>25$\,\GeV\ and within
$|\eta|<2.5$, and \gem\ represents the reconstruction efficiency, i.e.
the probability that the two leptons are reconstructed and pass all the
identification and isolation requirements. A fiducial cross-section \xfid\ 
can then be defined as $\xfid=\aem\xtt$, and measured by replacing
$\xtt\epsem$ with $\xfid\gem$ in Eq.~(\ref{e:tags}), leaving the
background terms unchanged. Measurement of the fiducial cross-section avoids
the systematic uncertainties associated with \aem, i.e. the extrapolation
from the measured lepton phase space to the full phase space populated by
inclusive \ttbar\ production. In this analysis, these come mainly from
knowledge of the PDFs and the QCD scale uncertainties. Since the analysis
technique naturally corrects for the fraction of jets which are 
outside the kinematic acceptance through the fitted value of \epsb, no 
restrictions on jet kinematics are imposed in the
definition of \xfid. In calculating \aem\ and \gem\ from the various 
\ttbar\ simulation samples, the lepton four-momenta were taken after 
final-state radiation, and including the four-momenta of any photons
within a cone of size $\Delta R=0.1$ around the lepton direction, excluding 
photons
from hadron decays or produced in interactions with detector material.
The values of \aem\ are about 1.4\,\% (including the 
$\ttbar\rightarrow e\mu\nu\nubar\bbbar$ branching ratio), and those of 
\gem\ about 55\,\%, at both centre-of-mass energies.

The measured fiducial cross-sections at \sxwt\ and \sxvt, for leptons with
$\pt>25$\,\GeV\ and $|\eta|<2.5$, are shown in the first row of 
Table~\ref{t:fidres}. The relative uncertainties are shown in the lower
part of Table~\ref{t:syst}; the PDF uncertainties are substantially reduced 
compared to the inclusive cross-section measurement, and the QCD scale 
uncertainties are reduced to a negligible level. The \ttbar\ modelling
uncertainties, evaluated from the difference between {\sc Powheg+Pythia}
and {\sc MC@NLO+Herwig} samples increase slightly, though the differences
are not significant given the sizes of the simulated samples. Overall,
the analysis systematics on the fiducial cross-sections are 6--11\,\% smaller
than those on the inclusive cross-section measurements.

\begin{table*}[htp]
\caption{\label{t:fidres}Fiducial cross-section measurement results at
\sxw\ and \sxv, for different requirements on the minimum lepton \pt\
and maximum lepton $|\eta|$, and with or without the inclusion of leptons from
$W\rightarrow\tau\rightarrow\ell$ decays.
In each case, the first uncertainty is statistical, 
the second due to analysis systematic effects, the third due to the
integrated luminosity and the fourth due to the LHC beam energy.}
\centering

\begin{tabular}{ccccc}
\hline\noalign{\smallskip}
$p^\ell_{\rm T}$\,(\GeV) & $|\eta^\ell|$ & $W\rightarrow\tau\rightarrow\ell$ &
\sxw\ (pb) & \sxv\ (pb) \\
\noalign{\smallskip}\hline\noalign{\smallskip}
$> 25$ & $<2.5$ & yes  & $ 2.615\pm  0.044 \pm  0.056 \pm  0.052 \pm  0.047$ & $ 3.448\pm  0.025 \pm  0.069 \pm  0.107 \pm  0.059$ \\
$> 25$ & $<2.5$ & no  & $ 2.305\pm  0.039 \pm  0.049 \pm  0.046 \pm  0.041$ & $ 3.036\pm  0.022 \pm  0.061 \pm  0.094 \pm  0.052$ \\
$> 30$ & $<2.4$ & yes  & $ 2.029\pm  0.034 \pm  0.043 \pm  0.040 \pm  0.036$ & $ 2.662\pm  0.019 \pm  0.054 \pm  0.083 \pm  0.046$ \\
$> 30$ & $<2.4$ & no  & $ 1.817\pm  0.031 \pm  0.039 \pm  0.036 \pm  0.033$ & $ 2.380\pm  0.017 \pm  0.048 \pm  0.074 \pm  0.041$ \\
\hline\noalign{\smallskip}
\end{tabular}
\end{table*}

Simulation studies predict that $11.9\pm 0.1$\,\% of \ttbar\ events in the 
fiducial region have at least one lepton produced via 
$W\rightarrow\tau\rightarrow\ell$ decay.
The second row in Table~\ref{t:fidres} shows the fiducial cross-section
measurements scaled down to remove this contribution. The third and
fourth rows show the measurements scaled to a different lepton fiducial 
acceptance of $\pt>30$\,\GeV\ and $|\eta|<2.4$, a common phase space accessible 
to both the ATLAS and CMS experiments.

\subsection{Top quark mass determination}\label{ss:mtop}

The strong dependence of the theoretical prediction for \xtt\ on \mtop\ 
offers the possibility of interpreting measurements of \xtt\ as measurements
of \mtop. The theoretical calculations use the pole mass \mtpole, 
corresponding to the definition of the mass of a free particle, whereas the 
top quark mass measured through direct reconstruction of the top decay products
\cite{mtopmeas} may differ from the pole mass by $O(1$\,\GeV) 
\cite{buckleymoch}.
It is therefore interesting to compare the values of \mtop\ determined from
the two approaches, as explored previously by the D0 \cite{d0mtopxsec} and
CMS \cite{cmsmtopxsec} collaborations.

The dependence of the cross-section predictions (calculated as described in 
Sect.~\ref{s:theo}) on \mtpole\ is shown in Fig.~\ref{f:xsecmt} at both
\sxw\ and \sxv. The calculations were fitted to the parameterisation proposed
in Ref. \cite{topxtheot}, namely:
\begin{equation}\label{e:mpar}
\xtttheo(\mtpole)=\sigma(\mtref)\left(\frac{\mtref}{\mtpole}\right)^4 
(1+a_1x+a_2x^2)
\end{equation}
where the parameterisation constant 
$\mtref=172.5$\,\GeV, $x=(\mtpole-\mtref)/\mtref$, and $\sigma(\mtref)$,
$a_1$ and $a_2$ are free parameters. This function was used to parameterise
the dependence of \xtt\ on \mtop\ separately for each of the NNLO PDF sets
CT10, MSTW and NNPDF2.3, together with their uncertainty envelopes.

\begin{figure}
\singlefigurexw{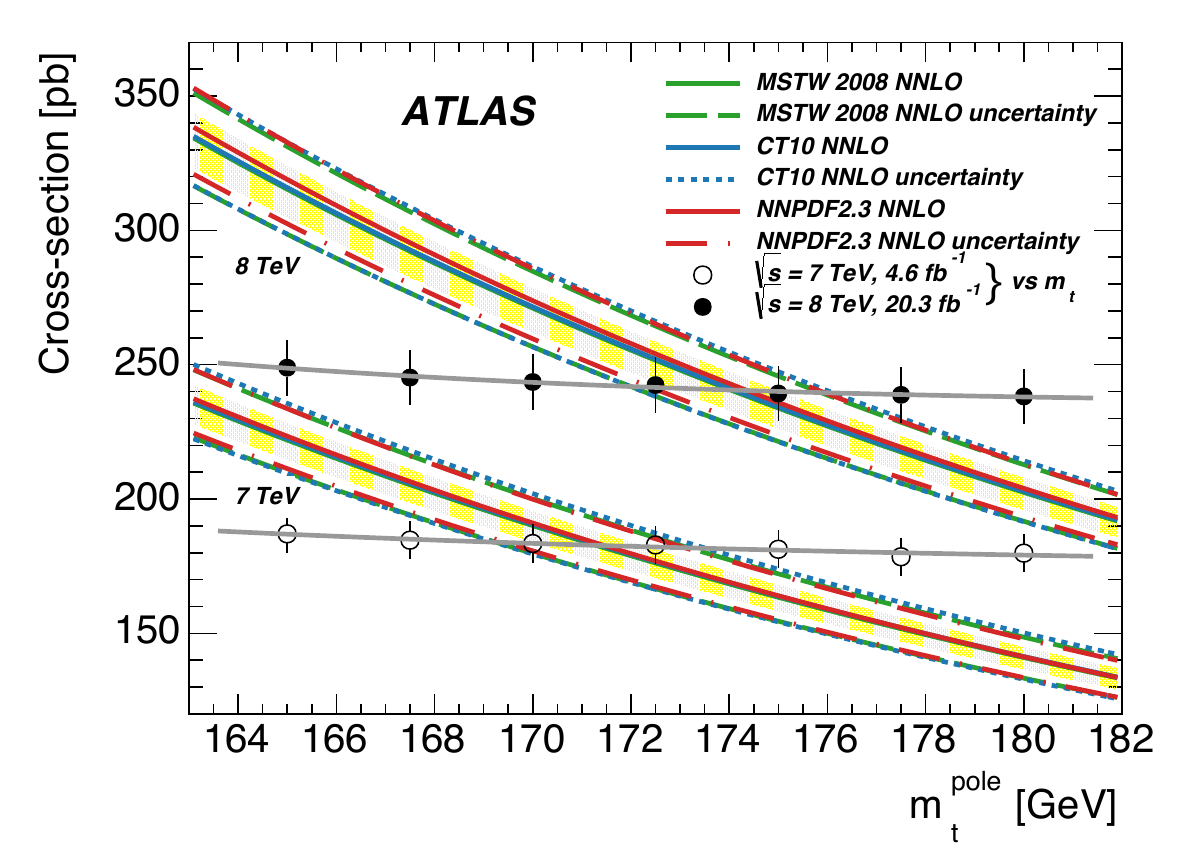}{92mm}{-3mm}
\caption{\label{f:xsecmt}Predicted NNLO+NNLL \ttbar\ production cross-sections
at \sxw\ and \sxv\ as a function of \mtpole, showing the central values (solid
lines) and total uncertainties (dashed lines) with several PDF sets. 
The yellow band shows the
QCD scale uncertainty. The measurements of \xtt\ are also
shown, with their dependence on the assumed value of \mtop\ through 
acceptance and background corrections parameterised using 
Eq.~(\ref{e:mpar}).}
\end{figure}

Figure~\ref{f:xsecmt} also shows the small dependence of the experimental 
measurement of \xtt\ on the assumed value of \mtop, arising from variations
in the acceptance and $Wt$ single top background, as discussed in 
Sect.~\ref{s:syst}. This dependence was also parameterised using 
Eq.~(\ref{e:mpar}), giving a derivative of 
$\dmtop=\dmtopval\pm\dmtoperr$\,\%/\GeV\
at 172.5\,\GeV\ for both centre-of-mass energies, where the uncertainty is
due to the limited size of the simulated samples. Here, \mtop\ represents the 
top quark mass used in the Monte Carlo generators, corresponding to that 
measured in direct reconstruction, rather than the pole mass. However, since 
this experimental dependence is small, differences between the two masses of
up to 2\,\GeV\ have a negligible effect ($<0.2$\,\GeV) on the pole mass 
determination. A comparison of the theoretical and experimental curves shown 
in Fig.~\ref{f:xsecmt} therefore allows an unambiguous extraction of the
top quark pole mass.

The extraction is performed by maximising the following Bayesian likelihood 
as a function of the top quark pole mass \mtpole:
\begin{eqnarray}\label{e:mlike}
\lefteqn{{\cal L}(\mtpole)=} \\
 & &  \int \gauss{\xttp}{\xtt(\mtpole)}{\xerrexp} \cdot
\gauss{\xttp}{\xtttheo(\mtpole)}{\xerrtheo\,} {\rm d}\xttp . \nonumber
\end{eqnarray}
Here, $\gauss{x}{\mu}{\rho}$ represents a Gaussian probability density in the
variable $x$ with mean $\mu$ and standard deviation $\rho$. The first Gaussian
term represents the experimental measurement \xtt\ with its dependence on 
\mtpole\ and uncertainty \xerrexp. The second Gaussian term 
represents the theoretical prediction given by Eq.~(\ref{e:mpar}) 
with its asymmetric uncertainty \xerrtheo\ obtained from the quadrature sum of 
PDF+$\alpha_{\rm s}$ and QCD scale uncertainties evaluated as
discussed in Sect.~\ref{s:theo}. The likelihood in Eq.~(\ref{e:mlike}) was 
maximised separately for each PDF set and centre-of-mass energy to give the 
\mtpole\ values shown in 
Table~\ref{t:mtoplike}. A breakdown of the
contributions to the total uncertainties is given for the CT10 PDF results
in Table~\ref{t:mtopsys}; it can be seen that the theoretical contributions
are larger than those from the experimental measurement of \xtt.
A single \mtpole\ value was derived for each centre-of-mass energy by
defining an asymmetric Gaussian  theoretical probability density in 
Eq.~(\ref{e:mlike}) with mean equal to the CT10 prediction, and 
a $\pm 1$ standard deviation uncertainty envelope
which encompasses the $\pm 1$ standard deviation uncertainties from each
PDF set following the PDF4LHC prescription \cite{pdflhc}, giving:
\begin{eqnarray*}
\mtpole & = &  171.4\pm  2.6\,\GeV \ (\sxw) {\rm\ and} \\
\mtpole & = &  174.1\pm  2.6\,\GeV \ (\sxv).
\end{eqnarray*}
Considering only uncorrelated experimental uncertainties, the two values
are consistent at the level of 1.7 standard deviations. 
The top pole mass was also extracted using a frequentist
approach, evaluating the likelihood for each \mtpole\ value as the 
Gaussian compatibility between the theoretically predicted and experimentally 
measured values, and fixing the theory uncertainties to those at 
$\mtpole=172.5$\,\GeV. The results differ from those of the Bayesian
approach by at most 0.2\,\GeV.

\begin{table}
\caption{\label{t:mtoplike} Measurements of the top quark pole mass determined
from the \ttbar\
cross-section measurements at \sxw\ and \sxv\ using various PDF sets.}
\centering
\begin{tabular}{lcc}
\hline\noalign{\smallskip}
 & \multicolumn{2}{c}{\mtpole (\GeV) from \xtt} \\
PDF & \sxw\ & \sxv \\
\noalign{\smallskip}\hline\noalign{\smallskip}
CT10 NNLO & $ 171.4\pm 2.6$ & $ 174.1\pm 2.6$ \\
MSTW 68\,\% NNLO & $ 171.2\pm 2.4$ & $ 174.0\pm 2.5$ \\
NNPDF2.3 5f FFN & $ 171.3^{+2.2}_{-2.3}$ & $ 174.2\pm 2.4$ \\
\hline\noalign{\smallskip}
\end{tabular}
\end{table}

\begin{table}
\caption{\label{t:mtopsys} Summary of experimental and theoretical 
uncertainty contributions
to the top quark pole mass determination at \sxw\ and \sxv\ with the CT10
PDF set.}
\centering
\begin{tabular}{lcc}
\hline\noalign{\smallskip}
$\Delta\mtpole$ (\GeV) & \sxw & \sxv \\
\noalign{\smallskip}\hline\noalign{\smallskip}
Data statistics & 0.6 & 0.3 \\
Analysis systematics & 0.8 & 0.9\\
Integrated luminosity & 0.7 & 1.2\\
LHC beam energy & 0.7 & 0.6\\
\noalign{\smallskip}\hline\noalign{\smallskip}
PDF+$\alpha_s$ &1.8 & 1.7\\
QCD scale choice & $^{+0.9}_{-1.2}$ & $^{+0.9}_{-1.3}$\\
\hline\noalign{\smallskip}
\end{tabular}
\end{table}

Finally, \mtpole\ was extracted from the combined \sxwt\ and \sxvt\ dataset
using the product of likelihoods (Eq.~(\ref{e:mlike})) for each 
centre-of-mass energy and accounting for correlations via nuisance 
parameters. The same set of experimental uncertainties
was considered correlated as for the cross-section ratio measurement, and
the uncertainty on \xtttheo\ was considered fully correlated between the two 
datasets. The resulting value using the envelope of all three considered PDF
sets is
\[
\mtpole = \mtopxs ^{+\emtopxsp}_{-\emtopxsm} \GeV
\]
and has only a slightly smaller uncertainty than the individual results
at each centre-of-mass energy, due to
the large correlations, particularly for the theoretical predictions.
The results are shown in Fig.~\ref{f:mtopsum}, together with
previous determinations using similar techniques from D0 \cite{d0mtopxsec}
and CMS \cite{cmsmtopxsec}. All extracted values are 
consistent with the average of measurements from kinematic reconstruction
of \ttbar\ events of $173.34\pm 0.76$\,\GeV\ \cite{mtopwa}, showing good 
compatibility of top quark masses extracted using very different techniques
and assumptions. 

\begin{figure}
\singlefigurexw{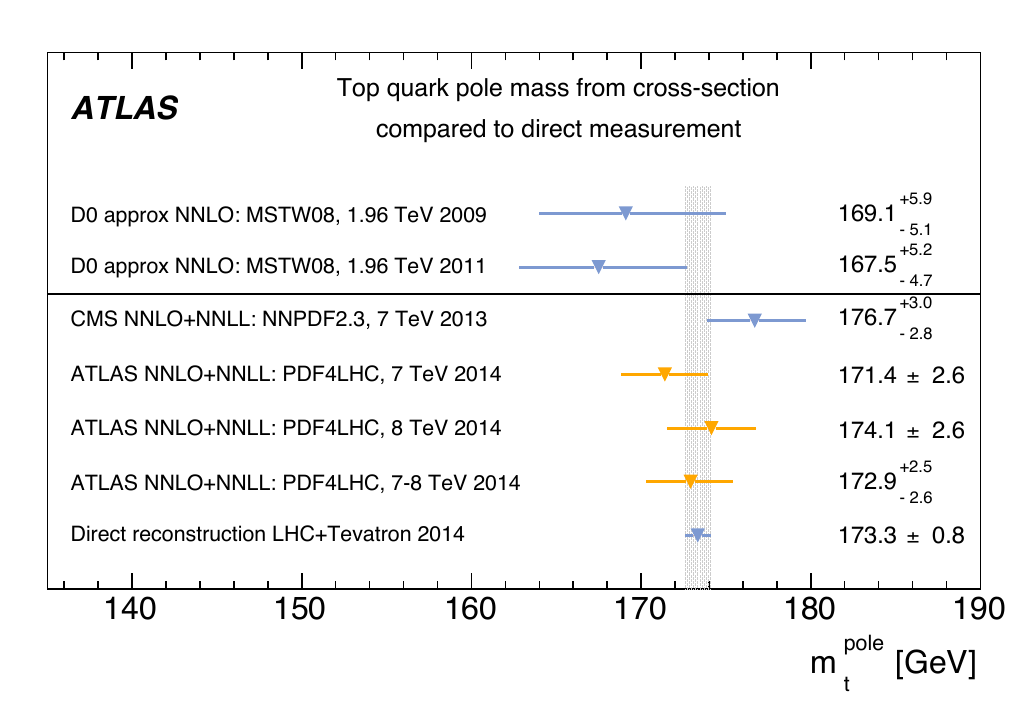}{96mm}{-6mm}
\caption{\label{f:mtopsum}Comparison of top quark pole mass values determined
from this and previous cross-section measurements \cite{d0mtopxsec,cmsmtopxsec}.
The average of top mass measurements from direct reconstruction 
\cite{mtopwa} is also shown.}
\end{figure}

\subsection{Constraints on stop-pair production}\label{ss:susy}

Supersymmetry (SUSY) theories predict new bosonic partners for the Standard
Model fermions and fermionic partners for the bosons. In the framework
of a generic $R$-parity conserving minimal supersymmetric extension of the
SM \cite{mssm}, SUSY particles are produced in pairs and the lightest
supersymmetric particle is stable. If SUSY is realised in
nature and responsible for the solution to the hierarchy problem, naturalness
arguments suggest that the supersymmetric partners of the top quark---the top
squarks or stops---should have mass close to \mtop\ in order to effectively 
cancel
the top quark loop contributions to the Higgs mass \cite{susytheo}. In this
scenario, the lighter top squark mass eigenstate \stopi\ would be produced in
pairs, and could decay via $\stopi\rightarrow t\xiz$ if $\mstop>\mtop+\mxiz$,
where \xiz, the lightest 
neutralino, is the lightest supersymmetric particle and is therefore stable. 
Stop-pair production could therefore give rise to $\ttbar\xiz\xiz$ intermediate
states, appearing like \ttbar\ production with additional missing transverse 
momentum carried away by the escaping neutralinos. The predicted cross-sections
at \sxvt\ are about 40\,pb for $\mstop=175$\,\GeV, falling to 20\,pb for 
200\,\GeV. If the top squark mass 
\mstop\ is smaller than about 200\,\GeV, such events would look very similar to
SM QCD \ttbar\ production, making traditional searches exploiting kinematic
differences very difficult, but producing a small excess in the measured
\ttbar\ cross-section, as discussed {\em e.g.} in Refs. \cite{stopexcess}.

The potential stop-pair signal yield was studied for top squark masses in 
the range 175--225\,\GeV\ and neutralino masses in the range 
1\,\GeV$<\mxiz<\mstop-\mtop$ using simulated samples generated with 
{\sc Herwig++} \cite{herwigpp} with the CTEQ6L1 PDFs \cite{ctsixpdf},
and NLO+NLL production cross-sections \cite{stopxsec}. The mixing matrices
for the top squarks and the neutralinos were chosen such that the top quark
produced in the $\stopi\rightarrow t\xiz$ decay has a right-handed 
polarisation in 95\,\% of the decays. Due to
the slightly more central $|\eta|$ distribution of the leptons from the
subsequent $t\rightarrow Wq$, $W\rightarrow\ell\nu$ decay,
the preselection
efficiency \epsem\ for these events is typically 10--20\,\% higher than for
SM QCD \ttbar, increasing with \mstop. However, the fraction of preselected
events with one or two $b$-tagged jets is very similar to the SM case. The 
effect of a small admixture of stop-pair production in addition to the SM
\ttbar\ production is therefore to increase the measured cross-section
by $\rstop\xstop$, where \rstop is the ratio of \epsem\ values for 
stop-pair and SM \ttbar\ production, and \xstop\ is the stop-pair
production cross-section.

Limits were set on stop-pair production by fitting the effective production 
cross-section $\rstop\xstop$ multiplied by a signal strength $\mu$ to the 
difference between the measured cross-sections (\xtt) and the 
theoretically predicted SM QCD production cross-sections (\xtttheo).
The two datasets were fitted simultaneously, assuming values of 
$\xtttheo=177.3^{+11.5}_{-12.0}$\,pb for \sxwt\ and 
$252.9^{+15.3}_{-16.3}$\,pb for \sxvt,
including the uncertainty due to a $\pm 1$\,\GeV\ variation
in the top quark mass. The limits were determined
using a profile likelihood ratio in the asymptotic limit \cite{asimov}, using
nuisance parameters to account for correlated theoretical and 
experimental uncertainties.

The observed and expected limits on $\mu$ at the 95\,\% confidence 
level (CL) were extracted using the CLs prescription \cite{cls} and are shown 
in Fig.~\ref{f:susy}. Due to the rapidly decreasing stop-pair production 
cross-section with increasing \mstop, the analysis is most sensitive below
180\,\GeV. Adopting the convention of reducing the estimated SUSY production
cross-section by one standard deviation of its theoretical uncertainty 
(15\,\%, coming from PDFs and QCD scale uncertainties \cite{susyunc}), 
stop masses between the top
mass threshold and \mstopexm\,\GeV\ are excluded, assuming 100\,\% branching 
ratio for $\stopi\rightarrow t\xiz$ and $\mxiz=1$\,\GeV. 
The limits from considering the
\sxw\ and \sxv\ datasets separately are only slightly weaker, due to the
large correlations in the systematic uncertainties between beam energies,
particularly for the theoretical predictions. At each energy, they 
correspond to excluded stop-pair production cross-sections of 25--27\,pb at
95\,\% CL.

\begin{figure}
\singlefigurex{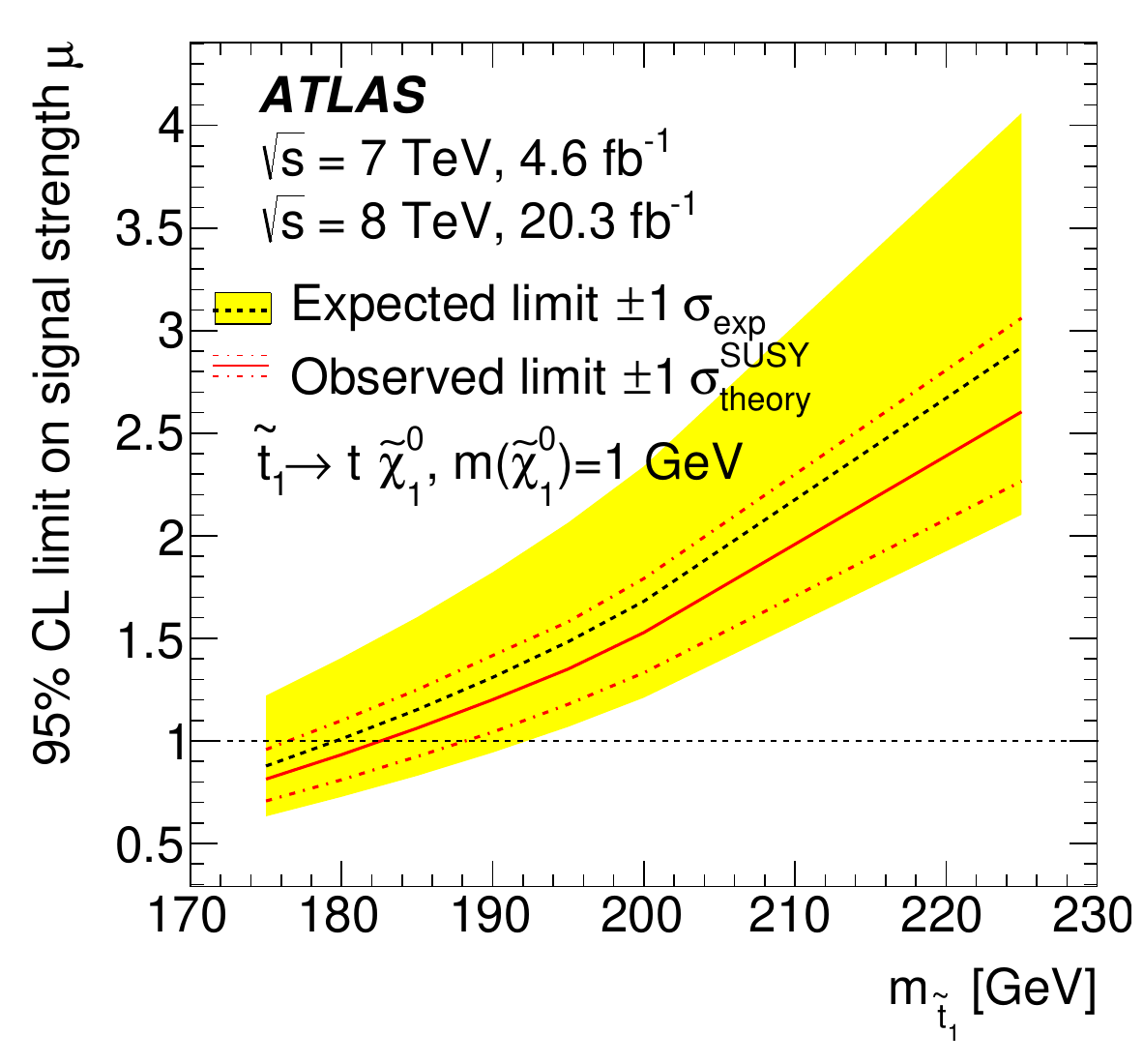}
\caption{\label{f:susy}Expected and observed limits at 95\,\% CL on
the signal strength $\mu$ as a function of \mstop, for pair produced top
squarks \stopi\ decaying with 100\,\% branching ratio via 
$\stopi\rightarrow t\xiz$ to predominantly right-handed top quarks, 
assuming $\mxiz=1$\,\GeV. 
The black dotted line shows the expected limit
with $\pm 1\sigma$ contours, taking into account all uncertainties 
except the theoretical cross-section uncertainties on the signal. The red
solid line shows the observed limit, with dotted lines indicating the changes
as the nominal signal cross-section is scaled up and down by its theoretical
uncertainty.}
\end{figure}

The combined cross-section limits depend
only slightly on the neutralino mass, becoming e.g. about 3\,\% weaker
at $\mstop=200$\,\GeV\ for $\mxiz=20$\,\GeV. However, they depend more 
strongly on the assumed top quark polarisation; in a 
scenario with $\mstop=175$\,\GeV\ and $\mxiz=1$\,\GeV, and squark mixing 
matrices chosen such that the top quarks are produced with full left-handed
polarisation, the limits are 4\,\% weaker than the predominantly right-handed
case, rising to 14\,\% weaker for $\mstop=200$\,\GeV. In this scenario, top
squarks with masses from \mtop\ to \mstopexmlh\,\GeV\ can be excluded. 
Although the analysis has some sensitivity to three-body top squark decays
of the form $\stopi\rightarrow bW\xiz$ for $\mstop<\mtop$, the $b$-jets
become softer than those from SM \ttbar\ production, affecting the
determination of \epsb. Therefore, no limits are set for scenarios
with $\mstop<\mtop$.

\section{Conclusions}\label{s:conc}

The inclusive \ttbar\ production cross-section has been measured 
at the LHC using
the full ATLAS 2011--2012 $pp$ collision data sample of \intlumiw\,\ifb\
at \sxwt\ and \intlumi\,\ifb\ at \sxvt, in the 
dilepton $\ttbar\rightarrow e\mu\nu\nubar\bbbar$ decay channel. The numbers
of opposite-sign $e\mu$ events with one and two $b$-tagged jets were
counted, allowing a simultaneous determination of the \ttbar\ cross-section
\xtt\ and the probability to reconstruct and $b$-tag a jet from a \ttbar\
decay. Assuming a top quark mass of $\mtop=172.5$\,\GeV, the results are:
\begin{eqnarray*}
\xtt & = & \ttxvalw\pm\ttxstatw\pm\ttxsystw\pm\ttxlumiw\pm\ttxebeamw\,\rm pb\ (\sxw) {\rm\ and} \\
\xtt & = & \ttxval\pm\ttxstat\pm\ttxsyst\pm\ttxlumi\pm\ttxebeam\,\rm pb\ (\sxv),\end{eqnarray*}
where the four uncertainties arise from data statistics, experimental and
theoretical systematic effects, knowledge of the integrated luminosity, 
and of the LHC beam 
energy, giving total uncertainties of \ttxtotw\,pb (\ttxrelw) and 
\ttxtot\,pb (\ttxrel) at \sxwt\ and \sxvt. The dependence of the results on 
the assumed value of \mtop\ is $\dmtop=\dmtopval$\,\%/\GeV, 
and the associated uncertainty is not 
included in the totals given above. The results are consistent with recent
NNLO+NNLL QCD calculations, and have slightly
smaller uncertainties than the theoretical predictions.
The ratio of the two cross-sections, and 
measurements in fiducial ranges corresponding to the experimental acceptance, 
have also been reported.

The measured \ttbar\ cross-sections have been used to determine the top quark
pole mass via the dependence of the predicted cross-section on \mtpole, giving
a value of $\mtpole=\mtopxs^{+\emtopxsp}_{-\emtopxsm}$\,\GeV,
compatible with the mass measured from kinematic reconstruction of
\ttbar\ events. 

The results have also been used to search for 
pair-produced supersymmetric top squarks decaying to top quarks and light
neutralinos. Assuming 100\,\% branching ratio for the decay 
$\stopi\rightarrow t\xiz$, and the production of predominantly right-handed 
top quarks, top squark masses between the top quark mass
and \mstopexm\,\GeV\ are excluded at 95\,\% CL.


\section{Acknowledgements}

We thank CERN for the very successful operation of the LHC, as well as the
support staff from our institutions without whom ATLAS could not be
operated efficiently.

We acknowledge the support of ANPCyT, Argentina; YerPhI, Armenia; ARC,
Australia; BMWFW and FWF, Austria; ANAS, Azerbaijan; SSTC, Belarus; CNPq and FAPESP,
Brazil; NSERC, NRC and CFI, Canada; CERN; CONICYT, Chile; CAS, MOST and NSFC,
China; COLCIENCIAS, Colombia; MSMT CR, MPO CR and VSC CR, Czech Republic;
DNRF, DNSRC and Lundbeck Foundation, Denmark; EPLANET, ERC and NSRF, European Union;
IN2P3-CNRS, CEA-DSM/IRFU, France; GNSF, Georgia; BMBF, DFG, HGF, MPG and AvH
Foundation, Germany; GSRT and NSRF, Greece; ISF, MINERVA, GIF, I-CORE and Benoziyo Center,
Israel; INFN, Italy; MEXT and JSPS, Japan; CNRST, Morocco; FOM and NWO,
Netherlands; BRF and RCN, Norway; MNiSW and NCN, Poland; GRICES and FCT, Portugal; MNE/IFA, Romania; MES of Russia and ROSATOM, Russian Federation; JINR; MSTD,
Serbia; MSSR, Slovakia; ARRS and MIZ\v{S}, Slovenia; DST/NRF, South Africa;
MINECO, Spain; SRC and Wallenberg Foundation, Sweden; SER, SNSF and Cantons of
Bern and Geneva, Switzerland; NSC, Taiwan; TA\-EK, Turkey; STFC, the Royal
Society and Leverhulme Trust, United Kingdom; DOE and NSF, United States of
America.

The crucial computing support from all WLCG partners is acknowledged
gratefully, in particular from CERN and the ATLAS Tier-1 facilities at
TRIUMF (Canada), NDGF (Denmark, Norway, Sweden), CC-IN2P3 (France),
KIT / GridKA (Germany), INFN-CNAF (Italy), NL-T1 (Netherlands), PIC (Spain),
ASGC (Taiwan), RAL (UK) and BNL (USA) and in the Tier-2 facilities
worldwide.

%

\clearpage

\section*{Addendum}\label{s:addendum}

The most precise measurement of the \ttbar\ cross-section (\xtt) 
in proton--proton
collisions at \sxvt\ from the ATLAS Collaboration was made using events with
an opposite-charge electron--muon pair and one or two $b$-tagged jets
\cite{TOPQ-2013-04}, and used a preliminary calibration of the integrated 
luminosity. The luminosity calibration has been finalised since
\cite{DAPR-2013-01} with a total uncertainty of 1.9\,\%, corresponding
to a substantial
improvement on the previous uncertainty of 2.8\,\%. Since the uncertainty on
the integrated luminosity contributed 3.1\,\% of the total 4.3\,\% uncertainty
on the \xtt\ measurement reported in \cite{TOPQ-2013-04}, a significant 
improvement in the measurement is possible by using the new 
luminosity calibration, as documented in this Addendum.

The new calibration corresponds to an integrated luminosity of 
\LLintlumi\,fb$^{-1}$ for
the \sxvt\ sample, a decrease of 0.2\,\%. The cross-section was 
recomputed taking into account the effects on both the conversion of the
\ttbar\ event yield to a cross-section, and the background estimates, giving
a result of:
\[
\xtt = \LLttxval\pm\LLttxstat\pm\LLttxsyst\pm\LLttxlumi\pm\LLttxebeam\,\rm pb,
\]
where the four uncertainties arise from data statistics, experimental and
theoretical systematic effects, knowledge of the integrated luminosity, 
and of the LHC beam 
energy, giving a  total uncertainty of \LLttxtot\,pb (\LLttxrel). The result
is consistent with the theoretical prediction of 
$252.9^{+13.3}_{-14.5}$\,pb, calculated at next-to-next-to-leading-order
with next-to-next-to-leading-logarithmic soft gluon terms with the 
{\tt top++ 2.0} program \cite{toppp} as discussed in detail in 
Ref. \cite{TOPQ-2013-04}.

The updated value of the ratio of cross-sections 
$\rxtt=$ \xtt(8\,\TeV)/\xtt(7\,\TeV) is:
\[
\rxtt=1.328\pm 0.024 \pm 0.015 \pm 0.038 \pm 0.001,
\]
with uncertainties defined as above, adding in quadrature to a total of 
0.047. The largest uncertainty comes from the uncertainties on the integrated
luminosities, considered to be uncorrelated between
the \sxwt\ and \sxvt\ datasets. 
This result is $2.1\sigma$ below the expectation of $1.430\pm 0.013$
calculated from {\tt top++ 2.0} as discussed in Ref. \cite{TOPQ-2013-04}.

The updated fiducial cross-sections, for a \ttbar\ decay producing an 
$e\mu$ pair within a given fiducial region,
are shown in Table~\ref{t:fidres}, updating Table~5
of Ref. \cite{TOPQ-2013-04}. The results are given both for the 
analysis requirements of $\pt>25\,\GeV$ and $|\eta|<2.5$ for both leptons,
and for a reduced acceptance of $\pt>30\,\GeV$ and $|\eta|<2.4$. They are given
separately for the two cases where events with either one or both leptons coming
from $t\rightarrow W\rightarrow\tau\rightarrow\ell$ rather than the
direct decay $t\rightarrow W\rightarrow\ell$ ($\ell=e$ or $\mu$) are 
included, or where the contributions involving $\tau$ decays are subtracted. The
results shown for the \sxwt\ data sample are unchanged with respect to those
in Ref. \cite{TOPQ-2013-04}. The results for the top quark pole mass and
limits on light supersymmetric 
top squarks presented in Ref. \cite{TOPQ-2013-04} are derived
from \sxwt\ and \sxvt\ cross-section measurements taken together, and 
would be only slightly improved by the luminosity update described here.

\begin{table*}[ht]
\centering

\begin{tabular}{ccc|c|c}\hline
$p^\ell_{\rm T}$ & $|\eta^\ell|$ & $W\rightarrow\tau$ &
\sxw\ (pb) & \sxv\ (pb) \\
(GeV) & & $\rightarrow\ell$ & & \\
\hline
$> 25$ & $<2.5$ & yes  & $ 2.615\pm  0.044 \pm  0.056 \pm  0.052 \pm  0.047$ & $ 3.455\pm  0.025 \pm  0.070 \pm  0.073 \pm  0.059$ \\
$> 25$ & $<2.5$ & no  & $ 2.305\pm  0.039 \pm  0.049 \pm  0.046 \pm  0.041$ & $ 3.043\pm  0.022 \pm  0.061 \pm  0.064 \pm  0.052$ \\
$> 30$ & $<2.4$ & yes  & $ 2.029\pm  0.034 \pm  0.043 \pm  0.040 \pm  0.036$ & $ 2.667\pm  0.019 \pm  0.054 \pm  0.056 \pm  0.046$ \\
$> 30$ & $<2.4$ & no  & $ 1.817\pm  0.031 \pm  0.039 \pm  0.036 \pm  0.033$ & $ 2.385\pm  0.017 \pm  0.048 \pm  0.050 \pm  0.041$ \\
\end{tabular}
\caption{\label{t:fidres}Fiducial cross-section measurement results at
\sxw\ and \sxv, for different requirements on the minimum lepton \pt\
and maximum lepton $|\eta|$, and with or without the inclusion of leptons from
$W\rightarrow\tau\rightarrow\ell$ decays, with the final 2012 luminosity
calibration. In each case, the first uncertainty is statistical, 
the second due to analysis systematic effects, the third due to the
integrated luminosity and the fourth due to the LHC beam energy.}
\end{table*}


\onecolumn
\clearpage
\begin{flushleft}
{\Large The ATLAS Collaboration}

\bigskip

G.~Aad$^{\rm 84}$,
B.~Abbott$^{\rm 112}$,
J.~Abdallah$^{\rm 152}$,
S.~Abdel~Khalek$^{\rm 116}$,
O.~Abdinov$^{\rm 11}$,
R.~Aben$^{\rm 106}$,
B.~Abi$^{\rm 113}$,
M.~Abolins$^{\rm 89}$,
O.S.~AbouZeid$^{\rm 159}$,
H.~Abramowicz$^{\rm 154}$
% ...
\bigskip
\\
% ...
$^{11}$ Institute of Physics, Azerbaijan Academy of Sciences, Baku, Azerbaijan\\
% ...
$^{84}$ CPPM, Aix-Marseille Universit{\'e} and CNRS/IN2P3, Marseille, France\\
% ...
$^{89}$ Department of Physics and Astronomy, Michigan State University, East Lansing MI, United States of America\\
% ...
$^{106}$ Nikhef National Institute for Subatomic Physics and University of Amsterdam, Amsterdam, Netherlands\\
% ...
$^{112}$ Homer L. Dodge Department of Physics and Astronomy, University of Oklahoma, Norman OK, United States of America\\
$^{113}$ Department of Physics, Oklahoma State University, Stillwater OK, United States of America\\
% ...
$^{116}$ LAL, Universit{\'e} Paris-Sud and CNRS/IN2P3, Orsay, France\\
% ...
$^{152}$ Institute of Physics, Academia Sinica, Taipei, Taiwan\\
% ...
$^{154}$ Raymond and Beverly Sackler School of Physics and Astronomy, Tel Aviv University, Tel Aviv, Israel\\
% ...
$^{159}$ Department of Physics, University of Toronto, Toronto ON, Canada\\
% ...
$^{*}$ Deceased
\end{flushleft}



\begin{thebibliography}{}
%
\bibitem{topxtheo}\npb{M. Beneke et al.}{Hadronic top-quark pair production with NNLL threshold resummation}{855}{2012}{695}{1109.1536};\\
\plb{M.~Cacciari et al.}{Top-pair production at hadron 
colliders with next-to-next-to-leading logarithmic soft-gluon resummation}
{710}{2012}{612}{1111.5869};\\
\prl{P.~B\"arnreuther et al.}{Percent Level Precision Physics at the Tevatron: First Genuine NNLO QCD Corrections to $\qqbar\rightarrow \ttbar$}{109}{2012}{132001}{1204.5201};\\
\jhep{M.~Czakon and A.~Mitov}{NNLO corrections to top-pair production at 
hadron colliders: the all-fermionic scattering channels}{1212}{2012}{054}
{1207.0236};\\
\jhep{M.~Czakon and A.~Mitov}{NNLO corrections to top pair production at hadron
colliders: the quark-gluon reaction}{1301}{2013}{080}{1210.6832}.

\bibitem{topxtheot}
\prl{M.~Czakon, P.~Fiedler and A.~Mitov}{The total top quark pair production
cross-section at hadron colliders through $\mathcal{O}(\alpha_S^4)$}
{110}{2013}{252004}{1303.6254}.

\bibitem{toppp}
\cpc{M. Czakon and A. Mitov}{Top++: a program for the calculation of the 
top-pair cross-section at hadron colliders}{182}{2014}{2930}{1112.5675}.

\bibitem{pdflhc}\arxiv{M. Botje et al.}{The PDF4LHC Working Group Interim 
Recommendations}{1101.0538}.

\bibitem{mstwnnlo}\epjc{A.D. Martin et al.}{Parton distributions for the LHC}
{63}{2009}{189}{0901.0002};\\
\epjc{A.D Martin et al.}{Uncertainties on $\alpha_s$ in global PDF analyses and 
implications for predicted hadronic cross sections}{64}{2009}{653}{0905.3531}.

\bibitem{cttenpdf}\prd{H.L.~Lai et al.}{New parton distributions for collider
physics}{82}{2010}{074024}{1007.2241}.

\bibitem{cttennnlo}\prd{J. Gao et al.}{The CT10 NNLO Global Analysis of QCD}
{89}{2014}{033009}{1302.6246}.

\bibitem{nnpdfffn}\npb{R. D.~Ball et al.}{Parton distributions with LHC data}
{867}{2013}{244}{1207.1303}.

\bibitem{hathor}\cpc{M.~Aliev et al.}{HATHOR---Hadronic top and heavy
quarks cross-section calculator}{182}{2011}{1034}{1007:1327}.

\bibitem{atlasdet}\ajinst{The ATLAS Experiment at the CERN Large Hadron
Collider}{3}{2008}{S08003}.

\bibitem{simul}\aepjc{The ATLAS Simulation Infrastructure}{70}{2010}{823}{1005.4568}.

\bibitem{geant4}\nima{S.~Agostinelli et al.}{GEANT4: A simulation toolkit }{506}{2003}{250}.

\bibitem{fastsim}\apubref{The simulation principle and performance of the 
ATLAS fast calorimeter simulation FastCaloSim}{2010-13}{1300517}.

\bibitem{pythia6}\jhep{T.~Sj\"ostrand, S.~Mrenna and P.~Skands}{PYTHIA 6.4 physics
and manual}{0605}{2006}{026}{hep-ph/0603175}.

\bibitem{pythia8}\cpc{T.~Sj\"ostrand, S.~Mrenna and P.~Skands}
{A brief introduction to PYTHIA 8.1}{178}{2008}{852}{0710.3820}.

\bibitem{powheg}\jhep{P.~Nason}{A new method for combining NLO QCD with shower Monte Carlo algorithms}{0411}{2004}{040}{hep-ph/0409146};\\
\jhep{S.~Frixione, P.~Nason and G.~Ridolfi}{A positive-weight 
next-to-leading-order Monte Carlo for heavy flavour hadroproduction}{0709}{2007}{126}{0707.3088};\\
\jhep{S.~Frixione, P.~Nason and C.~Oleari}{Matching NLO QCD computations with 
Parton Shower simulations: the POWHEG method}{0711}{2007}{070}{0709.2092}.

\bibitem{perugia}\prd{P.Z.~Skands}{Tuning Monte Carlo Generators: The Perugia 
Tunes}{82}{2010}{074018}{1005.3457}.

\bibitem{wbrpdg} J.~Erler and P.~Langacker,
{\em Electroweak model and constraints on new physics} in 
\prdna{Particle Data group, J.~Beringer et al.}{Review of Particle Physics}{86}{2012}{010001}. 

\bibitem{mcatnlo}\jhep{S.~Frixione and B.~Webber}{Matching NLO QCD computations
and parton shower simulations}{0206}{2002}{029}{hep-ph/0204244};\\
\jhep{S.~Frixione, P.~Nason and B.~Webber}{Matching NLO QCD and parton
showers in heavy flavour production}{0308}{2003}{007}{hep-ph/0305252}.

\bibitem{herwig}\jhep{G.~Corcella et al.}{HERWIG 6: An event generator for 
hadron emission reactions with interfering gluons (including supersymmetric
processes)}{0101}{2001}{010}{hep-ph/0011363}.

\bibitem{jimmy}\zphysc{J.M.~Butterworth, J.R.~Forshaw and M.H.~Seymour}
{Multiparton interactions in photoproduction at HERA}{72}{1996}{637}{hep-ph/9601371}.

\bibitem{auet}\apubref{New ATLAS event generator tunes to 2010 data}{2011-008}{1345343}.

\bibitem{alpgen}\jhep{M.L.~Mangano et al.}{ALPGEN, a generator for hard 
multiparton processes in hadronic collisions}{0307}{2003}{001}{hep-ph/0206293}.

\bibitem{ctsixpdf}\jhep{J.~Pumplin et al.}{New generation of parton distributions with uncertainties from global QCD analysis}{0207}{2002}{012}{hep-ph/0201195}.

\bibitem{powwtdr}\epjc{E.~Re}{Single-top $Wt$-channel production matched 
with parton showers using the POWHEG method}{71}{2011}{1547}{1009.2450}.

\bibitem{Wttheoxsec}\prd{N.~Kidonakis}{Two-loop anomalous dimensions for 
single top quark associated production with a $W^-$ or $H^-$}{82}{2010}{054018}{1005.4451}.

\bibitem{fewz}\cpc{R.~Gavin, Y.~Li, F.~Petriello and S.~Quackenbush}{FEWZ 2.0: A code for hadronic $Z$ production at next-to-next-to-leading order}{182}{2011}{2388}{1011.3540}.

\bibitem{dibmcfm}\npprocsup{J.M.~Campbell and R.K.~Ellis}{MCFM for the Tevatron
and the LHC}{205}{2010}{10}{1007.3492}.

\bibitem{madgraph}\jhep{J.~Alwall et al.}{MadGraph 5: Going Beyond}{1106}{2011}{128}{1106.0522}.

\bibitem{nlottwz}\jhep{J.M.~Campbell and R.K.~Ellis}{$\ttbar W^\pm$ 
production and decay at NLO}{1207}{2012}{052}{1204.5678};\\
\jhep{M.~Garzelli et al.}{$\ttbar W$ and $\ttbar Z$ Hadroproduction at NLO
accuracy in QCD with Parton Shower and Hadronization effects}{1211}{2012}{056}{1208.2665}.

\bibitem{sherpa}\jhep{T.~Gleisberg et al.}{Event generation with Sherpa 1.1}{0902}{2009}{007}{0811.4622}.

\bibitem{acer}\cpc{B.P.~Kersevan and E.~Richter-W\c{a}s}{The Monte Carlo 
event generator AcerMC version 2.0 with interfaces to PYTHIA 6.2 and HERWIG 6.5}{184}{2013}{919}{hep-ph/0405247}.

\bibitem{elecperf}\aepjc{Electron reconstruction and identification efficiency
measurements with the ATLAS detector using the 2011 LHC proton-proton 
collision data}{74}{2014}{2941}{1404.2240}.

\bibitem{muperf}\aepjc{Muon reconstruction efficiency
and momentum resolution of the ATLAS experiment in proton-proton collisions
at \sxv\ in 2010}{74}{2014}{3034}{1404.4562}.

\bibitem{mumini}\jhep{K.~Rethermann and B.~Tweedie}{Efficient identification 
of boosted semileptonic top quarks at the LHC}{1103}{2011}{059}{1007.2221}.

\bibitem{antikt}\plb{M.~Cacciari and G.P.~Salam}{Dispelling the $N^3$ myth for the
$k_t$ jet-finder}{641}{2006}{57}{hep-ph/0512210}; \\
\jhep{M.~Cacciari, G.P.~Salam and G.~Soyez}{The anti-$k_t$ jet clustering 
algorithm}{0804}{2008}{063}{0802.1189}.

\bibitem{jesx}\aepjc{Jet energy measurement with the ATLAS detector in 
proton-proton collisions at $\sqrt{s}=7$\,TeV}{73}{2013}{2304}{1112.6426}.

\bibitem{jetpile}\aconfref{Pileup subtraction and suppression for jets in ATLAS}{2013-083}{1570994}.

\bibitem{btagcom}\aconfref{Commissioning of the ATLAS high-performance
$b$-tagging algorithms in the 7\,TeV collision data}{2011-102}{1369219}.

\bibitem{btagptrel}\aconfref{Measurement of the $b$-tag efficiency in 
a sample of jets containing muons with 5\,\ifb\ of data from the ATLAS
detector}{2012-043}{1435197}.

\bibitem{atlasifsr}\aepjc{Measurement of \ttbar\ production with a veto on 
additional central jet activity in $pp$ collisions at $\sqrt{s}=7$\,TeV using
the ATLAS detector}{72}{2012}{2043}{1203.5015}.

\bibitem{wtinter}\jhep{C.~White, S.~Frixione, E.~Laenen and F.~Maltoni}
{Isolating $Wt$ production at the LHC}{0911}{2009}{074}{0908.0631}.

\bibitem{elecperfx}\aepjc{Electron performance measurements with the ATLAS
detector using the 2010 LHC proton-proton collision data}{72}{2012}{1909}
{1110.3174}.


\bibitem{jesxi}\arxiv{ATLAS Collaboration}{Jet energy scale and its 
systematic uncertainty in proton-proton collisions at $\sqrt{s}=7$\,TeV with 
the ATLAS detector}{1406.0076}, submitted to Eur.\ Phys.\ J.\ C.

\bibitem{jetres}\aepjc{Jet energy resolution in proton-proton collisions at 
$\sqrt{s}=7$\,TeV recorded in 2010 with the ATLAS detector}{73}{2013}{2306}
{1210.6210}.

\bibitem{systemviii}\aconfref{$b$-tagging efficiency calibration using the 
System8 method}{2011-143}{1386703}.

\bibitem{btagccal}\aconfref{$b$-jet tagging calibration on $c$-jets containing
$D^{*+}$ mesons}{2012-039}{1435193}.

\bibitem{btagmiscal}\aconfref{Measurement of the Mistag Rate of $b$-tagging
algorithms with 5\,\ifb\ of Data collected by the ATLAS Detector}{2012-040}{1435194}.

\bibitem{lumi}\aepjc{Improved luminosity determination in $pp$ collisions 
at $\sqrt{s}=7$\,TeV using the ATLAS detector at the LHC}{73}{2013}{2518}
{1302.4393}.

\bibitem{lhcenergy}J.~Wenninger, {\em Energy Calibration of the LHC Beams at 4\,TeV},
CERN-ATS-2013-40,\\ \url{http://cds.cern.ch/record/1546734}.


\bibitem{atlasxll}\ajhep{Measurement of the cross section for top-quark pair production in $pp$ collisions at $\sqrt{s}=7$\TeV\ with the ATLAS detector using
final states with two high-$p_T$ leptons}{1205}{2012}{059}{1202.4892}.

\bibitem{aida}\arxiv{ATLAS Collaboration}{Simultaneous measurements of the top 
quark pair, $W^+W^-$, and $Z/\gamma^{*}\rightarrow\tau\tau$ production 
cross-sections in $pp$ collisions with the ATLAS detector at \sxw}{1407.0573},
submitted to Phys.\ Rev.\ D.

\bibitem{cmsxw}
\jhep{CMS Collaboration}{Measurement of the \ttbar\ cross section in the dilepton channel in $pp$ collisions at \sxw}{1211}{2012}{067}{1208.2671}.

\bibitem{cmsxv}
\jhep{CMS Collaboration}{Measurement of the \ttbar\ production cross section in the dilepton channel in $pp$ collisions at \sxv}{1402}{2014}{024}{1312.7582}.

\bibitem{mtopmeas}
\prd{D0 Collaboration, V.~Abazov et al.}{Precise measurement of the top-quark
mass from lepton+jets events at D0}{84}{2011}{032004}{1105.6287};\\
\aepjc{Measurement of the top quark mass with the template
method in the top antitop $\rightarrow$ lepton+jets channel using ATLAS data}{72}{2012}{2046}{1203.5755};\\
\plb{CDF Collaboration, T.~Aaltonen et al.}{Precision Top-Quark Mass Measurements at CDF}{109}{2012}{152003}{1207.6758};\\
\jhep{CMS Collaboration}{Measurement of the top-quark mass in \ttbar\ events 
with lepton+jets final states in $pp$ collisions at \sxw}{1212}{2012}{105}{1209.2319}. 

\bibitem{buckleymoch}\phrep{A.~Buckley et al.}{General-purpose event generators for LHC physics}{504}{2011}{145}{1101.2599};\\
\arxiv{S.~Moch et al.}{High precision fundamental constants at the TeV scale}{1405.4781}.

\bibitem{d0mtopxsec}\prd{D0 Collaboration, V.~Abazov et al.}{Combination of 
\ttbar\ cross section measurements and constraints on the mass of the top 
quark and its decays into charged Higgs bosons}{80}{2009}{071102}{0903.5525};\\
\plb{D0 Collaboration, V.~Abazov et al.}{Determination of 
the pole and MSbar masses of the top quark from the \ttbar\ cross section}{703}{2011}{422}{1104.2887}.

\bibitem{cmsmtopxsec}
\plb{CMS Collaboration}{Determination of the top-quark
pole mass and strong coupling constant from the \ttbar\ production cross 
section in $pp$ collisions at \sxw}{728}{2014}{496}{1307.1907v4}
and Corrigendum, Phys.\ Lett.\ B {\bf 738} (2014) 526.

\bibitem{mtopwa}\arxiv{ATLAS, CDF, CMS and D0 Collaborations}{First combination
of Tevatron and LHC measurements of the top-quark mass}{1403.4427}.

\bibitem{mssm}\plbna{P.~Fayet}{Supersymmetry and Weak, Electromagnetic and Strong Interactions}{64}{1976}{159};\\
\plbna{P.~Fayet}{Spontaneously Broken Supersymmetric Theories of Weak, Electromagnetic and Strong Interactions}{69}{1977}{489};\\
\plbna{G.R.~Farrar and P.~Fayet}{Phenomenology of the Production, Decay, and Detection of New Hadronic States Associated with Supersymmetry}{76}{1978}{575};\\
\plbna{P.~Fayet}{Relations Between the Masses of the Superpartners of Leptons and Quarks, the Goldstino Couplings and the Neutral Currents}{84}{1979}{416};\\
\npbna{S.~Dimopoulos and H.~Georgi}{Softly Broken Supersymmetry and SU(5)}{193}{1981}{150}.

\bibitem{susytheo}\npbna{R.~Barbieri and G.~Giudice}{Upper Bounds on Supersymmetric Particle Masses}{306}{1988}{63};\\
\plb{B.~de Carlos and J.A. Casas}{One loop analysis of the electroweak 
breaking in supersymmetric models and the fine tuning problem}{309}{1993}{320}{hep-ph/9303291}.

\bibitem{stopexcess}\prd{M.R.~Buckley, T.~Plehn, M.J.~Ramsey-Musolf}{Stop on Top}{90}{2014}{014046}{1403.2726};\\
\arxiv{M.~Czakon et al.}{Closing the stop gap}{1407.1043}, submitted to Phys.\ Rev.\ Lett.

\bibitem{herwigpp}\epjc{M.~Bahr et al.}{Herwig++ physics and manual}{58}{2008}{639}{0803.0883}.

\bibitem{stopxsec}\npb{W.~Beenakker et al.}{Stop production at hadron colliders}{515}{1998}{3}{hep-ph/9710451};\\
\jhep{W.~Beenakker et al.}{Supersymmetric top and bottom squark production at hadron colliders}{1008}{2010}{098}{1006.4771};\\
\ijmpa{W.~Beenakker et al.}{Squark and gluino hadroproduction}{26}{2011}{2637}{1105.1110}.


\bibitem{asimov}\epjc{G.~Cowen et al}{Asymptotic formulae for likelihood-based
tests of new physics}{71}{2001}{1554}{1007.1727}.

\bibitem{cls}\jpgna{A.~Read}{Presentation of search results: the CL$_s$ 
technique}{28}{2002}{2693}.

\bibitem{susyunc}\arxiv{M.~Kramer et al.}{Supersymmetry production cross 
sections in $pp$ collisions at \sxw}{1206.2892}.

\bibitem{TOPQ-2013-04}\aepjc{\rhtitle}{74}{2014}{3109}{1406.5375} (this document).

\bibitem{DAPR-2013-01}\arxiv{ATLAS Collaboration}{Luminosity determination in $pp$ collisions at $\sqrt{s} = 8\;\mbox{TeV}$ using the ATLAS detector at the LHC}{1608.03953}, accepted by Eur.\ Phys.\ J.\ C.

\end{thebibliography}
\end{document}